\documentclass[onecollarge,natbib]{svjour2}
\bibpunct{[}{]}{;}{n}{}{,} 
\journalname{EPJ} 
\smartqed  
\usepackage[T1]{fontenc}
\usepackage{latexsym}
\usepackage{graphicx}
\usepackage{hyperref}

\usepackage{bbm}
\usepackage{amsfonts}
\usepackage[centertags]{amsmath}		
\usepackage{booktabs}				
\usepackage{multicol}
\usepackage{xcolor}
\usepackage{subfigure}

\newcommand{\alphastarone}{\alpha^{{}^{\star\left(1\right)}}_}
\newcommand{\betastarone}{\beta^{{}^{\star\left(1\right)}}_}
\newcommand{\gammastarone}{\gamma^{{}^{\star\left(1\right)}}_}
\newcommand{\deltastarone}{\delta^{{}^{\star\left(1\right)}}_}
\newcommand{\epsstarone}{\epsilon^{{}^{\star\left(1\right)}}_}

\newcommand{\alphastartwo}{\alpha^{{}^{\star\left(2\right)}}_}

\newcommand{\alphastarAmtwo}{\alpha^{{}^{\star\left(A-2\right)}}_}
\newcommand{\betastarAmtwo}{\beta^{{}^{\star\left(A-2\right)}}_}
\newcommand{\gammastarAmtwo}{\gamma^{{}^{\star\left(A-2\right)}}_}
\newcommand{\deltastarAmtwo}{\delta^{{}^{\star\left(A-2\right)}}_}

\newcommand{\alphastarthree}{\alpha^{{}^{\star\left(3\right)}}_}

\newcommand{\alphastarAmthree}{\alpha^{{}^{\star\left(A-3\right)}}_}
\newcommand{\betastarAmthree}{\beta^{{}^{\star\left(A-3\right)}}_}

\newcommand{\alphastaronestarone}{\left(\alphastarone{}\right){}^{{}^{\!\!\!\star\left(1\right)}}_{}}

\newcommand{\gammastaronestarone}{\left(\gammastarone{}\right){}^{{}^{\!\!\!\star\left(1\right)}}_{}}
\newcommand{\deltastaronestarone}{\left(\deltastarone{}\right){}^{{}^{\!\!\!\star\left(1\right)}}_{}}

\newcommand{\alphastaronestartwo}{\left(\alphastarone{}\right){}^{{}^{\!\!\!\star\left(2\right)}}_{}}

\newcommand{\deltastaronestartwo}{\left(\deltastarone{}\right){}^{{}^{\!\!\!\star\left(2\right)}}_{}}

\newcommand{\alphastaronestarAmthree}{\left(\alphastarone{}\right){}^{{}^{\!\!\!\star\left(A-3\right)}}_{}}

\newcommand{\epsstaronestarAmthree}{\left(\epsstarone{}\right){}^{{}^{\!\!\!\star\left(A-3\right)}}_{}}

\newcommand{\Ntot}{\mathcal{N}}
\newcommand{\blockqn}{(\mathcal{N},J,T)}

\begin{document}

\title{Jacobi no-core shell model for $p$-shell nuclei}
\author{S. Liebig\footnote{s.liebig@fz-juelich.de} \and U.-G. Mei{\ss}ner\footnote{meissner@hiskp.uni-bonn.de} 
\and A. Nogga\footnote{a.nogga@fz-juelich.de}}


\institute{S. Liebig  \and  U.-G. Mei{\ss}ner  \and A. Nogga \at                    
              Institute for Advanced Simulation, Institut f\"ur Kernphysik, 
              and J\"ulich Center for Hadron Physics, Forschungszentrum J\"ulich, 
              D-52425 J\"ulich, Germany    
           \and   
           S. Liebig  \and  U.-G. Mei{\ss}ner  \and A. Nogga \at
           JARA -- High Performance Computing, Forschungszentrum J\"ulich, D-52425 J\"ulich, Germany              
           \and 
           U.-G. Mei{\ss}ner  \at
             Helmholtz-Institut f\"ur Strahlen- und Kernphysik and 
             Bethe Center for Theoretical Physics, Universit\"at Bonn, D-53115 Bonn, Germany}

\date{Received: \today / Accepted: date}

\maketitle

\begin{abstract}
We introduce an algorithm to obtain coefficients of fractional parentage  
for light $p$-shell nuclei. The coefficients enable to use Jacobi coordinates 
in no-core shell model calculations 
separating off the center-of-mass motion. Fully antisymmetrized basis states 
are given together with recoupling coefficients that allow one to apply 
two- and three-nucleon operators. As an example, we study the 
dependence on the harmonic oscillator frequency of $^3$H, $^4$He, $^6$He, $^6$Li and $^7$Li
and extract their binding and excitation energies. The coefficients will be made 
openly accessible as HDF5 data files. 

\end{abstract}

\section{Introduction}
\label{sec:intro}

One of the major goals of nuclear physics is to understand properties of nuclei based 
on nuclear two-, three- and maybe more-body interactions. To this aim, methods have to be devised 
that allow one to predict such properties based on these interactions. In the very light systems, 
calculations are often directly done in configuration or momentum space 
\cite{Glockle:1996ca,Nogga:2000bb,Lazauskas:2015hr,Deltuva:2007ga}. Calculations using 
special basis sets, e.g. hyperspherical harmonics \cite{Viviani:2005js,Barnea:2010bu}, 
Sturmians \cite{Caprio:2012hl} or harmonic oscillator (HO) states \cite{Barrett:2013hr}, are also able to 
provide accurate solutions for the light systems but become the tool of choice for systems larger 
then $A = 4$.  

Here we will concentrate on the no-core shell model (NCSM) that has become a standard method to 
perform nuclear structure calculations for $p$-shell nuclei  (for recent applications 
see e.g. \cite{Barrett:2013hr,Roth:2007bb,Forssen:2009jt})
and is based on an expansion in terms of HO states. Although the Gaussian long distance behavior of the 
HO states is not particularly well suited for the description of the long distance 
behavior of nuclear wave functions \cite{Stetcu:2005bx}, the basis enables one to separate out 
the center-of-mass (CM) motion exactly and, which will be important below, to perform exact transformation 
between different choices of coordinates  within a finite set of HO states. Binding energies and especially 
excitation energies do not depend strongly on the long range behavior of the wave function and can  
therefore be predicted with high accuracy except for states that are dominated by HO excitations as for 
example $\alpha$-cluster states. Other schemes, like nuclear lattice calculations, are more suited for such
states \cite{Meissner:2015vp}. Nevertheless, due to its flexibility with respect to interactions, the NCSM became 
particularly useful for the study of chiral nucleon-nucleon (NN) and three-nucleon (3N) interactions
\cite{Nogga:2006jc,Navratil:2007kr,Binder:2015va}. Using an importance truncation scheme, the 
extension to more complex nuclei is possible \cite{Roth:2007bb}. 

Whereas $s$-shell nuclei are usually calculated using Jacobi coordinates within the NCSM 
\cite{Navratil:2000bc}, more complex systems have so far been mostly calculated using the so-called 
$m$-scheme basis where all nucleons are described by single particle states. This avoids the difficult 
antisymmetrization of states expressed in Jacobi coordinates. The price for this simplification is that 
the CM motion cannot be explicitly separated out anymore leading to much larger dimensions 
of the linear equations to be solved and. Furthermore, expensive transformations of interaction matrix elements 
from relative coordinates to single particle coordinates are necessary. Often these transformations have to be 
performed on-the-flight since matrix elements of the interactions cannot be stored in the single particle 
basis due to memory constraints. These constraints are especially relevant since it is clear 
by now that, for accurate calculations, chiral nuclear interactions of high order in chiral expansion 
are required \cite{Epelbaum:2009hy,Machleidt:2011gh} which implies that even four-nucleon interactions might be relevant 
\cite{Epelbaum:2006ee,Epelbaum:2007if,Nogga:2010ix}. The on-the-flight transformation of 
such interactions will be tremendously more difficult. 

The NCSM describes many-body systems containing $A$ point-like
non-relativistic nucleons in the HO basis where all $A$ nucleons of
the system are considered to be active \cite{Barrett:2013hr}. This HO basis 
allows one to represent the full complexity of nuclear interactions efficiently. 
But in order to reach converged results in practical calculations, the 
interactions have to be soft and should not include the strong repulsion
which is part of most nuclear interaction models. In order to be able to 
study nuclear systems, most of the standard interactions are only the starting point 
for obtaining a soft effective interaction. Early NCSM calculations relied 
on a decoupling formulated specifically for HO spaces (see e.g. \cite{Nogga:2006jc} for 
a summary of this approach). In this case, the effective interactions depend on the 
HO frequency and model space  size and are useful only for NCSM calculations. As within 
all approaches to effective interactions, many-body forces are induced. These 
have been included up to the level of three-nucleon forces (3NFs) \cite{Navratil:2002db}
which is sufficient to obtain converged results. But this approach has several disadvantages. 
The most important one is probably, that the effective interaction cannot be used elsewhere,
e.g. in Faddeev-Yakubovsky or even coupled-cluster calculations so that it is difficult 
to benchmark results and to check that induced many-body forces do not have 
large effects on other observables. The convergence pattern for these interactions is also 
more complicated since convergence for binding energies can be reached from above and below. 
These disadvantages can be circumvented using interactions that either constrain the interactions 
to low momenta as in the case of $V_{lowk}^{}$ \cite{Bogner:2003hx} or decouple low- and 
high momentum components using the similarity renormalization group (SRG) \cite{Bogner:2007hn}. 
In both cases, the interactions become soft enough so that converged results can be obtained. 
In recent years, SRG has become the tool of choice since it is also possible to obtain 
induced 3NFs \cite{Jurgenson:2009hq}. Our test calculations below are therefore also 
based on this approach. 

The subject of this work is to come back to the development of a Jacobi relative coordinate NCSM 
started in \cite{Navratil:2000bc} and extend it towards 
$p$-shell nuclei. The main difficulty is to built up an antisymmetrized set of nuclear HO states using 
Jacobi relative coordinates. Our algorithm to obtain these states is described in Section~\ref{sec:antisym}. 
These states alone are still not useful for applications. In order to be able to calculate matrix elements of 
two-body operators, we also need recoupling coefficients that separate out NN states from the 
$A$-body system. In Section~\ref{sec:nnrecoupl}, it is summarized how these transitions can be done. 
This is then extended to transition coefficients that separate out 3N clusters 
in Section~\ref{sec:3nrecoupl}. Such coefficients will be important in future to apply 3N interactions 
within this scheme. Using the new antisymmetrized states and the transitions to states that single 
out an NN subsystem allows us to do first example calculations for the binding and excitation 
energies for light nuclei in Section~\ref{sec:energies}. We use a new, mostly automatized 
scheme to extract binding energies and estimates of the numerical error from our results, which 
can be particularly easy performed based on the Jacobi basis states since 
binding energy calculations can be done for a wide range of HO frequencies. 
Once the antisymmetrized basis states are obtained,  the individual binding energy calculations are not computationally 
expensive anymore since the states are independent of the interaction and the HO frequency. In the appendices,
we summarize the implications of the conventions used for HO wave functions and list the sets of 
antisymmetrized HO states and transition coefficients that have been generated.

\section{Antisymmetrized HO states in a Jacobi basis}
\label{sec:antisym}

The Schr\"odinger equation for the internal motion of the $A$-nucleon system reads 
 \begin{eqnarray}
    H_A^{}
     & = &  \sum_{i=1}^{A} \frac{k_i^2}{2m} +   \sum_{i<j=1}^{A} V_{ij}^{}+\sum_{i<j<k=1}^{A} V_{ijk}^{}  -\frac{\vec P^2}{2M}  \cr 
     & = &   \sum_{i<j=1}^{A}  \frac{2}{A} \frac{ \vec{p}_{ij}^{\; 2} }{m}  +   \sum_{i<j=1}^{A} V_{ij}^{}+\sum_{i<j<k=1}^{A} V_{ijk}^{}  \ .
   \label{eq:HHO_jac}
  \end{eqnarray}
 Here the CM kinetic  energy $\frac{\vec P^2}{2M}$, where M is the total mass of the $A$-nucleon system and $\vec P$ the CM momentum, 
 is substracted to obtain the internal energy. We included 2N interactions of the pair $(ij)$ ($V_{ij}^{}$) and 3N interactions of the triplet $(ijk)$ 
 ($V_{ijk}^{}$).  The expression is rewritten such that  the individual momenta $\vec k_i$ of the nucleons are replaced by pair 
 momenta $\vec p_{ij}^{} = \frac{1}{2} (\vec k_i^{} - \vec k_j^{})$. $m$ is the mass of the nucleon. Here, we neglect the small difference of proton and neutron
 mass. 
 
 We will solve this equation in a basis $|  \, \alpha \, \rangle$ of antisymmetrized HO states
\begin{eqnarray}
   \langle \, \alpha \, | 
    H_A^{}
   | \,\beta\, \rangle 
   \langle \,\beta\, |\, \Psi\, \rangle
   & = & E \; \langle \,\alpha\, |\, \Psi\, \rangle  
   \label{eq:SE}
  \end{eqnarray}
  where a sum over all these HO states $|  \, \beta \, \rangle$ is implied. The difficulty is to define 
  the set of antisymmetrized states in Jacobi coordinates.

\begin{figure}
\centering
\includegraphics[scale=0.5]{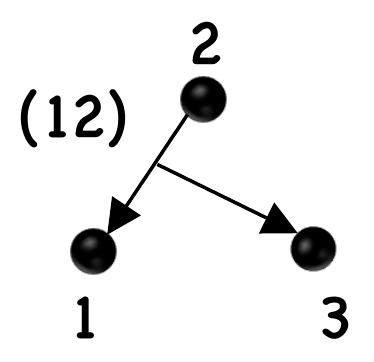}   
\hspace{2cm} \includegraphics[scale=0.5]{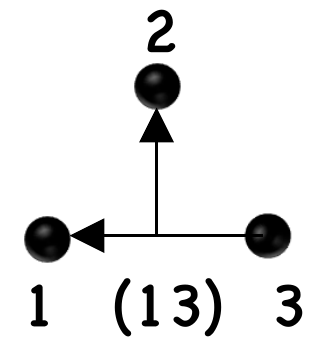} 

\caption{Three-cluster Jacobi coordinates. The left hand side singles out the third particle as spectator. The right hand side singles out the second 
one. The arrow defines the direction of corresponding relative momenta or positions. This direction defines the phases depending on the corresponding angular momenta. \label{fig:gen3body}}
\end{figure}

 \subsection{General set of Jacobi coordinates for three clusters}
   
  In order to find these antisymmetrized states, we start from a general set of Jacobi coordinates 
  for a system of three clusters $1$, $2$ and $3$. Each of the clusters is characterized by its mass, total angular momentum 
  and total isospin, $m_{1,2,3}^{}$, $s_{1,2,3}^{}$ and $t_{1,2,3}^{}$. The motion is then described by the motion within the pair (i.e. $(12)$) 
  and the motion of the spectator $3$. For an HO basis,  the corresponding basis reads 
  \begin{equation}
   \left| \ n_{12}^{} n_{3}^{}  \left(  \left( l_{12}^{} \, (s_1^{} s_2^{}) S_{12}^{} \right) J_{12}^{} \;
                    (l_3^{} s_3^{}) I_3^{}   \right) J; \,   \ 
                    \left( (t_1^{} t_2^{}) T_{12}^{} \, t_3^{} \right) T \  \right\rangle 
   \label{eq:jacobi12}
  \end{equation}
  where $n_{12}^{}$ ($n_3^{}$)is the HO quantum number for the relative motion of $1$ and $2$ (of the spectator $3$), $l_{12}^{}$ ($l_3^{}$) the corresponding 
  orbital angular momenta, $S_{12}^{}$, $J_{12}^{}$ and $I_3^{}$ are the pair spin, pair total angular momentum and the total angular momentum of the spectator and $J$ is the total 
  angular momentum of the system. The isospins of the pair couple to $T_{12}^{}$ which combines with the isospin of the spectator particle to 
  the total isospin $T$.  The states are eigenstates of HOs in the relative coordinates 
  \begin{equation}
  \label{eq:horel}
  H_{HO,rel}^{} = \frac{\vec p_{12}^2}{2\mu_{12}^{}} 
                + \frac{\vec p_{3}^2}{2\mu_{3}^{}} 
                + \frac{1}{2} \mu_{12}^{} \omega^2 \vec r_{12}^2  
                + \frac{1}{2} \mu_{3}^{} \omega^2 \vec R_{3}^2 \ .
   \end{equation}
  where the reduces masses  are defined as 
  \begin{equation}
  \mu_{12}^{}=\frac{m_1^{} m_2^{}}{ m_1^{}  + m_2^{} }\  ,     \  \  \   \mu_{3}^{}=\frac{(m_1^{}+m_2^{})m_3^{}}{ m_1^{}  + m_2^{} + m_3^{}}  
  \end{equation} 
  and the relative coordinates in terms of single cluster coordinates (momenta) $\vec r_i$ ($\vec k_i$) are given by 
  \begin{eqnarray}
  \vec r_{12}^{}= \vec r_1^{} - \vec r_2^{}\  ,   
    & \qquad & \vec p_{12}^{}=\frac{m_2^{}}{m_1^{}+m_2^{}}  \vec k_1^{} - \frac{m_1^{}}{m_1^{}+m_2^{}} \vec k_2^{}  \nonumber \\[10pt]
  \vec R_{3}^{}= \vec r_3^{} - \frac{m_1^{} \vec r_1^{}  + m_2^{} \vec r_2^{} }{m_1^{}+m_2^{}}\  , 
    & \qquad  & \vec p_{3}^{}=\frac{m_1^{}+m_2^{}}{m_1^{}+m_2^{}+m_3^{}}  \vec k_3^{} - \frac{m_3^{}}{m_1^{}+m_2^{}+m_3^{}}( \vec k_1^{}+\vec k_2^{})  \ . 
  \end{eqnarray}
  Here we omitted the internal state of the clusters since it will only become relevant later. 
  This kind of Jacobi coordinate is depicted on the left hand side of Fig.~\ref{fig:gen3body}.  Note that the direction of the arrows defines 
  the direction of corresponding relative positions or momenta as given above. 
   
  For such a general set of Jacobi coordinates, we need to perform a coordinate transformation to 
  the Jacobi coordinates depicted on the right hand side of Fig.~\ref{fig:gen3body}. Such a transformation does not change the 
  internal motion of the clusters and the total parity, angular momentum and isospin.  
  For an HO basis,  the corresponding states read 
   \begin{equation}
  \left| \ n_{13}^{}  n_2^{} \left(  \left( l_{13}^{} \, (s_1^{} s_3^{}) S_{13}^{} \right) J_{13}^{} \;
                     (l_2^{} s_2^{}) I_2^{}   \right) J \,   
                     \ \left( (t_1^{} t_3^{}) T_{13}^{} \, t_2^{} \right) T \  \right\rangle 
    \label{eq:jacobi13}
  \end{equation}
  and singles out the second particle as the spectator with corresponding definitions of the relative coordinates and momenta. 
  A special property of HO states is that also the 
  total HO energy quantum number 
  \begin{equation}
  \Ntot = 2n_{12}^{}+l_{12}^{} +2n_{3}^{}+l_{3}^{}  =  2n_{13}^{}+l_{13}^{}   + 2n_{2}^{}+l_{2}^{}   
  \end{equation}
 is conserved.
 
  In order to relate the spatial part of the transitions to Talmi-Moshinsky brackets \cite{Talmi:1952js,Moshinsky:1959eb}, 
  we introduce dimensionless relative coordinates using the oscillator lengths  $b_{12}^{} = \sqrt{\frac{1}{\mu_{12}^{}\,\omega}}$ and 
  $b_{3}^{} = \sqrt{\frac{1}{\mu_{3}^{}\,\omega}}$
  \begin{equation}
    \vec{\rho}_{12}^{} = \frac{\vec r_{12}^{}}{b_{12}^{}}\  ,   \ \ \ \vec{\rho}_{3}^{} = \frac{\vec R_{3}^{}}{b_{3}^{}} \ . 
  \end{equation}  
  The coordinate transformation can then be put into the form   of Ref.  \cite{Kamuntavicius:2001ig}
  \begin{equation}
   \left(\begin{array}{c}
          \vec{\rho}_{13}^{}\\
          -\vec{\rho}_2^{}
         \end{array}
   \right)
   = 
   \left(\begin{array}{cc}
          \sqrt{\frac{d}{1+d}} & \sqrt{\frac{1}{1+d}}\\
          \sqrt{\frac{1}{1+d}} &-\sqrt{\frac{d}{1+d}}
         \end{array}
   \right)\,
        \left(\begin{array}{c}
                \vec{\rho}_{12}^{}\\
                -\vec{\rho}_3^{}
               \end{array}\right) 
\end{equation}               
where $\displaystyle d  =  \frac{m_2^{}\, m_3^{}}{m_1^{}\left(m_1^{}+m_2^{}+m_3^{}\right)}$. 
Note that the additional minus signs required in front of $\vec \rho_2^{}$ and $\vec \rho_3^{}$ 
need to be taken into account  by an extra phase factor $(-)^{l_2^{}+l_3^{}}$. 
The spatial part is therefore given by the corresponding HO bracket. The spin and isospin part just
requires recoupling. One therefore finds for the general coordinate transformation 
\begin{eqnarray}
& & \!\!\!\! \left\langle  \ n_{13}^{} n_{2}^{} \left(  \left( l_{13}^{} \, (s_1^{} s_3^{}) S_{13}^{} \right) J_{13}^{} \:
                        (l_2^{} s_2^{}) I_2^{}   \right)J \, \left( (t_1^{} t_3^{}) T_{13}^{} \, t_2^{} \right) T       
           \right. \left| \ n_{12}^{} n_{3}^{}   \left(  \left( l_{12}^{} \, (s_1^{} s_2^{}) S_{12}^{} \right) J_{12}^{} \:
                        (l_3^{} s_3^{}) I_3^{}   \right)J \, \left( (t_1^{} t_2^{}) T_{12}^{} \, t_3^{} \right) T  \right\rangle \nonumber \\[10pt]
 & & \quad  = 
   \hat J_{13}^{} \, \hat I_2^{} \, 
   \hat J_{12}^{} \, \hat I_3^{} \;
   \sum_{LS} \; 
    \hat L^2_{} \, \hat S^2_{} 
    \left\{ \begin{array}{ccc}
          l_{13}^{} & S_{13}^{} & J_{13}^{} \\
          l_{2}^{} & s_2^{} & I_2^{} \\
          L & S & J
         \end{array}
    \right\}        
    \left\{ \begin{array}{ccc}
          l_{12}^{} & S_{12}^{} & J_{12}^{} \\
          l_{3}^{} & s_3^{} & I_3^{} \\
          L & S & J    
         \end{array}     \right\} \nonumber \\[10pt]
&&  \qquad        
           \ (-1)^{l_2^{}+l_3^{}}\;
  \langle n_{13}^{}\; l_{13}^{}\, ,
      \;  n_2^{}\; l_2^{}\: : \, L \, | 
      \,  n_{12}^{}\; l_{12}^{}\, ,
      \;  n_3^{}\, l_3^{}\: : \, L
  \rangle_{d}  \nonumber \\[10pt]
& & \qquad \left(-1\right)^{S_{13}^{}+s_2^{}+S_{12}^{}+s_3^{}}
        \hat S_{13}^{} \, \hat S_{12}^{}\,
     \left\{\begin{array}{ccc}
            s_2^{} & s_1^{} & S_{12}^{}\\
            s_3^{} & S & S_{13}^{}
            \end{array}
     \right\}  \ \left(-1\right)^{T_{13}^{}+t_2^{}+T_{12}^{}+t_3^{}}
        \hat T_{13}^{} \, \hat T_{12}^{}\,
     \left\{\begin{array}{ccc}
            t_2^{} & t_1^{} & T_{12}^{}\\
            t_3^{} & T & T_{13}^{}
            \end{array}
     \right\}  \ . 
  \label{eq:gencoordchange}   
\end{eqnarray}
For quantum numbers, we use the abbreviation $\hat l = \sqrt{2l+1}$. The HO 
bracket $ \langle n_{13}^{}\; l_{13}^{}\, ,
      \;  n_2^{}\, l_2^{}\: : \, L \, | 
      \,  n_{12}^{}\; l_{12}^{}\, ,
      \;  n_3^{}\, l_3^{}\: : \, L
  \rangle_{d}^{}$ follows the conventions of \cite{Kamuntavicius:2001ig} and the mass ratio $d$ is given above. 
  In Appendix~\ref{app:Rnl}, we summarize explicitly which configuration and momentum space HO wave functions 
  are implied by these conventions.  

For the case that clusters/particles 2 and 3 are identical, the coordinate transformations are 
equivalent to transposition operators   
\begin{eqnarray}
\label{eq:genpermutation}
 & & _{(13)2}^{} \left\langle  \ n_{13}^{} n_{2}^{}  \left(  
                       \left( l_{13}^{} \, (s_1^{} s_3^{}) S_{13}^{} \right) J_{13}^{} \:
                            (l_2^{} s_2^{}) I_2^{}   \right) J \,   
                            \left( (t_1^{} t_3^{}) T_{13}^{} \, t_2^{} \right) T      
           \right| \nonumber \\[10pt]
  & &   \qquad \qquad    \left| \ n_{12}^{} n_{3}^{}   \left(  
                       \left(l_{12}^{} \, (s_1^{} s_2^{}) S_{12}^{} \right) J_{12}^{} \:
                            (l_3^{} s_3^{}) I_3^{}   \right) J \,   
                            \left( (t_1^{} t_2^{}) T_{12}^{} \, t_3^{} \right) T  \right\rangle _{(12)3}^{} \nonumber \\[10pt]
 & \equiv &  _{(12)3}^{}  \left\langle  \ n_{13}^{} n_{2}^{}   \left(  
                               \left(l_{13}^{} \, (s_1^{} s_2^{}) S_{13}^{}   \right) J_{13}^{} \:
                                    (l_2^{} s_3^{}) I_2^{}   \right) J \,    
                                    \left( (t_1^{} t_2^{}) T_{13}^{} \, t_3^{} \right) T       
           \right|  \nonumber \\[10pt]
  & &       \qquad \qquad        \mathcal{P}_{23}^{}   
                         \left| \ n_{12}^{}  n_{3}^{}  \left(  
                       \left(l_{12}^{} \, (s_1^{} s_2^{}) S_{12}^{} \right) J_{12}^{} \:
                            (l_3^{} s_3^{}) I_3^{}   \right) J \,   
                            \left( (t_1^{} t_2^{}) T_{12}^{} \, t_3^{} \right) T   \right\rangle _{(12)3}^{}   \ .
\end{eqnarray}
We added subscripts ${(ij)k}$ to the states to make the clusters involved in the subsystem and the spectator explicit. Note that, for the right hand side, this 
implies that the quantum number $S_{13}^{}$ and $T_{13}^{}$ are total spins and isospins related to $s_1^{}$, $s_3^{}$ , $t_1^{}$ and $t_3^{}$. Correspondingly, the 
labels of the other quantum numbers are related to the $(12)$ subsystem and spectator $3$ even if the labels of the quantum numbers are different. Below,
 we will explicitly show how quantum numbers of the $A$-nucleon system are related to the the quantum numbers in Eq.~(\ref{eq:gencoordchange}).
     
\subsection{Antisymmetrization of  $A$-body states}

\begin{table}
\caption{\label{tab:list-of-states}Labeling and graphical representation of different sets of coordinates for the $A$-body system.}
  \centering
  \setlength{\tabcolsep}{0.3cm}
  \begin{tabular}{|c|c|c|c|}
   \hline
   &&&\\
     \raisebox{1.5ex}{label} 
   & \raisebox{1.5ex}{graphical rep.} 
   & \raisebox{1.5ex}{subsystems }
   & \raisebox{1.5ex}{dimensionality}
   \\
%
   \hline
   &&&\\[-6pt]
   \raisebox{1.5ex}{$\alphastarone{}$} 
   & \raisebox{0.5ex}{$\includegraphics[scale=0.22]{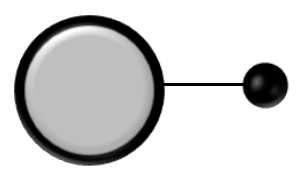}$} 
   & \raisebox{1.5ex}{$\alpha_{{}_{A-1}}^{}+ \rm{N}$}
   & \raisebox{1.5ex}{$A\times$}\hspace{-0.1cm}
     \raisebox{0.5ex}{\includegraphics[scale=0.24]{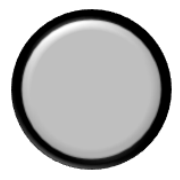}}
   \\
%
   \hline
   &&&\\[-6pt]
   \raisebox{1.5ex}{$\alphastaronestarone$} 
   & \raisebox{0.5ex}{$\includegraphics[scale=0.2]{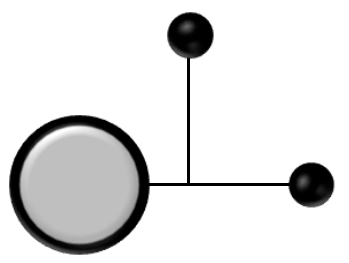}$} 
   & \raisebox{1.5ex}{$\alphastarone{{}_{A-1}} + \rm{N}$}
   & \raisebox{1.5ex}{$\,\left(A-1\right)\times$}\hspace{-0.1cm}
     \raisebox{0.7ex}{\includegraphics[scale=0.22]{alphastar1.pdf}}
   \\
   &&& $= \left(A-1\right)\!\times \!A\times$ \hspace{-0.2cm}
     \raisebox{-1.0ex}{\includegraphics[scale=0.24]{alpha.pdf}}
   \\[3pt]
%
   \hline
   &&&\\[-6pt]
   \raisebox{1.5ex}{$\alphastartwo{}$} 
   & \raisebox{0.5ex}{$\includegraphics[scale=0.2]{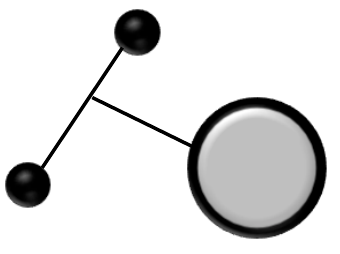}$} 
   & \raisebox{1.5ex}{$\alpha_{{}_{A-2}}^{} + \alpha_{{}_{12}}^{}$}
   & \raisebox{1.5ex}{$\frac{1}{2} \,\times$} \hspace{-0.18cm}
     \raisebox{1.0ex}{\includegraphics[scale=0.2]{alphastar1-star1.pdf}}
   \\[2pt]
   &&& $= \frac{\left(A-1\right)}{2} \,\times$  \hspace{-0.18cm}
     \raisebox{-0.7ex}{\includegraphics[scale=0.22]{alphastar1.pdf}}
   \\[9pt]
   &&& $= \frac{A\left(A-1\right)}{2} \,\times$ \hspace{-0.18cm}
     \raisebox{-0.8ex}{\includegraphics[scale=0.24]{alpha.pdf}}
   \\[4pt]
%
   \hline
   &&&\\[-6pt]
   \raisebox{1.5ex}{$\alphastaronestartwo$} 
   & \raisebox{0.5ex}{$\includegraphics[scale=0.18]{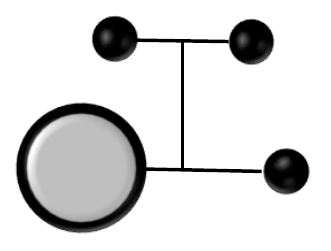}$} 
   & \raisebox{1.5ex}{$\alphastarone{{}_{A-2}} + \alpha_{{}_{12}}^{}$}
   & \raisebox{1.5ex}{$\,\left(A-2\right)\times$} \hspace{-0.18cm}
     \raisebox{-0.1ex}{\includegraphics[scale=0.2]{alphastar2.pdf}}
   \\[3pt]
   &&& $= \frac{A\left(A-1\right)\left(A-2\right)}{2} \,\times$ \hspace{-0.2cm}
     \raisebox{-1.0ex}{\includegraphics[scale=0.24]{alpha.pdf}}
   \\[4pt]
%
   \hline
   &&&\\[-6pt]
   \raisebox{1.5ex}{$\alphastaronestarAmthree$} 
   & \raisebox{0.5ex}{$\includegraphics[scale=0.18]{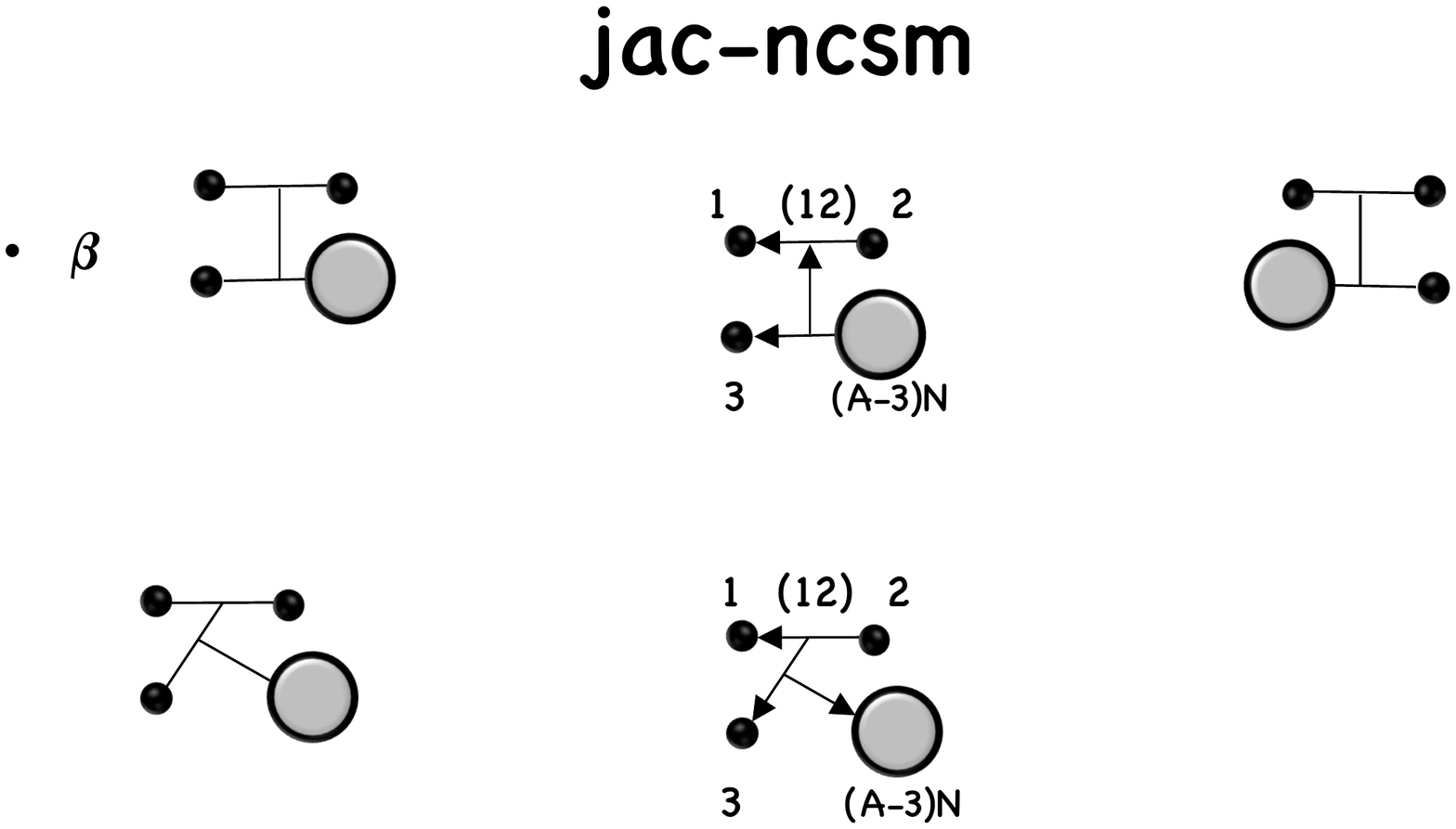}$} 
   & \raisebox{1.5ex}{$\alphastarone{{}_3} + \alpha_{{}_{A-3}}^{}$}
   & \raisebox{2.2ex}{$\approx$} \hspace{-0.1cm}
     \raisebox{1.0ex}{\includegraphics[scale=0.18]{alphastar1-star2.pdf}} \\
   &&& \raisebox{1.5ex}{$= \left(A-2\right)\times$} \hspace{-0.18cm}
     \raisebox{-0.1ex}{\includegraphics[scale=0.2]{alphastar2.pdf}}
   \\[3pt]
   &&& $= \frac{A\left(A-1\right)\left(A-2\right)}{2} \,\times$ \hspace{-0.2cm}
     \raisebox{-1.0ex}{\includegraphics[scale=0.24]{alpha.pdf}}
   \\[4pt]
%
   \hline
   &&&\\[-6pt]
   \raisebox{1.5ex}{$\alphastarthree{}$} 
   & \raisebox{0.5ex}{$\includegraphics[scale=0.18]{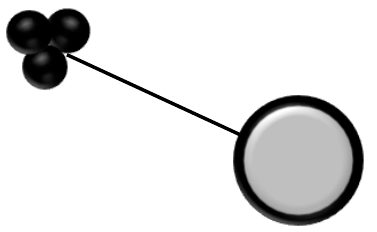}$} 
   & \raisebox{1.5ex}{$\alpha_{{}_{A-3}}^{} + \alpha_{{}_{3}}^{}$}
   & \raisebox{1.5ex}{$\frac{1}{3} \,\times$} \hspace{-0.18cm}
     \raisebox{1.0ex}{\includegraphics[scale=0.18]{alphastar1-starA-3.pdf}}
   \\[2pt]
   &&& $= \frac{\left(A-2\right)}{3} \,\times$  \hspace{-0.18cm}
     \raisebox{-0.7ex}{\includegraphics[scale=0.2]{alphastar2.pdf}}
   \\[9pt]
   &&& $= \frac{A\left(A-1\right)\left(A-2\right)}{6} \,\times$ \hspace{-0.2cm}
     \raisebox{-1.0ex}{\includegraphics[scale=0.24]{alpha.pdf}}
   \\[4pt]
   \hline
  \end{tabular}
  \end{table}

These relations are the basis of all following expressions.  The calculations often require 
to represent basis states in terms of states of subclusters. In Table~\ref{tab:list-of-states}, we 
have summarized the labeling of such states. In short, states are labeled by a greek letter
that indexes all possible states for a set of given quantum numbers.  A superscript $*(i)$ 
indicates that an $i$-particle subcluster has been separated off from the rest of the $A$-body system. 
The relative distance or momentum of the two clusters point here towards the $i$-nucleon cluster. 
This operation can be repeated to form states with a special subclustering. The graphical representation 
given in the table should clarify the clusters involved.  Since we are going to obtain the basis states 
recursively starting from $A=3$, $A$-body cluster states are labeled by the index of the 
($A$--$x$)N-clusters. The contributing indices are given in the third column. The number of particles of the subclusters 
is here given as a subscript. We assume that the complete state 
and the clusters are antisymmetrized which is not the case anymore for the states that explicitly 
single out clusters. This implies that more states are required to cover the physical Hilbert space completely. 
The last column of the table gives first estimates of the relations of the dimensionalities.

It is now the aim to express the completely antisymmetric states in terms of  $| \alphastarone{}  \rangle$. 
In the first step, we therefore need to obtain the antisymmetrization operator $\mathcal{A}$ in this basis. 
Assuming antisymmetry for the ($A$--$1$)-nucleon system, the matrix of 
$\mathcal{A}$ for $A$ nucleons can be written in terms of the transposition operator of the outer two 
nucleons $\mathcal{P}_{A-1,A}^{}$ as
  \begin{eqnarray}
   \langle \alphastarone{} |
    \mathcal{A}
    | \betastarone{} \rangle 
   & = & 
   \frac{1}{A}
   \langle \alphastarone{} | 
     \left( \mathbbm{1} - 
           \left(A-1\right)
           \mathcal{P}_{A-1,A}^{}
      \right)
    | \betastarone{} 
   \rangle  \ . 
   \label{eq:general-antisym-mat}
  \end{eqnarray}
The antisymmetric $A$-body states are eigenstates of $\mathcal{A}$ 
for the eigenvalue $\lambda=1$, e.g. are solutions of 
  \begin{eqnarray}
   \langle \alphastarone{} | 
     \mathcal{A}
    | \gammastarone{} \rangle
   \langle 
    \gammastarone{} | 
    \beta 
   \rangle 
   & = &
   \lambda \, 
   \langle 
    \alphastarone{} | 
    \beta 
   \rangle  = 
   \langle \, 
    \raisebox{-0.5ex}{\includegraphics[scale=0.22]{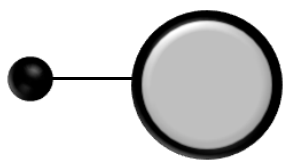}} \, | \,
    \raisebox{-0.8ex}{\includegraphics[scale=0.24]{alpha.pdf}} \,   
   \rangle \ . 
    \label{eq:book-eigenvalue-problem}
   \end{eqnarray}
Here, the graphical representation of the states is added to simplify the notation and a sum over 
$\gammastarone{}$-states is implied.  The matrix elements  $ \langle \, 
 \raisebox{-0.5ex}{\includegraphics[scale=0.22]{betastar1.pdf}} \, | \,
    \raisebox{-0.8ex}{\includegraphics[scale=0.24]{alpha.pdf}} \,  
   \rangle  $ are the well-known coefficients of fractional parantage (cfp) \cite{Racah:1943gh} which 
define the antisymmetric $A$-body state in terms of antisymmetric ($A$--1)-body states  
in relative motion with respect to the $A$-th nucleon. We will obtain these states by diagonalization 
of $\mathcal{A}$. The problem is therefore reduced to the calculations of the matrix 
$\langle \alphastarone{} |  \mathcal{A}   | \gammastarone{} \rangle$.  
To this aim, we need to explicitly define the coupling 
scheme for states $ |\alphastarone{} \rangle$ given by 
  \begin{eqnarray}
    |\alphastarone{}
    \rangle 
     & = &
     |  \, \alpha_{A-1}^{} \;\;\; n_A^{} 
      \left(l_A^{} \, s_A^{}  \right)I_A^{} \; t_A^{}\, ;
      \, \left(J_{A-1}^{} \: I_A^{}\right)\!J \;
      \left(T_{A-1}^{} \: t_A^{}\right)\!T \,
     \big\rangle .
    \label{eq:def-alphastar1-AN}
   \end{eqnarray}   
The states are based on complete antisymmetrized states $  |  \, \alpha_{A-1}^{} \rangle $ 
with well defined total angular momentum $J_{A-1}^{}$ and isospin $T_{A-1}^{}$ 
and total HO energy quantum number $\mathcal{N}_{A-1}^{}$. Note that we dropped 
the last quantum number in Eq.~(\ref{eq:def-alphastar1-AN}) to simplify the notation. The motion of the $A$-th nucleon 
is given by its HO quantum number $n_A^{} $, orbital angular momentum $l_A^{}$, spin $s_A^{}=\frac{1}{2}$, total angular momentum 
$I_A^{}$ and isospin $t_A^{}=\frac{1}{2}$. In order to end up with a well-defined total angular momentum $J$ and isospin $T$ of the $A$-body 
system, we finally couple the individual angular momenta and isospins as indicated.  

\begin{figure}
\centering
\includegraphics[scale=0.5]{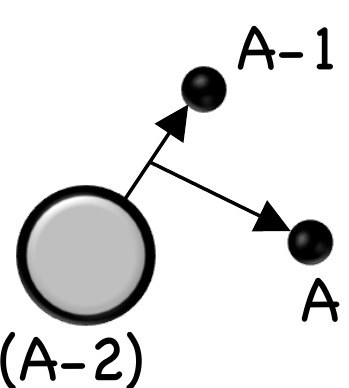}   
\hspace{2cm} \includegraphics[scale=0.5]{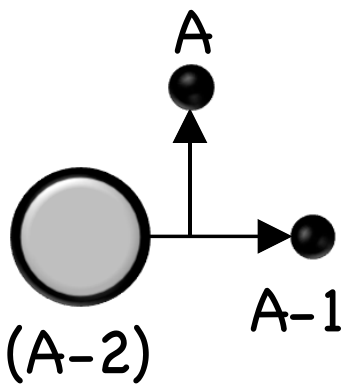} 

\caption{Two representations of $ | \alphastaronestarone \rangle$ coordinates used for the antisymmetrization 
operator. The arrangement matches the general coordinates shown in Fig.~\ref{fig:gen3body}. Note that the direction of 
the coordinates differs for the subsystems. \label{fig:genAbody}}
\end{figure}

\begin{table}
\caption{ \label{tab:qnrelPA1A}Identification of quantum numbers in states of Eq.~(\ref{eq:alphastaronestarone}) to the 
   ones of the permutation operators in Eqs.(\ref{eq:gencoordchange}) and (\ref{eq:genpermutation}).
   Given are only relations to the quantum numbers of  Eq.~(\ref{eq:jacobi12}) since the relation to 
   Eq.~(\ref{eq:jacobi13}) is a simple generalization.}
  \centering
  \def\arraystretch{1.5}
  \setlength{\tabcolsep}{0.3cm}
  \begin{tabular}{|c|c|c|c|c|c|c|c|c|c|c|c|c|c|c|}
   \hline 
 $n_{12}^{}$  & $n_3^{}$      &  $l_{12}^{}$   & $s_1^{}$          & $s_2^{}$            \\
 $n_{A-1}^{}$ &  $ n_{A}^{} $& $l_{A-1}^{}$ &  $J_{A-2}^{}$   & $  \frac{1}{2} $         \\   
 \hline \hline   
 $J_{12}^{} $  &  $l_{3}^{}$ & $s_3^{}$        & $I_{3}^{}$  & $J$    \\
 $ J_{A-1}^{} $&  $l_A^{} $  & $ \frac{1}{2}$ & $I_A^{} $   & $ J$ \\
 \hline \hline
 $t_1^{}$           & $t_2^{}$          & $T_{12}^{}$  & $t_3^{}$ & $T$   \\
 $ T_{A-2}^{} $  & $ \frac{1}{2}  $ & $ T_{A-1}^{} $  & $\frac{1}{2}$ & $T$ \\
 \hline \hline
     $S_{12}^{}$        &   &   &   &   \\
   sum   &   &   &   &      \\
 \hline
  \end{tabular}
\end{table}

The antisymmetrization operator is given by $\mathcal{P}_{A-1,A}^{}$. In the next step, we therefore need to use the known cfp of the 
($A$--1)-nucleon system to disentangle the ($A$--1)-th nucleon from the antisymmetric cluster. We end up with states  
  \begin{eqnarray}
   | \alphastaronestarone \rangle
   & = &   | \alphastarone{A-1}   \; n_A^{}  \left(l_A^{} \, s_A^{}  \right) I_A^{} \; t_A^{}\, ;
      \, \left(J_{A-1}^{} \: I_A^{}\right)\!J \;
      \left(T_{A-1}^{} \: t_A^{}\right)\!T \,
     \big\rangle \ .
  \end{eqnarray}
or more explicitly by reinserting the definition Eq.~(\ref{eq:def-alphastar1-AN})
 \begin{eqnarray}
    \label{eq:alphastaronestarone}
   | \alphastaronestarone \rangle
        & = &   | \alpha_{A-2}^{}   \; n_{A-1}^{}  \left(l_{A-1}^{} \, s_{A-1}^{}  \right) I_{A-1} \; t_{A-1}^{}\, ,
        \; n_A^{}   \left(l_A^{} \, s_A^{}  \right)I_A^{} \; t_A^{}\, ;  \cr  
    & & \quad    \, \left(    \, \left(J_{A-2}^{} \: I_{A-1}^{}\right)\!  J_{A-1}^{} \: I_A^{}\right)\!J \; 
        \left(    \left(T_{A-2}^{} \: t_{A-1}^{}\right)\! T_{A-1}^{} \: t_A^{}\right)\!T \, \big\rangle \  
    =  \!|\! \raisebox{-0.7ex}{ \includegraphics[scale=0.25]{alphastar1-star1.pdf} } \!\!\rangle .
 \end{eqnarray}    
Again, the graphical representation is given to simplify the expression. The complete coupling scheme will 
however be important to explicitly obtain the matrix element of  $\mathcal{P}_{A-1,A}^{}$. The directions of the momenta 
(or coordinates) are given in the more detailed Fig.~\ref{fig:genAbody}.  
In order to match the states of Eqs.~(\ref{eq:jacobi12}) and (\ref{eq:jacobi13}) with  the ones of Eq. (\ref{eq:alphastaronestarone}), we
first identify the clusters 2 and 3 with the nucleons $A$--1 and $A$. Comparing the directions given in Fig.~\ref{fig:gen3body} and 
\ref{fig:genAbody}, it is obvious that the position vectors of the spectator 
 agree with the one of  $A$-th nucleon, but the relative positions of the ($A$--2)-nucleon cluster and the ($A$--1)-th nucleon are opposite implying  
additional phases $(-)^{l_{A-1}^{}}$ for each of the states. The coupling of the angular momentum quantum numbers of the subsystem is also 
different to the general three cluster expression. 
We need an additional $6j$ coefficient and extra phase to recouple from 
\begin{equation}
\label{eq:recouplstaronestarone}
\left(  J_{A-2}^{} \, \left( l_{A-1}^{} \; s_{A-1}^{} \right)   I_{A-1}^{}  \right)  J_{A-1}^{}
   \ \ \mbox{ to } \ \ \left(  l_{A-1}^{} \,    \left(  J_{A-2}^{} \; s_{A-1}^{}\right)   S_{12}  \right)  J_{A-1}^{}  \ \  .
\end{equation}
$S_{12}$ is a new quantum number we need to sum over. Its name is chosen to match Eq.~(\ref{eq:jacobi12}), for 
Eq.~(\ref{eq:jacobi13}) $S_{13}$ is more natural. 

After this recoupling the quantum numbers can be identified to the ones of Eq.~(\ref{eq:gencoordchange}) as shown in Table~\ref{tab:qnrelPA1A} leading to the 
matrix element of the permutation operator 
  \begin{eqnarray}
  &&
   \langle \gammastaronestarone | 
    \mathcal{P}_{A-1,A}^{} 
    | \deltastaronestarone \rangle   =
   \langle \!
    \raisebox{-0.7ex}{
    \includegraphics[scale=0.25]{alphastar1-star1.pdf} }
    \!|\!
    \raisebox{-1.0ex}{
    \includegraphics[scale=0.25]{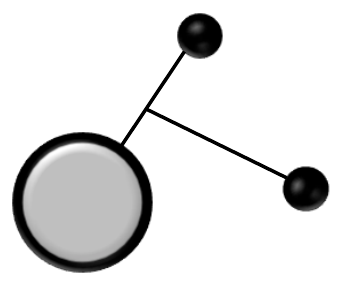} }
   \!\!\rangle  = 
   \left(-1\right)^{2J_{A-2}^{}
                    +I_{A-1}^{\delta}+T_{A-1}^{\delta}+l_A^{\delta}
                    +I_{A-1}^{\gamma}+T_{A-1}^{\gamma}+l_A^{\gamma}} 
   \cr
   &&\cr
   &&
   \phantom{=}
   \times \, 
    \hat I_{A-1}^{\gamma} \: 
    \hat J_{A-1}^{\gamma} \: 
    \hat T_{A-1}^{\gamma} \:
    \hat I_A^{\gamma}    \:\:
    \hat I_{A-1}^{\delta} \:
    \hat J_{A-1}^{\delta} \:
    \hat T_{A-1}^{\delta} \:
    \hat I_A^{\delta} 
  \, 
   \sum_{S_{{}_{12}}^{} S_{{}_{13}}^{}}
    \hat S_{13}^{ \,2} \: \hat S_{12}^{ \,2} \:  
    \left\{\begin{array}{ccc}
            J_{A-2}^{} &  \frac{1}{2} &  S_{13}^{}\\[2pt]
            l_{A-1}^{\gamma} & J_{A-1}^{\gamma} & I_{A-1}^{\gamma} 
           \end{array}
    \right\}          
    \left\{\begin{array}{ccc}
            J_{A-2}^{} & \frac{1}{2} & S_{12}^{}\\[2pt]
            l_{A-1}^{\delta} & J_{A-1}^{\delta} & I_{A-1}^{\delta}
           \end{array}
    \right\} 
   \cr
   &&\cr
   &&
   \;\;
   \times \sum_{LS}\hat L_{}^2 \, \hat S_{}^2 
    \left\{\begin{array}{ccc}
            \frac{1}{2} & J_{A-2}^{} & S_{13}^{} \\[2pt]
            \frac{1}{2} & S & S_{12}^{}
           \end{array}
    \right\}
    \left\{\begin{array}{ccc}
            \frac{1}{2} & T_{A-2}^{} & T_{A-1}^{\gamma} \\[2pt]
            \frac{1}{2} & T & T_{A-1}^{\delta}
           \end{array}
    \right\} \; 
    \left\{\begin{array}{ccc}
            l_{A-1}^{\gamma} & S_{13}^{} & J_{A-1}^{\gamma} \\[2pt]
            l_{A}^{\gamma}  & \frac{1}{2} & I_A^{\gamma} \\[2pt]
            L & S & J    
           \end{array}
    \right\}       
    \left\{\begin{array}{ccc}
            l_{A-1}^{\delta} & S_{12}^{} & J_{A-1}^{\delta} \\[2pt]
            l_{A}^{\delta}  & \frac{1}{2} & I_A^{\delta} \\[2pt]
            L & S & J    
           \end{array}
    \right\}       
   \cr
   &&\cr
   &&
   \;\;
   \times
    \langle n_{A-1}^{\gamma}\; l_{A-1}^{\gamma}\, ,
            \; n_A^{\gamma}\; l_A^{\gamma}\: : 
            \, L \, | 
            \, n_{A-1}^{\delta}\; l_{A-1}^{\delta}\, ,
            \; n_A^{\delta}\, l_A^{\delta}\: : 
            \, L
    \rangle_{d=\frac{1}{A\left(A-2\right)}}^{}\,   \  \  .
   \label{eq:antisym-trafo-AN}
  \end{eqnarray}
The left and right hand side states are labeled by superscripts $\gamma$ and $\delta$. We omitted these labels for 
quantum numbers that are conserved. Kronecker $\delta$'s for these quantum numbers are implied. 
To complete the antisymmetrization operator of Eq.~(\ref{eq:general-antisym-mat}), 
we only need to use the cfp obtained before for the ($A$--1)N-system to relate the $\gammastaronestarone$ to $\betastarone{}$ 
states
 \begin{eqnarray}
   \langle 
    \deltastaronestarone |
    \betastarone{}
   \rangle 
   & = &
   \langle \!
    \raisebox{-1.0ex}{
    \includegraphics[scale=0.25]{betastar1-star1.pdf} }
    \!|\!
    \raisebox{-0.7ex}{
    \includegraphics[scale=0.28]{alphastar1.pdf}  }
   \!\! \rangle = \delta_{spectator}^{}
   \langle  
    \raisebox{2.8ex}{\includegraphics[scale=0.22,angle=240]{betastar1.pdf}} | \,
    \raisebox{-0.8ex}{\includegraphics[scale=0.24]{alpha.pdf}} \,
   \rangle_{A-1}^{}
   = \delta_{spectator}^{} \langle 
      \deltastarone{} | 
      \beta
     \rangle_{A-1}^{}
   \label{eq:antisym-book--show-recursion}
  \end{eqnarray}
The Kronecker symbol $\delta_{spectator}^{}$ represents the conservation of all spectator quantum numbers, 
the total angular momentum, isospin and HO energy quantum numbers for the ($A$--1)-body subsystem and the $A$-body 
system. The permutation operator of Eq.~(\ref{eq:general-antisym-mat}) can then be represented by 
\begin{eqnarray}
   \langle \alphastarone{} |
    \mathcal{P}_{bc}^{} 
   | \betastarone{} \rangle 
   &=&
   \langle 
    \alphastarone{} |
    \gammastaronestarone 
   \rangle
   \langle \gammastaronestarone | 
    \mathcal{P}_{A-1,A}^{} 
    | \deltastaronestarone 
   \rangle
   \langle
    \deltastaronestarone |
    \betastarone{}
   \rangle
   \cr
   &&\cr       
   & =&
   \langle \!\!
    \raisebox{-0.7ex}{
     \includegraphics[scale=0.28]{betastar1.pdf} }
     \!|\!
    \raisebox{-0.7ex}{
     \includegraphics[scale=0.25]{alphastar1-star1.pdf} }
   \!\!\rangle 
   \langle \!
    \raisebox{-0.7ex}{
    \includegraphics[scale=0.25]{alphastar1-star1.pdf} }
    \!|\!
    \raisebox{-1.0ex}{
    \includegraphics[scale=0.25]{betastar1-star1.pdf} }
   \!\!\rangle
   \langle \!
    \raisebox{-1.0ex}{
    \includegraphics[scale=0.25]{betastar1-star1.pdf} }
    \!|\!
    \raisebox{-0.7ex}{
    \includegraphics[scale=0.28]{alphastar1.pdf}  }
   \!\!\rangle
   \label{eq:def-sum-antisym-book}
  \end{eqnarray}
where a sum over intermediate states is implied.  
Because the total angular momentum $J$, isospin $T$ and HO energy quantum number $\mathcal{N}$ 
is conserved  by $\langle \alphastarone{}  | \mathcal{P}_{bc}^{}   | \betastarone{} \rangle $, the antisymmetrized states can be obtained 
for each $J$,  $T$ and $\mathcal{N}$  separately. To this aim, we implemented a parallelized code that generates 
the antisymmetrization matrix elements of Eq.~(\ref{eq:general-antisym-mat}) for each block $J$,  $T$ and $\mathcal{N}$
and performs a diagonalization using the parallelized eigenvector packages SCALAPACK \cite{Blackford:1997ta} and ELPA \cite{Auckenthaler:2011fy}. The starting point of the recursive procedure is the $A=3$ system
where we impose antisymmetry of the (12)-subsystem via the condition $\left(-1\right)_{}^{l_{12}^{}+S_{12}^{}+T_{12}^{}}$. Here, we use the basis set
Eq.~(\ref{eq:jacobi12}) directly to represent the antisymmetrized states of the three nucleons without further recoupling. 
Based on the diagonalization of the  permutation operator as given in  Eq.~(\ref{eq:genpermutation}), antisymmetrized states 
are found that are then used to recursively proceed to $A>3$.  

\begin{table}
\caption{\label{tab:dim_antisym}Dimensions of selected sets of $\alpha$, $\alphastarone{} $, $\alphastaronestarone$ and $\alphastarAmtwo{}$ states for blocks 
with given total angular momentum, isospin and HO energy quantum number for an $A$-nucleon system. }
\begin{center}
\def\arraystretch{2.5}
\setlength{\tabcolsep}{0.3cm}
\begin{tabular}{|cccc|cccc|}
\hline
$A$ & $J$ & $T$ & $\mathcal{N}$ & dim($\alpha$) & dim($\alphastarone{}$) & dim($\alphastaronestarone$) & dim($\alphastarAmtwo{}$) \\
   4   &    0     &     0      &     10      &      217    &       791    &      2373   &     1225 \\
   4   &    0     &     0      &     12      &      417    &      1551    &      4648   &     2380 \\
   4   &    4     &     0      &     12      &     2123    &      8370    &     25110   &      --  \\
   4   &    4     &     1      &     12      &     3104    &     12516    &     37626   &      --  \\
   7   &   1/2    &    1/2     &      7      &     1269    &      9957    &     65369   &    32190 \\
   7   &   1/2    &    1/2     &      9      &     8963    &     67453    &    429132   &   212318 \\
   7   &   5/2    &    1/2     &      9      &    18839    &    142535    &    910342   &      --  \\
   7   &   5/2    &    3/2     &      9      &    16629    &    130896    &    861394   &      --  \\
\hline   
\end{tabular}
\end{center}
\end{table}

Explicit calculations confirmed 
that the antisymmetrization operator $\mathcal{A}$ has only two eigenvalues $\lambda=0$ and $1$. The dimension of the 
eigenspace for $\lambda=1$ is approximately by a factor $\frac{1}{A}$ smaller then the total dimensionality of the space spanned by 
$ | \alphastarone{} \rangle $. The normalized eigenvectors are the cfp. We note that fully antisymmetrized states (and eigenvalues $\lambda=0,1)$ 
are only obtained when the complete block of intermediate states for $J$,  $T$ and $\mathcal{N}$ was included. 
In this first study, we also included all states of the ($A$--2)N-subsystem that 
can be combined with the two outer nucleons to $J$,  $T$ and $\mathcal{N}$. The dimensions of selected sets of 
$\alpha$, $\alphastarone{} $ and $\alphastaronestarone$ states are given in Table~\ref{tab:dim_antisym}. The sets generated so far 
are tabulated in Appendix~\ref{app:ready}. They will be made available in the machine independent HDF5 format \cite{hdf5}.

\section{$2$N+($A$--$2$)N states for 2N operators}
\label{sec:nnrecoupl}

For the representation of two-nucleon operators as for example the NN interaction, the complete 
antisymmetrized states are not suitable. The most efficient way to obtain matrix elements for these operators is 
to change to a basis that singles out two nucleons from the $A$-nucleon system. Following the notation of 
Table~\ref{tab:list-of-states}, such states are given by $  | \alphastarAmtwo{} \rangle$.
It is the aim of this section to calculate the overlap  $\langle \alpha | \betastarAmtwo{} \rangle$. Again these
transition coefficients will be independent of the HO frequency and will conserve total $A$-body $J$, $T$ 
and $\cal{N}$.  Since the two-body states are directly linked to the matrix elements of  any two-nucleon 
operator, it will be straightforward to apply these operator to any $A$-body state once the transition matrix 
elements are known.

\begin{figure}
\centering
\includegraphics[scale=0.5]{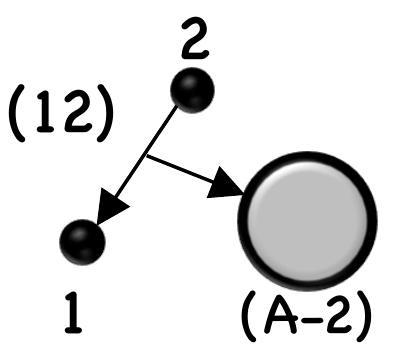}   
\hspace{2cm} \includegraphics[scale=0.5]{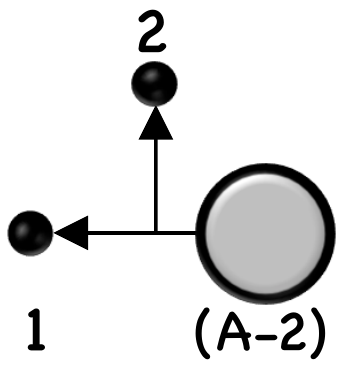} 

\caption{Left hand side: more detailed definition of $| \alphastarAmtwo{} \rangle $ states including the 
              direction of momenta. Right hand side: the same for $| \alphastaronestarone \rangle $ states.\label{fig:2NFstates}  }
\end{figure}

Explicitly, the  $  | \alphastarAmtwo{} \rangle$ states are given by
  \begin{eqnarray}
   | \alphastarAmtwo{} \rangle
   & = &
   |\alpha_{12}^{} \;\, n_{\lambda}^{} \lambda
    \;\, \alpha_{A-2}^{} \, ;
   \; \left( \left( l_{12}^{} \, \left(s_1^{} s_2^{}\right)S_{12}^{} \right) J_{12}^{}
   \, \left(\lambda \; J_{A-2}^{}\right)I_\lambda^{} \right) J^{} \; 
    \left(\left(t_1^{} t_2^{}\right)T_{12}^{} \: T_{A-2}^{}
    \right)T
   \rangle   \  \  \  .
   \label{eq:def-alphastarA-2}
  \end{eqnarray}
  The state of the two-nucleon subsystem is labeled here with $\alpha_{12}^{}$. As usual this combined label 
  corresponds to the HO quantum number $n_{12}^{}$, the orbital angular momentum $l_{12}^{}$ and spin $S_{12}^{}$ 
  that are coupled to the total angular momentum $J_{12}^{}$ and the total isospin $T_{12}^{}$ of the two-nucleon subsystem. 
  The relative motion of this cluster with respect to the ($A$--$2$)N rest system is described by the HO quantum number $n_\lambda$
  and the orbital angular momentum $\lambda$. The quantum numbers  $\alpha_{A-2}^{}$, $J_{A-2}^{}$ and  $T_{A-2}^{}$ have already   
  been defined above. In order to be able to define the total angular momentum, one intermediate quantum number 
  $I_{\lambda}^{}$ is necessary, which is given here by coupling $\lambda$ and $J_{A-2}^{}$. Note that the conventions 
  for the direction of momenta and orderings of couplings correspond indeed to $  | \alphastarAmtwo{} \rangle$ as can be seen 
  in Fig.~\ref{fig:2NFstates} and not, as one 
  might naively expect, to $| \alphastartwo{} \rangle$.  For shorter notation, we label the two nucleons separated out 
  with number 1 and 2 here, but whenever we are referring to  $|  \alphastaronestarone \rangle$ we keep labeling them 
  nucleon $A-1$ and $A$ as done in the previous section. 
  
  \begin{table}
\caption{\label{tab:qnrel2Am2}Left hand side: Identification of quantum numbers of  $| \alphastarAmtwo{} \rangle$ states to the 
   ones of the cooordinate  transformation in Eq.(\ref{eq:gencoordchange}) used to obtain the transitions to $2$N+($A$--$2$) states.  
   Right hand side: the same for $|  \alphastaronestarone \rangle$ 
   states. }
  \centering
  \def\arraystretch{1.5}
  \setlength{\tabcolsep}{0.3cm}
  \begin{tabular}{|c|c|c|c|c|c|c|c|c|c|c|c|c|c|c|}
   \hline 
 $n_{12}^{}$  & $n_3^{}$                &  $l_{12}^{}$   & $s_1^{}$           & $s_2^{}$            \\
 $n_{12}^{}$  &  $ n_{\lambda}^{} $&  $l_{12}^{}$   &   $ \frac{1}{2}$  & $  \frac{1}{2} $         \\   
 \hline \hline   
 $J_{12}^{} $  &  $l_{3}^{}$    & $s_3^{}$         & $I_{3}^{}$             & $J$    \\
 $J_{12}^{} $  &  $\lambda $  & $ J_{A-2}^{}$  & $I_\lambda^{} $   & $ J$ \\
 \hline \hline
 $t_1^{}$           & $t_2^{}$          & $T_{12}^{}$  & $t_3^{}$ & $T$   \\
 $\frac{1}{2}$    & $ \frac{1}{2}  $ & $T_{12}^{}$  & $ T_{A-2}^{} $  & $T$ \\
 \hline \hline
     $S_{12}^{}$        &   &   &   &   \\
     $S_{12}^{}$        &   &   &   &      \\
 \hline
  \end{tabular}
  \hspace{2cm}
   \begin{tabular}{|c|c|c|c|c|c|c|c|c|c|c|c|c|c|c|}
   \hline 
 $n_{13}^{}$  & $n_2^{}$                &  $l_{13}^{}$   & $s_1^{}$           & $s_2^{}$            \\
 $n_{A-1}^{}$  &  $ n_{A}^{} $         &  $l_{A-1}^{}$   &   $ \frac{1}{2}$  & $  \frac{1}{2} $         \\   
 \hline \hline   
 $J_{13}^{} $  &  $l_{2}^{}$    & $s_3^{}$         & $I_{2}^{}$             & $J$    \\
 $J_{A-1}^{} $  &  $l_{A}^{}$   & $ J_{A-2}^{}$  & $I_{A}^{} $   & $ J$ \\
 \hline \hline
 $t_1^{}$           & $t_2^{}$          & $T_{13}^{}$  & $t_3^{}$ & $T$   \\
 $\frac{1}{2}$    & $ \frac{1}{2}  $ & $T_{A-1}^{}$  & $ T_{A-2}^{} $  & $T$ \\
 \hline \hline
     $S_{13}^{}$        &   &   &   &   \\
    sum                     &   &   &   &      \\
 \hline
  \end{tabular}
\end{table}

  The transition can be done most easily in two steps which mostly involve matrix elements already known. 
  We first use the cfp $\langle 
    \alpha | 
    \gammastarone{}   
   \rangle  = 
   \langle 
    \raisebox{-0.9ex}{
    \includegraphics[scale=0.3]{alpha.pdf}} 
    |\raisebox{-0.7ex}{
    \includegraphics[scale=0.28]{alphastar1.pdf} }
   \rangle $ obtained by solving Eq.~(\ref{eq:book-eigenvalue-problem}) and then 
   $\langle 
    \gammastarone{} |
    \deltastaronestarone
   \rangle
     =
   \langle
    \raisebox{-0.7ex}{
    \includegraphics[scale=0.28]{alphastar1.pdf}} 
    |\raisebox{-0.6ex}{
    \includegraphics[scale=0.25]{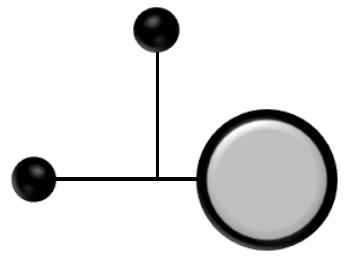} }
   \rangle$ that were already involved in the definition of the antisymmetrization 
   operator Eq.~(\ref{eq:def-sum-antisym-book}). The final step, the transition 
    $\langle 
    \deltastaronestarone |
    \betastarAmtwo{}
   \rangle
    =
   \langle \!\!
    \raisebox{-0.6ex}{
    \includegraphics[scale=0.25]{betastar1-star1_2NF.pdf}}
    |\!
    \raisebox{-1.0ex}{
    \includegraphics[scale=0.25]{alphastar2.pdf} }
   \rangle $, is new but can also be traced back to the general change of three-body 
  coordinates defined in Eq.~(\ref{eq:gencoordchange}). 
  We identify this time the clusters 1 and 2 with the nucleons $A$--1 and $A$ and the ($A$--2)N-subsystem with cluster 3. 
  For this choice, the position vectors of the spectator agree with the directions used in Fig.~\ref{fig:gen3body} 
  as can be easily seen by comparing to Fig.~\ref{fig:2NFstates}. In order to match  the 
  $| \alphastaronestarone \rangle$ states to the general expressions a recoupling is necessary again resulting in
  a sum over an intermediate spin quantum number $S_{13}^{}$, a $6j$-coefficient and phases which differ from the ones 
  in the previous section because the different identification of clusters used in $| \alphastaronestarone \rangle$
  here and in the previous section requires an opposite order for the coupling of $J_{A-2}$ and $\frac{1}{2}$ to $S_{13}$
  in order to match to Eq.~(\ref{eq:gencoordchange}). Then the quantum numbers can been identified 
  as summarized in  Table~\ref{tab:qnrel2Am2}. Altogether, we find 
    \begin{eqnarray}
   &&
   \langle 
    \deltastaronestarone |
    \betastarAmtwo{}
   \rangle
   \cr      
   && = 
   \langle \!\!
    \raisebox{-0.6ex}{
    \includegraphics[scale=0.25]{betastar1-star1_2NF.pdf}}
    |\!
    \raisebox{-1.0ex}{
    \includegraphics[scale=0.25]{alphastar2.pdf} } 
   \rangle
   \cr      
   &&\cr
   && = 
   \left(-1\right)^{3J_{A-2}^{}
                    +2T_{A-2}^{}
                    +I_{A-1}^{\delta}
                    +l_{A-1}^{\delta}
                    +l_A^{\delta}
                    +S_{12}^{\beta}
                    +T_{12}^{\beta}
                    +\lambda_{}^{\beta}}
   \cr
   &&\cr
   &&
   \phantom{=}
   \times \;
    \hat I_{A-1}^{\delta} \:
    \hat J_{A-1}^{\delta} \: 
    \hat T_{A-1}^{\delta} \: 
    \hat I_A^{\delta}  \: \:
    \hat S_{12}^{\beta} \: 
    \hat J_{12}^{\beta} \: 
    \hat T_{12}^{\beta}  \: 
    \hat I_\lambda^{\beta}
   \cr      
   &&\cr
   &&
   \phantom{=}
   \times
   \sum_{S_{13}^{}} 
    \left(-1\right)^{S_{13}^{}}\,
    \hat S_{13}^{{}^{\;2}}
    \left\{\begin{array}{ccc}
            J_{A-2}^{} & 
            \frac{1}{2} & 
            S_{13}^{}\\[2pt]
            l_{A-1}^{\delta} & 
            J_{A-1}^{\delta} & 
            I_{A-1}^{\delta}
           \end{array}
    \right\}  
   \cr      
   &&\cr
   &&
   \;\;\;\;\;\;\;\;\;\;\;\;
   \times
   \sum_{LS} \hat L_{}^2 \, \hat S_{}^2 \, 
    \left\{\begin{array}{ccc}
            \frac{1}{2} & \frac{1}{2} & S_{12}^{\beta}\\[2pt]
            J_{A-2}^{} & 
            S & 
            S_{13}^{}
           \end{array}
    \right\}
    \left\{\begin{array}{ccc}
            \frac{1}{2} & \frac{1}{2} & T_{12}^{\beta}\\[2pt]
            T_{A-2}^{} & 
            T & 
            T_{A-1}^{\delta}
           \end{array}
    \right\}
   \cr      
   &&\cr
   &&
   \;\;\;\;\;\;\;\;\;\;\;\;\;\;\;\;\;\;\;\;\;\;\;
   \times
    \left\{\begin{array}{ccc}
            l_{A-1}^{\delta} & 
            S_{13}^{} & 
            J_{A-1}^{\delta} \\[2pt]
            l_{A}^{\delta} & \frac{1}{2} & I_A^{\delta} \\[2pt]
            L & S & J
           \end{array}
    \right\}        
    \left\{\begin{array}{ccc}
            l_{12}^{\beta} & S_{12}^{\beta} & J_{12}^{\beta} \\[2pt]
            \lambda_{}^{\beta}  & 
            J_{A-2}^{} & 
            I_\lambda^{\beta} \\[2pt]
            L & S & J   
           \end{array}
    \right\}
   \cr      
   &&\cr
   &&
   \;\;\;\;\;\;\;\;\;\;\;\;\;\;\;\;\;\;\;\;\;\;\;
   \times \;
    \langle n_{A-1}^{\delta} \, 
            l_{A-1}^{\delta},
            \; n_A^{\delta}\, l_A^{\delta}\: : 
            \, L \, | 
            \, n_{12}^{\beta}\, l_{12}^{\beta}\, ,
            \; n_\lambda^{\beta} \, \lambda_{}^{\beta} \: : 
            \, L
    \rangle_{d=\frac{A-2}{A}}^{}   \  \  \  .
   \label{eq:2NF-trafo}
  \end{eqnarray}
for this third matrix element. 

Based on these three ingredients, the transition to  $  | \betastarAmtwo{} \rangle$ states is obtained by   
  \begin{eqnarray}
   \langle 
    \alpha | 
    \betastarAmtwo{}
   \rangle 
   &=&
   \langle
    \raisebox{-0.9ex}{
    \includegraphics[scale=0.3]{alpha.pdf} } 
    |\raisebox{-1.0ex}{
     \includegraphics[scale=0.25]{alphastar2.pdf}}  
   \rangle
   \cr
   &&\cr
   &=&   
   \langle 
    \, \alpha \,| 
    \gammastarone{}   
   \rangle
   \langle 
    \gammastarone{} |
    \deltastaronestarone
   \rangle
   \langle 
    \deltastaronestarone |
    \betastarAmtwo{}
   \rangle
   \cr
   &&\cr
   & = &
   \langle 
    \raisebox{-0.9ex}{
    \includegraphics[scale=0.3]{alpha.pdf}} 
    | \raisebox{-0.7ex}{
      \includegraphics[scale=0.28]{alphastar1.pdf} }
   \rangle 
   \langle \!\!
    \raisebox{-0.7ex}{
    \includegraphics[scale=0.28]{alphastar1.pdf}} 
    |\!
    \raisebox{-0.6ex}{
    \includegraphics[scale=0.25]{betastar1-star1_2NF.pdf} }
   \!\!\rangle \, 
   \langle \!\!
    \raisebox{-0.6ex}{
    \includegraphics[scale=0.25]{betastar1-star1_2NF.pdf}}
    |\!
    \raisebox{-1.0ex}{
    \includegraphics[scale=0.25]{alphastar2.pdf} } 
   \rangle
   \label{eq:def-sum-2NF-book}
  \end{eqnarray}
where summations over the intermediate states is implied. This has been implemented in 
two steps. We decided to first perfom the summation over  $\gammastarone{} $, 
store the intermediate result in core memory and then proceed to the  $\deltastaronestarone$ 
summation. In Table~\ref{tab:dim_antisym}, we also give the dimensions for $\alphastarAmtwo{}$ states 
for a few selected blocks.  The transition matrix element will be made publicly available in HDF5 format. 
The sets generated so far are also tabulated in Appendix~\ref{app:ready}.

\section{$3$N+($A$--$3$)N states for 3N operators}
\label{sec:3nrecoupl}
Although we have not used them in this first application, it will be important in future 
to apply also 3N operators, e.g. to take 3N interactions into account. As can be seen below, 
the calculation of  the pertinent transition coefficients can be done in three steps involving 
four kinds of matrix elements. Therefore, the calculation is not a direct extension of the
$2$N+($A$--$2$)N transitions discussed in the previous section. We note however that further 
extensions towards $4$N, $5$N, \ldots operators can be done using the same three steps 
as outlined now for the $3$N case. Also for this reason, we consider it interesting to explicitly 
give our results for the $3$N+($A$--$3$)N transitions here. 

\begin{figure}
\centering
\includegraphics[scale=0.5]{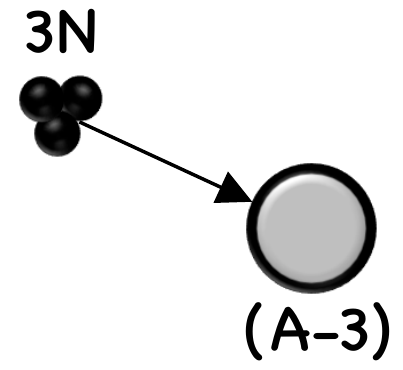}   
\hspace{2cm} \includegraphics[scale=0.5]{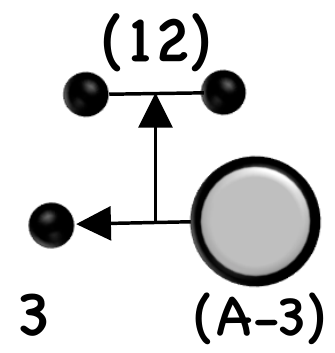} 
\hspace{2cm} \includegraphics[scale=0.5]{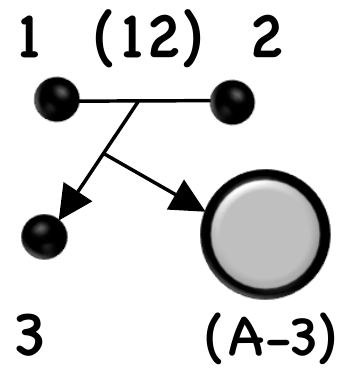} 
\caption{Left hand side: more detailed definition of $| \alphastarAmthree{} \rangle $ states including the 
              direction of momenta. Middle:  the same for $ | \alphastaronestartwo \rangle $ states.  
              Right hand side: the same for $| \alphastaronestarAmthree \rangle$ states. \label{fig:3NFstates} }
\end{figure}

For the application of the $3$N operators, we define states 
\begin{eqnarray}
   &&\cr
   | \alphastarAmthree{} \rangle
   & = & 
   |\alpha_{\,3}^{} \;\, n_{\lambda}^{} \lambda \;\, \alpha_{A-3}^{} \, ; \:
    \left( J_{3}^{} \, \left(\lambda \, J_{A-3}^{}\right)I_\lambda^{} \right) J \; 
    \left( T_{3}^{} \: T_{A-3}^{}\right)T^{}
   \rangle 
   \label{eq:def-alphastarA-3}
  \end{eqnarray}
which single out a three-nucleon cluster. 
The state of the three-nucleon subsystem is labeled here with $\alpha_{3}^{}$. This state is one 
of the antisymmetrized $3$N states obtained by diagonalizing the $3$N system for a given 
total 3N angular momentum $J_3^{}$, isospin $T_3^{}$ and HO quantum number ${\cal N}_3^{}$. 
The states are therefore only meaningfully defined in conjunction with an \textit{a priori} given set of 
cfp for the 3N system.  
The relative motion of this cluster with respect to the ($A$--$3$)N-subsystem is described by the HO quantum number $n_\lambda$
and the orbital angular momentum $\lambda$. The quantum numbers  $\alpha_{A-3}^{}$, $J_{A-3}^{}$ and  $T_{A-3}^{}$ 
label the antisymmetrized state of the ($A$--$3$)N-cluster.  Again, in order to be able to define the total angular momentum, 
one intermediate quantum number $I_{\lambda}^{}$ is necessary, which is given here by coupling $\lambda$ and $J_{A-3}^{}$. 
The conventions for the direction of momenta can be read off from  the left hand side of Fig.~\ref{fig:3NFstates}. 

  \begin{table}
\caption{\label{tab:qnrel3Am3}Left hand side: Identification of quantum numbers of  $| \alphastaronestarAmthree \rangle$  states to the 
   ones of the cooordinate  transformation in Eq.(\ref{eq:gencoordchange}) used to obtain the transitions to $3$N+($A$--$3$) states.  
   Right hand side: the same for $ | \alphastaronestartwo \rangle $
   states. }
  \centering
  \def\arraystretch{1.5}
  \setlength{\tabcolsep}{0.3cm}
  \begin{tabular}{|c|c|c|c|c|c|c|c|c|c|c|c|c|c|c|}
   \hline 
 $n_{12}^{}$  & $n_3^{}$                &  $l_{12}^{}$   & $s_1^{}$           & $s_2^{}$            \\
 $n_{3}^{}$  &  $ n_{\lambda}^{} $&  $l_{3}^{}$   &   $ \frac{1}{2}$  & $  J_{12}^{}$         \\   
 \hline \hline   
 $J_{12}^{} $  &  $l_{3}^{}$    & $s_3^{}$         & $I_{3}^{}$             & $J$    \\
 $J_{3}^{} $   &  $\lambda $  & $ J_{A-3}^{}$  & $I_\lambda^{} $   & $ J$ \\
 \hline \hline
 $t_1^{}$           & $t_2^{}$          & $T_{12}^{}$  & $t_3^{}$ & $T$   \\
 $\frac{1}{2}$    & $ T_{12}^{} $   & $T_{3}^{}$  & $ T_{A-3}^{} $  & $T$ \\
 \hline \hline
     $S_{12}^{}$        &   &   &   &   \\
     $S_{3}^{}$          &   &   &   &      \\
 \hline
  \end{tabular}
  \hspace{2cm}
   \begin{tabular}{|c|c|c|c|c|c|c|c|c|c|c|c|c|c|c|}
   \hline 
 $n_{13}^{}$   &   $n_2^{}$                        &  $l_{13}^{}$   & $s_1^{}$           & $s_2^{}$            \\
 $n_{A-2}^{}$  &  $ n_{\lambda}^{} $         &  $l_{A-2}^{}$   &   $ \frac{1}{2}$  & $ J_{12}^{} $         \\   
 \hline \hline   
 $J_{13}^{} $  &  $l_{2}^{}$                & $s_3^{}$         & $I_{2}^{}$             & $J$    \\
 $J_{A-2}^{} $  &  $\lambda$   & $ J_{A-3}^{}$  & $I_{\lambda} $          & $ J$ \\
 \hline \hline
 $t_1^{}$           & $t_2^{}$          & $T_{13}^{}$  & $t_3^{}$ & $T$   \\
 $\frac{1}{2}$    & $ T_{12}^{}  $  & $T_{A-2}^{}$  & $ T_{A-3}^{} $  & $T$ \\
 \hline \hline
     $S_{13}^{}$        &   &   &   &   \\
     $S_{A-2}^{}$      &   &   &   &      \\
 \hline
  \end{tabular}
\end{table}

To define the transition matrix elements 
$ \langle 
    \,\alpha\, | 
    \betastarAmthree{}
   \rangle      
   = 
   \langle \!\!
    \raisebox{-0.9ex}{
    \includegraphics[scale=0.3]{alpha.pdf} } 
    \!\!|\!\!
    \raisebox{-0.9ex}{
    \includegraphics[scale=0.22]{alphastar3.pdf}  }
   \!\!\rangle$, 
we need to introduce two further sets of intermediate states:
\begin{eqnarray}
   | \alphastaronestartwo \rangle
   & = &
   | \alphastarone{A-2} \;\, n_{\lambda}^{} \lambda \;\, \alpha_{12}^{} \, ;
   \cr
   &&
   \;\;
   \left( \left( J_{A-3}^{} \, \left(l_{A-2}^{} \, s_3^{}\right) I_{A-2}^{} \right) J_{A-2}^{} \, 
   \left(\lambda \: J_{12}^{}\right)I_\lambda^{} \right) J \; 
   \left( \left(T_{A-3}^{} \, t_3^{}\right) T_{A-2}^{} \: T_{12}^{} \right)T
   \rangle \cr  
&&\cr    
   | \alphastaronestarAmthree \rangle
   & = &
   | \alphastarone{\,3} \;\, n_{\lambda}^{} \lambda \;\, \alpha_{A-3}^{} \, ;
   \cr
   &&
   \;\;
   \left( \left( J_{12}^{} \, \left(l_{3}^{} \, s_3^{}\right)I_{3}^{} \right) J_{3}^{}  \, 
   \left(\lambda \: J_{A-3}^{}\right)I_\lambda^{} \right) J \; 
   \left( \left(T_{12}^{} \, t_3^{}\right) T_{3}^{} \: T_{A-3}^{} \right)T
   \rangle 
\end{eqnarray}
which are also depicted in Fig.~\ref{fig:3NFstates}. 
Three of the four involved matrix elements are already known from previous calculations. In the first step, we will 
need the transition coefficients to $2$N+($A$--$2$)N states 
 $ \langle 
     \,\alpha\, | 
     \gammastarAmtwo{} 
    \rangle   =  \langle \!\!
    \raisebox{-0.9ex}{
    \includegraphics[scale=0.3]{alpha.pdf} } \! | \!
    \raisebox{-1.0ex}{
    \includegraphics[scale=0.25]{alphastar2.pdf} } 
   \!\!\rangle $ and, in the final step, cfp for the 3N system 
   $ \langle
     \epsstaronestarAmthree | 
     \betastarAmthree{}
    \rangle =  \langle \!
    \raisebox{-1.0ex}{
    \includegraphics[scale=0.22]{alphastar1-starA-3.pdf} } \! | \!
    \raisebox{-0.9ex}{
    \includegraphics[scale=0.22]{alphastar3.pdf} } 
   \!\!\rangle$. Also the matrix elements 
   $ \langle 
     \gammastarAmtwo{} |\!
     \deltastaronestartwo
    \rangle  =   \langle  \!
    \raisebox{-1.0ex}{
    \includegraphics[scale=0.25]{alphastar2.pdf} } \! | \!
    \raisebox{-1.0ex}{
    \includegraphics[scale=0.22]{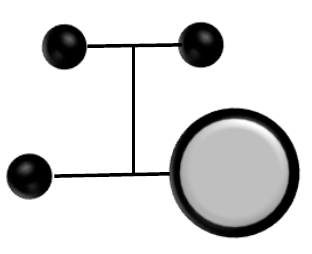} } 
   \!\!\rangle  $ are given by the known cfp of the ($A$--2)N-system.
The only new ingredient 
$ \langle 
     \deltastaronestartwo | \!
     \epsstaronestarAmthree
    \rangle =  \langle  \!
    \raisebox{-1.0ex}{
    \includegraphics[scale=0.22]{betastar1-star2.pdf} }\! |\!
    \raisebox{-1.0ex}{
    \includegraphics[scale=0.22]{alphastar1-starA-3.pdf} } 
   \!\!\rangle \, $ can again be traced back to the general change of three body 
  coordinates defined in Eq.~(\ref{eq:gencoordchange}). Comparing 
  the coordinates depicted in middle and right of Fig.~\ref{fig:3NFstates} 
  with the general coordinates of Fig.~\ref{fig:gen3body}, one easily identifies. 
  clusters 1, 2 and 3 with the third nucleon, the 2N-cluster and the ($A$--3)N-subsystem. 
  Also the direction of the coordinates agree in this case with the general ones.  
  But the coupling scheme of both kinds of states, 
  $ | \alphastaronestartwo \rangle$ and  $| \alphastaronestarAmthree \rangle$, do not fit to the 
  general expression. For the $ | \alphastaronestartwo \rangle$, we therefore recouple the angular momenta of the ($A$--2)N-subsystem from 
  \begin{equation}
      \left( J_{A-3}^{} \, \left(l_{A-2}^{} \, s_3^{}\right) I_{A-2}^{} \right) J_{A-2}^{} \ \ \mbox{ to } \ \  
      \left(  \, l_{A-2}^{} \, \left( s_3^{}  J_{A-3}^{} \right) S_{A-2}^{} \right) J_{A-2}^{}
  \end{equation}  
 whereby introducing the new intermediate  spin quantum number  $S_{A-2}^{} $. Similarly, the original coupling of the 
 $| \!
     \epsstaronestarAmthree
    \rangle $ states has to be recoupled from 
 \begin{equation}
   \left( J_{12}^{} \, \left(l_{3}^{} \, s_3^{}\right)I_{3}^{} \right) J_{3}^{}     \ \ \mbox{ to } \ \    \left( \, l_{3}^{} \,  \left(s_3^{} J_{12}^{} \right)S_{3}^{} \right) J_{3}^{}  
 \end{equation}    
 where $S_{3}^{}$ was introduced as a new spin quantum number. Both recouplings lead to $6j$-coefficients and phases. Then the 
 quantum numbers can be matched to the ones of Eq.~(\ref{eq:gencoordchange}) as shown in Table~\ref{tab:qnrel3Am3}. The complete 
 expression then reads 
  \begin{eqnarray}
    &&
    \langle 
     \deltastaronestartwo | \!
     \epsstaronestarAmthree
    \rangle
    =
   \langle  \!
    \raisebox{-1.0ex}{
    \includegraphics[scale=0.22]{betastar1-star2.pdf} }\! |\!
    \raisebox{-1.0ex}{
    \includegraphics[scale=0.22]{alphastar1-starA-3.pdf} } 
   \!\!\rangle 
    \cr
    &&\cr
    && 
    =
    \left(-1\right)^{3J_{A-3}^{}
                     +l_{A-2}^{\delta}
                     +I_{A-2}^{\delta}
                     +2T_{A-2}^{\delta}
                     +l_3^{\epsilon}
                     +I_3^{\epsilon}
                     +1
                     +J_{12}^{}
                     +\lambda_{}^{\delta}
                     +\lambda_{}^{\epsilon}
                     }
    \cr
    &&\cr
    && 
    \phantom{=}
    \times \;
     \hat I_{A-2}^{\delta} \:
     \hat J_{A-2}^{\delta} \: 
     \hat T_{A-2}^{\delta} \: 
     \hat I_\lambda^{\delta} \:\:
     \hat I_3^{\epsilon} \:
     \hat T_3^{\epsilon} \:
     \hat J_3^{\epsilon} \: 
     \hat I_\lambda^{\epsilon}
    \cr
    &&\cr
    &&
    \phantom{=}
    \times 
     \sum_{S_{A-2}^{} S_3^{}} 
      \left(-1\right)^{S_{A-2}^{}+S_{3}^{}} \,
      \hat{S}_{A-2}^{} \, \hat{S}_{3}^{\,^{2}}
    \cr
    &&
    \;\;\;\;\;\;\;\;\;\;\;\;\;\;\;\;\;\;\;\;
    \times
     \left\{\begin{array}{ccc}
             J_{A-3}^{} & 
             \frac{1}{2} & 
             S^{}_{A-2}\\[2pt]
             l_{A-2}^{\delta} & 
             J_{A-2}^{\delta} & 
             I_{A-2}^{\delta}
            \end{array}
     \right\}  
     \left\{\begin{array}{ccc}
             J_{12}^{} & \frac{1}{2} & S_{3}^{}\\[2pt]
             l_3^{\epsilon} & J_{3}^{\epsilon} & I_3^{\epsilon}
            \end{array}
     \right\}  
    \cr
    &&\cr
    &&
    \;\;\;\;\;\;\;\;\;\;\;\;\;\;\;\;\;\;\;\;
    \times \,\, \sum_{LS}\hat L_{}^2 \, \hat S_{}^2 
     \left\{\begin{array}{ccc}
             J_{12}^{} & \frac{1}{2} & S_{3}^{}\\[2pt]
             J_{A-3}^{} & 
             S & 
             S_{A-2}^{}
            \end{array}
     \right\}
     \left\{\begin{array}{ccc}
             T_{12}^{} & \frac{1}{2} & T_{3}^{\epsilon}\\[2pt]
             T_{A-3}^{} & 
             T & 
             T_{A-2}^{\delta}
            \end{array}
     \right\}
     \cr
     &&\cr
     && 
     \;\;\;\;\;\;\;\;\;\;\;\;\;\;\;\;\;\;
     \;\;\;\;\;\;\;\;\;\;\;\;\;\;\;\;\;
     \times\,
      \left\{\begin{array}{ccc}
              l_{A-2}^{\delta} & 
              S_{A-2}^{} & 
              J_{A-2}^{\delta} \\[3pt]
              \lambda^{\delta}_{} & 
              J_{12}^{} & I_\lambda^{\delta} \\[3pt]
              L & S & J     
             \end{array}
      \right\}        
      \left\{\begin{array}{ccc}
              l_{3}^{\epsilon} & S_{3}^{} & J_{3}^{\epsilon} \\[3pt]
              \lambda^{\epsilon}_{}  & 
              J_{A-3}^{} & 
              I_\lambda^{\epsilon}    \\[3pt]
              L      & S     & J    
             \end{array}
      \right\}
     \cr
     &&\cr
     &&
     \;\;\;\;\;\;\;\;\;\;\;\;\;\;\;\;\;\;
     \;\;\;\;\;\;\;\;\;\;\;\;\;\;\;\;\;
     \times\,
      \langle n_{A-2}^{\delta} \, l_{A-2}^{\delta}\, ,\;
              n_\lambda^{\delta} \, \lambda^{\delta}\, : L | \,
              n_{3}^{\epsilon}\, l_3^{\epsilon} ,\; 
              n_\lambda^{\epsilon} \, \lambda_{}^{\epsilon} \, :  L
      \rangle_{d=\frac{2\left(A-3\right)}{A}}^{}\; 
     \cr
     &&\cr
     &&
   \label{eq:equiv-3NF-trafo}
  \end{eqnarray}
We note that the extension to $4$N-($A$--4)N transitions will only require straightforward 
changes of this relation. 
Based on these three ingredients, the transition to  $  | \betastarAmthree{} \rangle$ states is obtained by   
  \begin{eqnarray}
   \langle 
    \,\alpha\, | 
    \betastarAmthree{}
   \rangle      
   & = &
   \langle \!\!
    \raisebox{-0.9ex}{
    \includegraphics[scale=0.3]{alpha.pdf} } 
    \!\!|\!\!
    \raisebox{-0.9ex}{
    \includegraphics[scale=0.22]{alphastar3.pdf}  }
   \!\!\rangle
   \cr
   &&\cr
   & = &
    \langle 
     \,\alpha\, | 
     \gammastarAmtwo{} 
    \rangle 
    \langle 
     \gammastarAmtwo{} |\!
     \deltastaronestartwo
    \rangle 
    \langle 
     \deltastaronestartwo | \!
     \epsstaronestarAmthree
    \rangle 
    \langle
     \epsstaronestarAmthree | 
     \betastarAmthree{}
    \rangle
   \cr
   &&\cr
   & = &
   \langle \!\!
    \raisebox{-0.9ex}{
    \includegraphics[scale=0.3]{alpha.pdf} } \! | \!
    \raisebox{-1.0ex}{
    \includegraphics[scale=0.25]{alphastar2.pdf} } 
   \!\!\rangle
   \langle  \!
    \raisebox{-1.0ex}{
    \includegraphics[scale=0.25]{alphastar2.pdf} } \! | \!
    \raisebox{-1.0ex}{
    \includegraphics[scale=0.22]{betastar1-star2.pdf} } 
   \!\!\rangle  
   \langle  \!
    \raisebox{-1.0ex}{
    \includegraphics[scale=0.22]{betastar1-star2.pdf} }\! |\!
    \raisebox{-1.0ex}{
    \includegraphics[scale=0.22]{alphastar1-starA-3.pdf} } 
   \!\!\rangle \, 
   \langle \!
    \raisebox{-1.0ex}{
    \includegraphics[scale=0.22]{alphastar1-starA-3.pdf} } \! | \!
    \raisebox{-0.9ex}{
    \includegraphics[scale=0.22]{alphastar3.pdf} } 
   \!\!\rangle
   \label{eq:def-sum-3NF-book}
  \end{eqnarray}
where again sums over intermediate states are implied. Our implementation generates 
the complete expression in three steps where the results dependent on the 
intermediate quantum numbers. As can be seen from Table~\ref{tab:list-of-states}, the size 
of these sets of intermediate states are orders of magnitude larger than the set of completely 
antisymmetrized $A$-body states implying not only more floating point operations but also larger 
memory requirements. The parallelization on a distributed memory massively parallel computer 
therefore required a compromise of most efficient memory usage and minimalization of 
communication between the processes. 
The details of the technical implementation are discussed in more detail in \cite{Antisymmetrisationi:2013vg}.
We stress again that an extension to more complex operators can be done using the same algorithms 
in future. 

\section{Results}
  \label{sec:energies}

As a first application of the cfp and transition coefficients, we are now presenting binding energies
for light nucei based on these  Jacobi HO states. For this test, we only use NN interactions. In order 
to be able to obtain converged results, we rely on SRG evolved interactions \cite{Bogner:2007hn}
starting from  the chiral interaction at next-to-next-to-next-to-leading order (N$^3$LO) 
from the Idaho group~\cite{Entem:2003hx} considering NN partial waves up to $J_{\rm{NN}}^{^{\;\rm{max}}}=6$.
The charge dependence of the nuclear force is taken into account 
by building an averaged NN interaction as outlined in \cite{Kamuntavicius:1999js}. The relative weight 
of proton-proton (pp), neutron-neutron (nn)  and neutron-proton (np) interactions in isospin $T_{12}=1$ states 
thereby depend on the nucleus considered. For pp and nn interactions, we added the 
electromagnetic interactions of AV18~\cite{Wiringa:1995co}.

For the solution of the Schr\"odinger equation and taking only NN interactions into account, 
we rewrite the matrix elements of the Hamiltonian in the antisymmetrized $A$-nucleon 
basis $|\, \alpha \, \rangle$ as  
\begin{equation}
   \langle \, \alpha \, | 
   H_A^{}
   | \,\beta\, \rangle 
   =
   \langle 
    \, \alpha \, | 
    {\gammastarAmtwo{}}
   \rangle
   \langle 
    {\gammastarAmtwo{}} |
    \sum_{i<j=1}^{A} 
    H_{ij}^{}
    | \deltastarAmtwo{}
   \rangle 
   \langle 
    \deltastarAmtwo{} |
    \,\beta\, 
   \rangle  \ \ .
   \label{eq:schroed-eq-NNF-only}  
\end{equation}   
The coefficients  $ \langle 
    \, \alpha \, | 
    {\gammastarAmtwo{}}
   \rangle$ are known from the previous sections, independent of the HO frequency $\omega$ and conserve 
total $J$,$T$ and ${\cal N}$. The two-nucleon matrix elements can be simplified making use 
of the identity of the nucleons 
\begin{equation}   
     \langle 
    {\gammastarAmtwo{}}^{} |
    \sum_{i<j=1}^{A} 
    H_{ij}^{}
    | \deltastarAmtwo{}
   \rangle 
   = 
   \delta_{\blockqn \gamma_{A-2}^{} \,
           \blockqn \delta_{A-2}^{} \,
           }^{}  \  
    \dbinom{A}{2} 
    \langle 
     \,\gamma^{}_{12} \,|
      \Big(\frac{2}{A}\, T^{}_{12} + V^{}_{12} \Big) 
     |\, \delta^{}_{12} \,
    \rangle \ \ .   
   \  \end{equation}
It is convenient to express the relative kinetic energy in terms of an NN operator. This matrix elements 
conserves $J$ and $T$ in our approximation. It will however not conserve ${\cal N}$. Nevertheless,
all quantum numbers of the ($A$--2)N-subsystem are conserved as indicated by the Kronecker $\delta$ 
symbols. As usual, the NN interaction is diagonal in $J_{12}$ and $T_{12}$. Therefore, the application of 
$H_{A}$ on an arbitrary $A$-body state can be separated in three steps that only evolve 
rather low dimensional operations. The use of Jacobi coordinates further reduces the dimensionality
since the problem can be solved for each $J$ and $T$ independently.  Therefore, once the cfp and transition 
coefficients are known, the calculations are much simpler and can be done quickly. In the following, we therefore 
map out the complete dependence on the HO frequency $\omega$ of the energy of each state for all model space 
size defined by the maximal HO energy $\cal N$.

\begin{figure}[tbp]
 \begin{center}
    \includegraphics[scale=0.36]{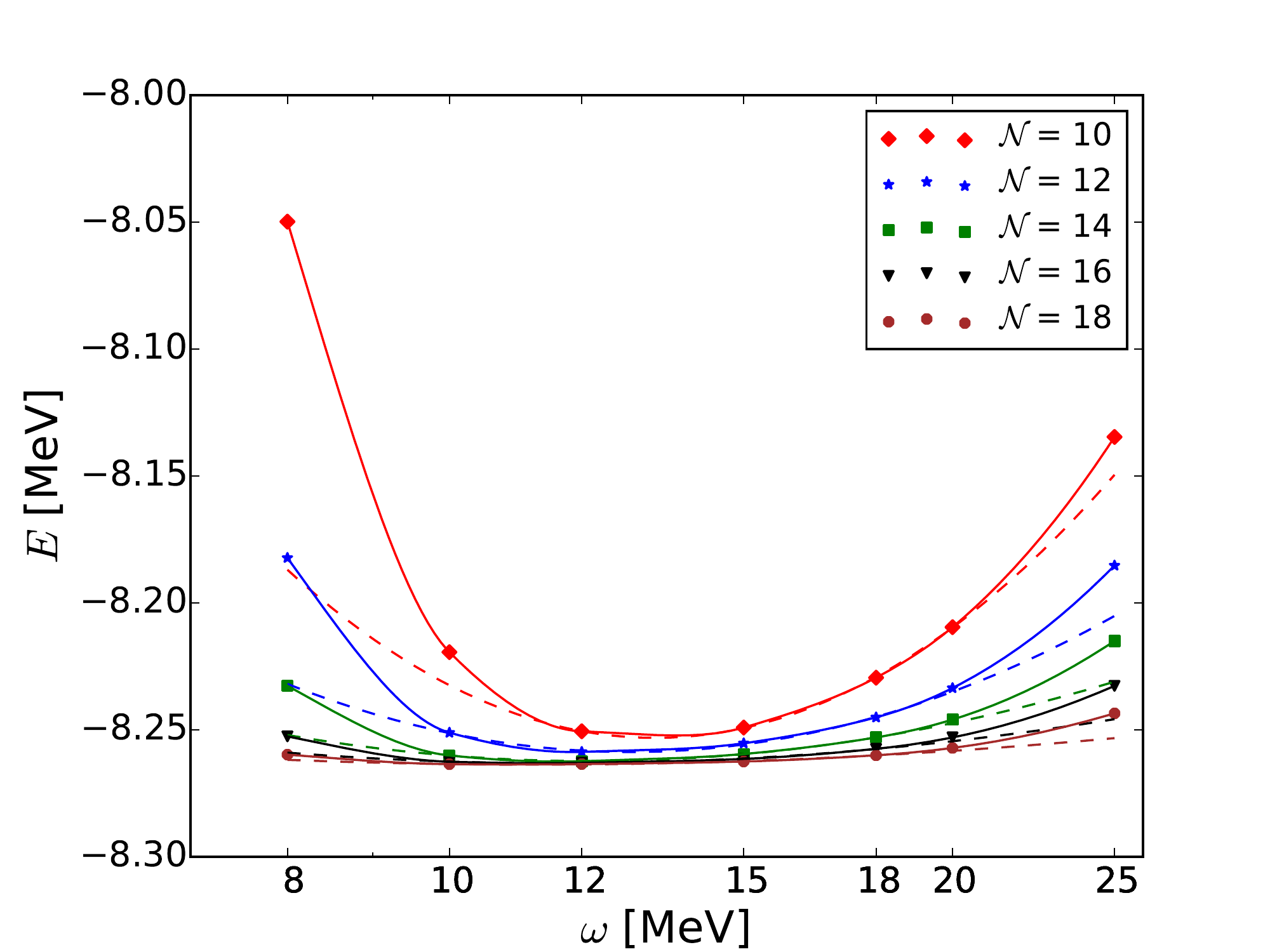}
    \includegraphics[scale=0.36]{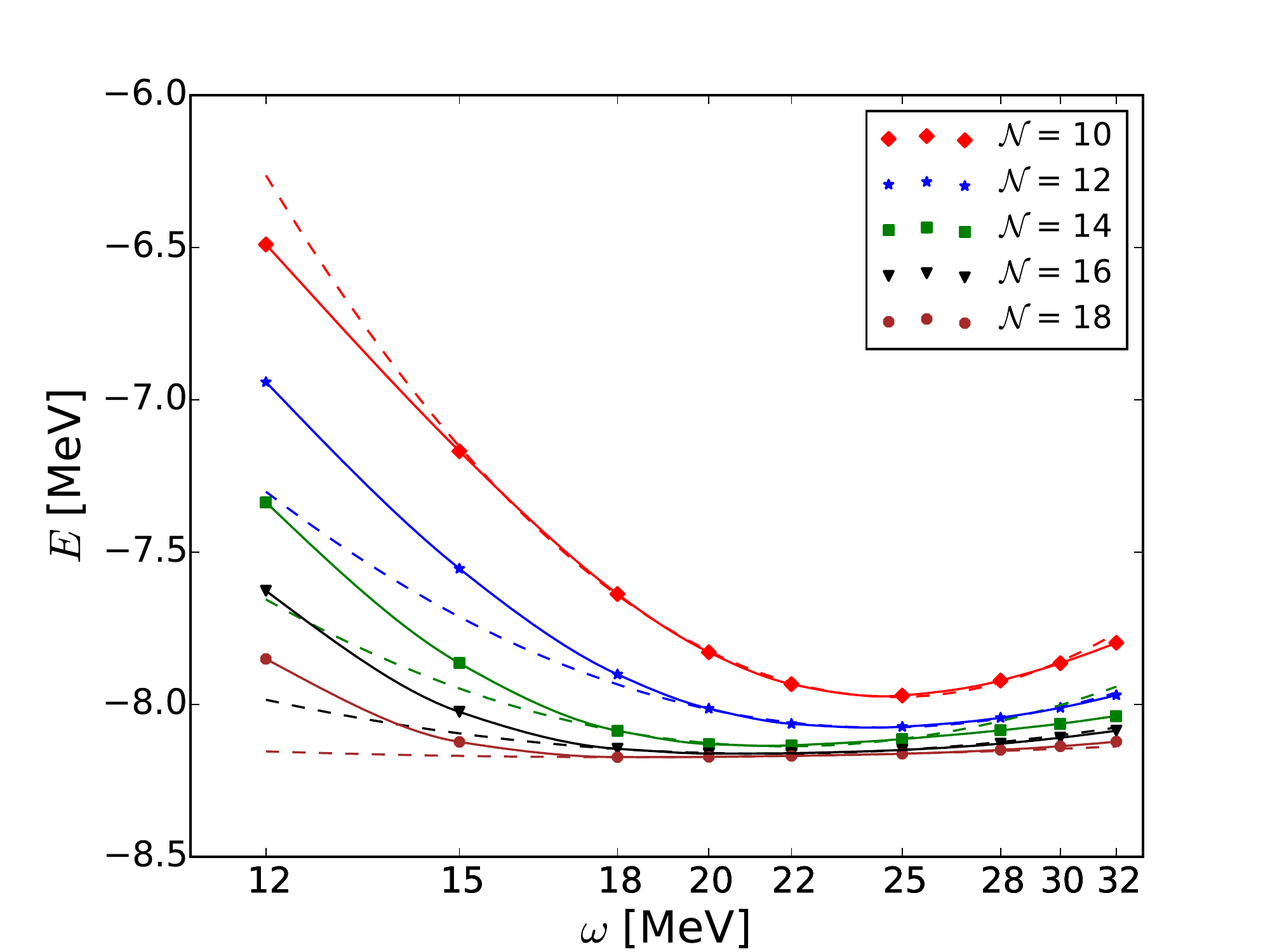}
  \end{center}  
    \caption{\label{fig:3Homdep}$\omega$-dependence of the $_{}^3$H binding energy for $\lambda=1.5$~fm$^{-1}$ (left) and 
      $\lambda=2.5$~fm$^{-1}$ (right). Results for different model space sizes can be distinguished 
      by the different markers and colors. The solid lines are added to guide the eye, the dashed lines are 
      obtained using Eq.~(\ref{eq:eneromdep}).}
\end{figure}

\subsection{Extrapolation procedure for the example of $_{}^3$H}
We found that in most cases the $\omega$-dependence around the optimal frequency $\omega_0$ can be well 
described by  the ansatz
 \begin{eqnarray}
   E_{b}^{}\left(\omega\right)
   & = &
   E_{\,\Ntot}^{}+ \kappa
    \left(\log\left(\omega\right)
         -\log\left(\omega_{opt}^{}\right)
    \right)^2 \ \   . 
    \label{eq:eneromdep}
    \end{eqnarray}
By a simple fit, the parameters $E_{\,\Ntot}^{}$, $\omega_0$ and $\kappa$ are extracted from the 
results for a given model space size $\cal N$ (in a limited regions around $\omega_0$). 
As an example, we show the $\omega$-dependence 
as solid lines for the case of $^3$H in Fig.~\ref{fig:3Homdep} for two different SRG cutoffs $\lambda=1.5$~fm$^{-1}$ and 
$\lambda=2.5$~fm$^{-1}$. Different lines correspond to different model space sizes. The result of the fit to Eq.~(\ref{eq:eneromdep}) 
is also shown by the  dashed lines. As expected the results become less $\omega$ dependent for larger model spaces. It is also clearly 
seen that the convergence is much faster for smaller $\lambda$. For this small system, it is still possible to 
obtain converged results for $\lambda=2.5$~fm$^{-1}$, this will however not be possible anymore for the more 
complex nuclei. For  $\lambda=1.5$~fm$^{-1}$, the convergence is fast enough that we will also be able to present 
converged numbers for $p$-shell nuclei for this first application.  Around 
the optimal $\omega=\omega_0$ at the minimum, Eq.~(\ref{eq:eneromdep}) is able to reproduce the $\omega$-dependence very well. 
The agreement of the calculations with the results even improves for larger model spaces. Therefore, we will extract our final 
result for a given model space using the fit result $E_{\,\Ntot}^{}$. 

   \begin{figure}[tbp]
   \begin{center}
    \includegraphics[scale=0.35]{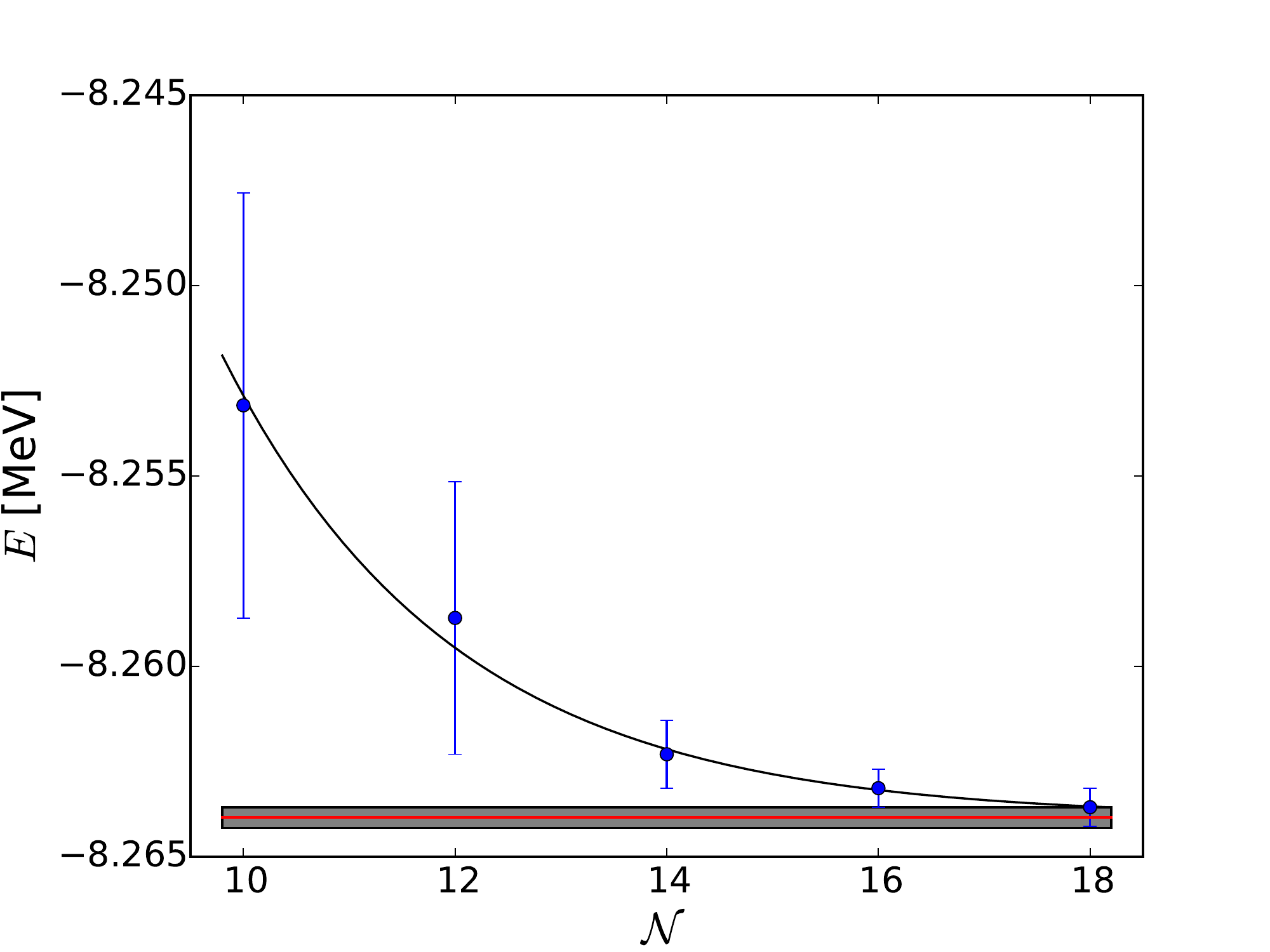}
    \includegraphics[scale=0.35]{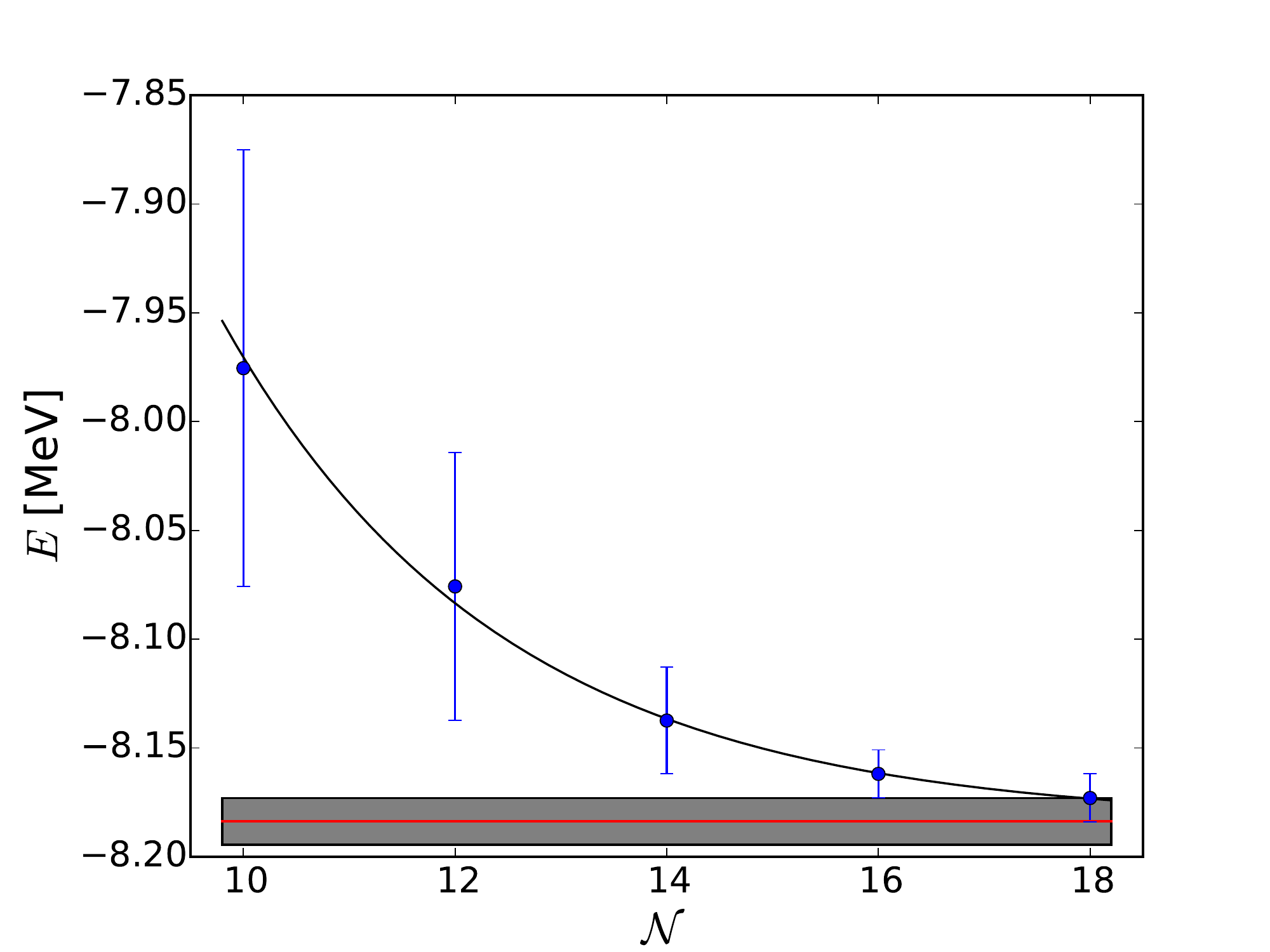}
    \end{center}
    \caption{\label{fig:3Hndep}$\cal N$-dependence of the $_{}^3$H binding energy for $\lambda=1.5$~fm$^{-1}$ (left) and 
      $\lambda=2.5$~fm$^{-1}$ (right). The black line is obtained using Eq.~(\ref{eq:enerNdep}). The 
      result of the exponential extrapolation os indicated by the red line. The shaded area indicates the estimated 
      uncertainty of the final result which is given by the difference of the result for the largest model 
      space and the exponential extrapolation.} 
   \end{figure}

 \begin{table}
  \centering
  \renewcommand*{\arraystretch}{1.5}
  \setlength{\tabcolsep}{0.3cm}
  \caption{\label{tab:energyres}Ground state and excitation energies 
  	       of $^3$H, $^4$He, $^6$He, $^6$Li and $^7$Li 
           for different cutoff-parameters $\lambda$
           in comparison to the experimental values \cite{Audi:2012dv,Tilley:2002hs}.
           $^*$ denotes $^6$He excitation  energies where the uncertainty estimate might not be 
           reliable. See text for further explanations.}
  \begin{tabular}{c|c|c|c|c|c}
   \multicolumn{4}{c}{}\\[5pt]
   \hline
   \hline
   $\lambda$ & 
   $^3$H & 
   $^4$He & 
   $^6$He & 
   $^6$Li & 
   $^7$Li \\ 
   $\left[{\rm{fm}}^{-1}\right]$ &
   [MeV]&
   [MeV]&
   [MeV]&
   [MeV]&
   [MeV]\\
   \hline
   &&&&&\\
   1.0 &
   -7.460 &
   -24.271 &
   -26.76(4)  &
   -29.70(1)  &
   -37.04(10)
   \\
   &&&
   2.21(4) $^*$ &
   3.10(2)  &
   0.133(2) 
   \\[5pt]
   1.2 &
   -7.929 &
   -26.549 &
   -28.76(10)  &
   -31.77(8)  &
   -39.68(28)
   \\
   &&&
   2.42(7) $^*$ &
   3.08(1)   &
   0.219(2) 
   \\[5pt]
   1.5 &
   -8.264 &
   -28.173  &
   -29.91(30)  &
   -32.90(29)  &
   -41.12(84)  
   \\
   &&&
   2.58(9) $^*$ &
   2.94(1)   &
   0.335(9) 
   \\[5pt]
   1.8 &
   -8.332 &
   -28.397(1) &
   -29.89(93)  &
   -32.83(96)  &
   -41.67(303)
   \\
   &&&
   2.78(8) $^*$ &
   2.81(2)  &
   0.420(27) 
   \\[5pt]
   2.0 &
   -8.314(4) &
   -28.189(3) &
   -29.80(191)  &
   -32.80(204)  &
   -41.50(505)  
   \\
   &&&
   2.72(7) $^*$ &
   2.76(3)   &
   0.428(33) 
   \\[5pt]
   2.2 &
   -8.269(6) &
   -27.890(10) &
   -30.35(384) &
   -33.68(434)  &
   -43.02(937)
   \\
   &&&
   2.65(6) $^*$ &
   2.68(7)  &
   0.421(39) 
   \\[5pt]
   2.5 &
   -8.184(11)  &
   -27.378(23)  &
   -34.26(1026)  &
   -38.68(1192)  &
   -51.72(2302)  
   \\
   &&&
   2.53(4) $^*$ &
   2.53(10)   &
   0.443(60) 
   \\[5pt]
   \hline
   exp.&
   -8.482  &
   -28.296  &
   -29.271(54)  & 
   -31.994  & 
   -39.245(7)  \\
   &&& 1.797(25) & 2.186(2)  & 0.478(3)  \\
   \hline
   \hline
  \end{tabular}
 \end{table}

Fig.~\ref{fig:3Hndep} summarizes these results again for $^3$H and $\lambda=1.5$~fm$^{-1}$ and $\lambda=2.5$~fm$^{-1}$. 
In order to extract the converged binding energy from the $\cal N$-dependence, we assume a simple exponential dependence
\begin{eqnarray}
E_{\,\Ntot}^{} = 
   E_{\infty}^{} 
    + A_{}^{} \; e^{-b_{}^{}\Ntot} \ \ .
   \label{eq:enerNdep}
  \end{eqnarray}
We note that other effective field theory motivated extrapolation schemes have been discussed 
in \cite{Jurgenson:2013jn,Furnstahl:2015gp}. They should be employed in forthcoming publications. 
But for our purpose here, the exponential interpolation was sufficiently accurate to determine the final 
binding energies. In order to determine $  E_{\infty}^{}$, $A$ and $b$, we first assign an uncertainty estimate 
to each $E_{\,\Ntot}^{}$. This uncertainty estimate will serve as a weight for the fit ensuring that 
automatically more weight is given to the larger model spaces for the determination of the parameters. It is not 
the aim to assign a realistic absolute uncertainty to each individual $E_{\,\Ntot}^{}$, but only to determine an
estimate of the relative uncertainties of the different $E_{\,\Ntot}^{}$. We therefore assigned 
$\Delta E_{\,\Ntot}^{} =  | E_{\,\Ntot}^{}-E_{\,\Ntot+2}^{} | $ for the uncertainty. For the largest model space $\Ntot_{\rm max}$
considered, we used the estimate of the previous model space. Therefore, the two largest model spaces 
contribute to the fit with equal weight. The errorbars obtained in this way are also shown in Fig.~\ref{fig:3Hndep}
together with the result of the fit. Our final binding energy is then given by $E_{\infty}^{}$. In order to obtain 
a conservative estimate of the uncertainty of this result, we assign 
$\Delta E_{\infty}^{} = | E_{\infty}^{} -  E_{\,\Ntot_{\rm max}}^{} | $.  $E_{\infty}^{}$  and $\Delta E_{\infty}^{}$ are 
shown in the figures as red line and the surrounding shaded area. 

Similar calculations have been performed for seven values of $\lambda$ between 1.0~fm$^{-1}$ and 2.5~fm$^{-1}$.
The results for $^3$H and all other nuclei considered in 
this work are summarized in Table~\ref{tab:energyres}. 
The energies for $^3$H have been obtained with very high accuracy since 
we were able to obtain the cfp for very large model spaces for this system. 
We also note that the results for $^3$H agree well with 
the results obtained by solving Faddeev equations \cite{Nogga:2002kw}. 

\subsection{$_{}^4$He}

The 3N system is an interesting test case since the convergence is quite slow because the 
nucleus is not very compact. So contributions from large $\Ntot$ can be tested. But since the cfp 
already separate the NN-subsystem, they do not test our transition coefficients to 2N-($A$--2)N-states. 
Therefore, we consider as a second test nucleus $^4$He. Here, for the first time, the transition 
matrix elements enter. We expect however much faster convergence with respect to $\Ntot$ 
since the nucleus is more compact. At the same time, solutions of 
Yakubovsky equations \cite{Nogga:2002kw} are available so that the results and the extrapolation procedure 
can be checked. 

\begin{figure}[tbp]
 \begin{center}
    \includegraphics[scale=0.36]{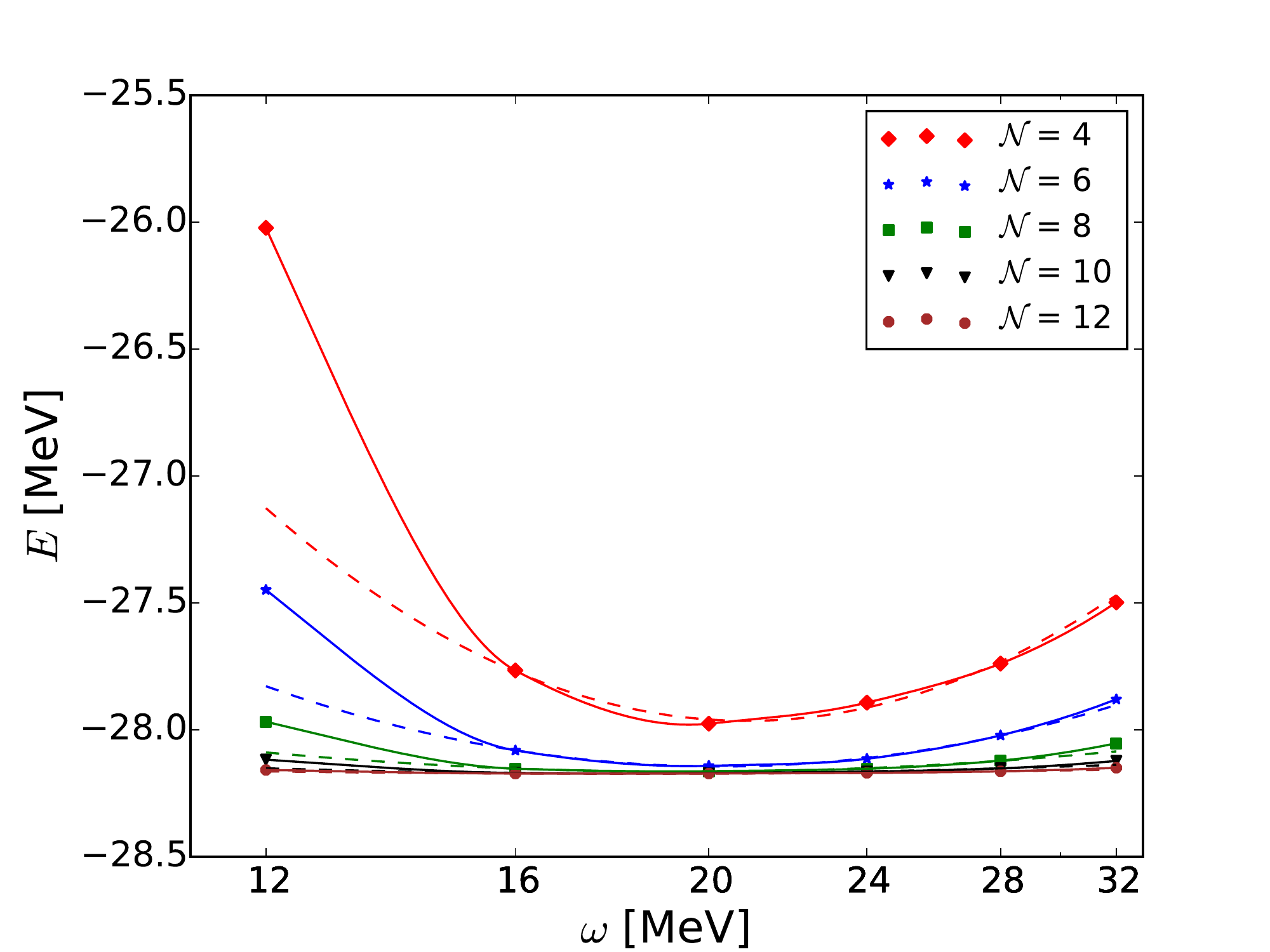}
    \includegraphics[scale=0.36]{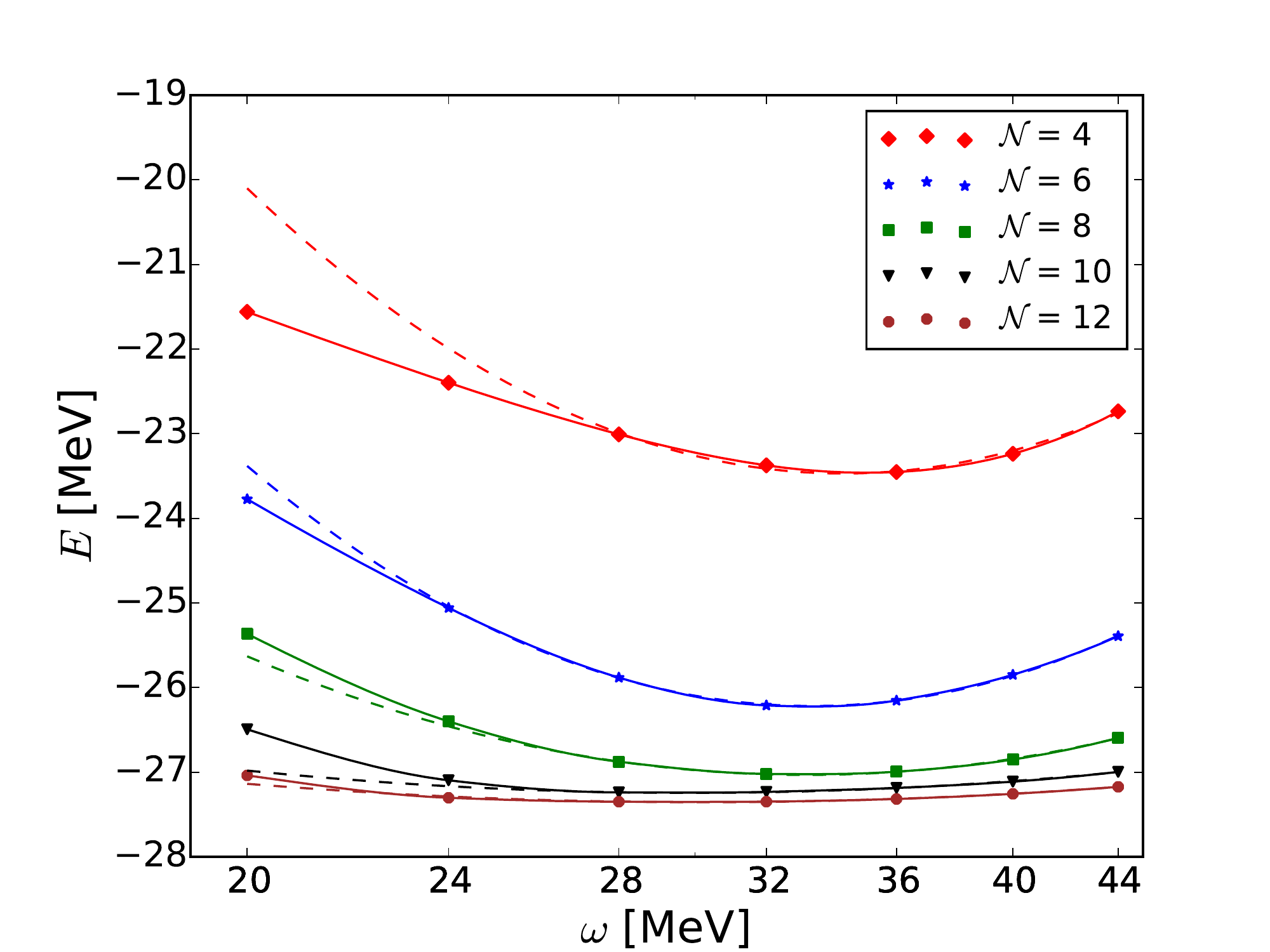}
  \end{center}  
    \caption{\label{fig:4Heomdep}$\omega$-dependence of the $_{}^4$He binding energy for $\lambda=1.5$~fm$^{-1}$ (left) and 
      $\lambda=2.5$~fm$^{-1}$ (right). Results for different model space sizes can be distinguished 
      by the different markers and colors. The solid lines are added to guide the eye, the dashed lines are 
      obtained using Eq.~(\ref{eq:eneromdep}).}
\end{figure}

\begin{figure}[tbp]
\begin{center}
    \includegraphics[scale=0.35]{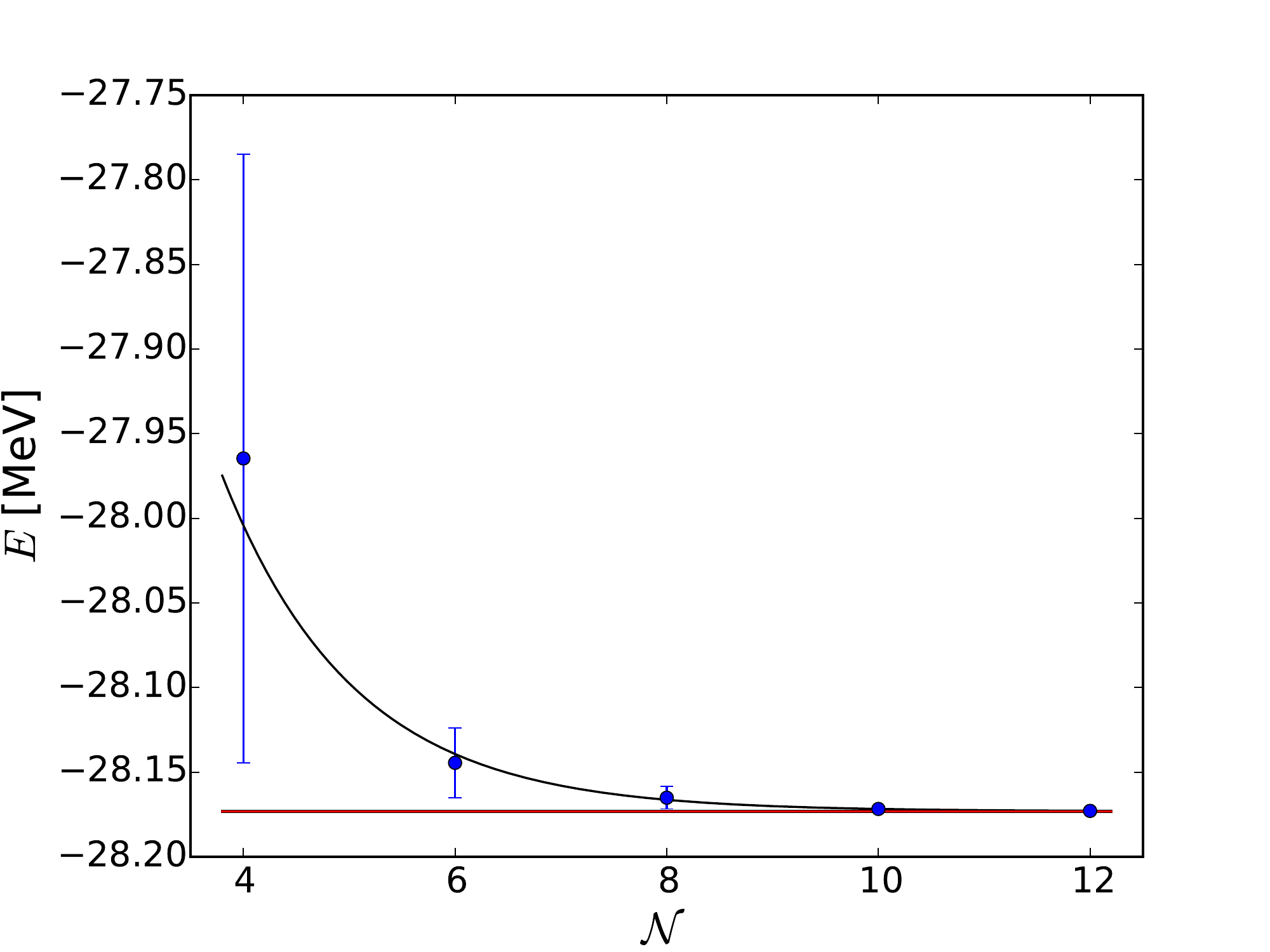}
    \includegraphics[scale=0.35]{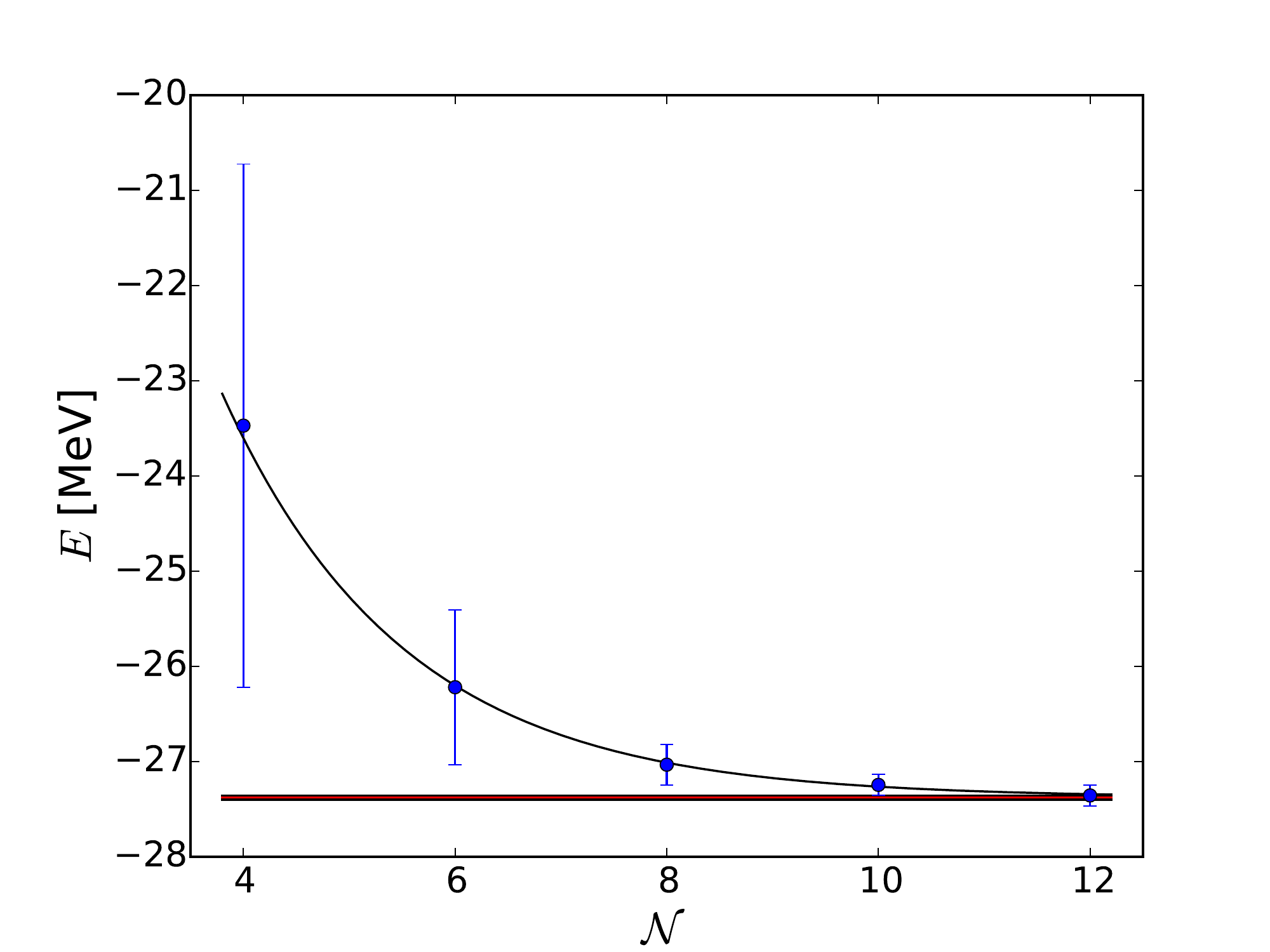}
 \end{center}   
    \caption{\label{fig:4Hendep}$\cal N$-dependence of the $_{}^4$He binding energy for $\lambda=1.5$~fm$^{-1}$ (left) and 
      $\lambda=2.5$~fm$^{-1}$ (right). The black line is obtained using Eq.~(\ref{eq:enerNdep}). The 
      result of the exponential extrapolation is indicated by the red line. The shaded area indicates the estimated 
      uncertainty of the final result which is given by the difference of the result for the largest model 
      space and the exponential extrapolation.} 
   \end{figure}

\begin{figure}[tbp]
\begin{center}
    \includegraphics[scale=0.35]{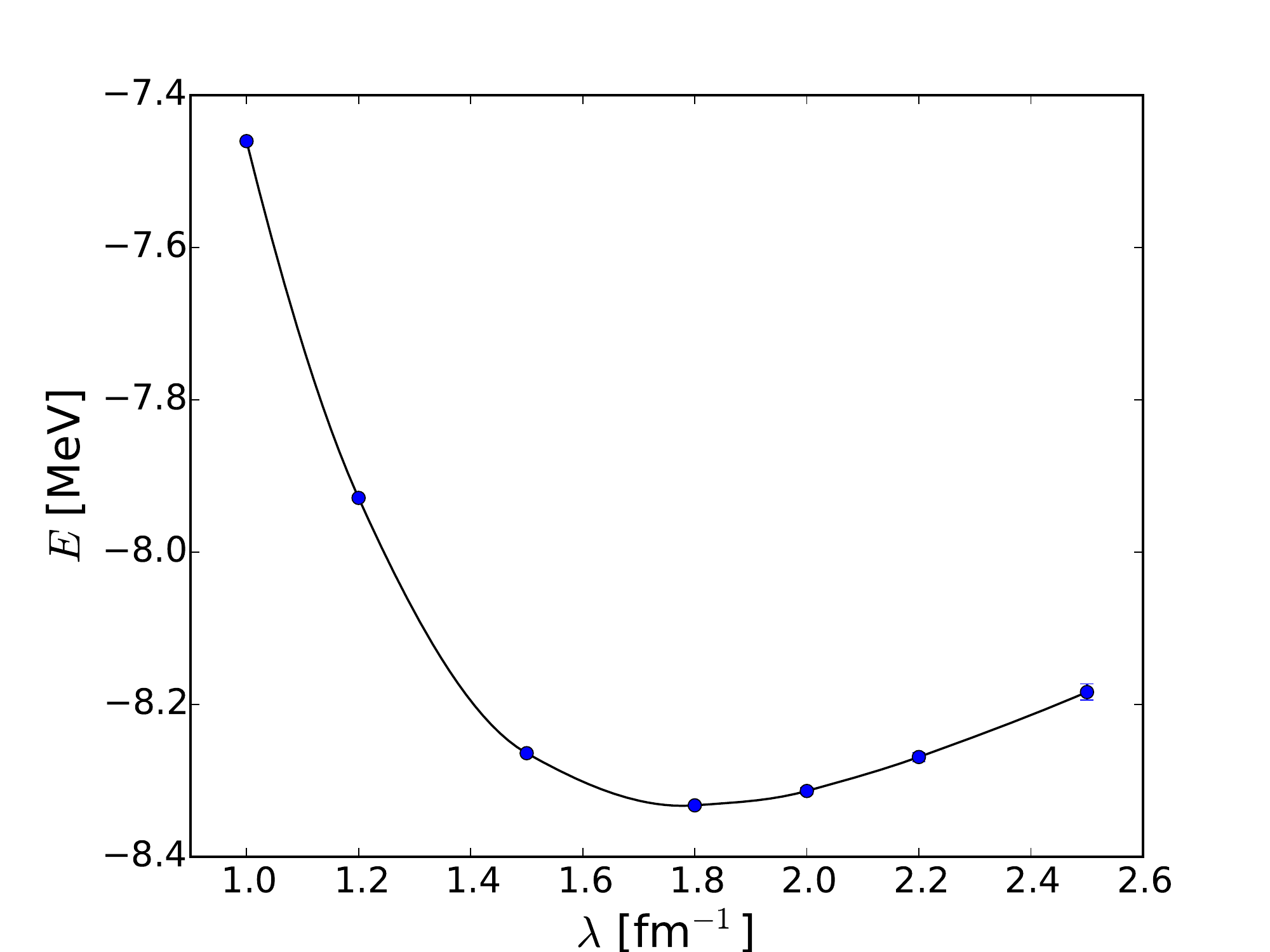}
    \includegraphics[scale=0.35]{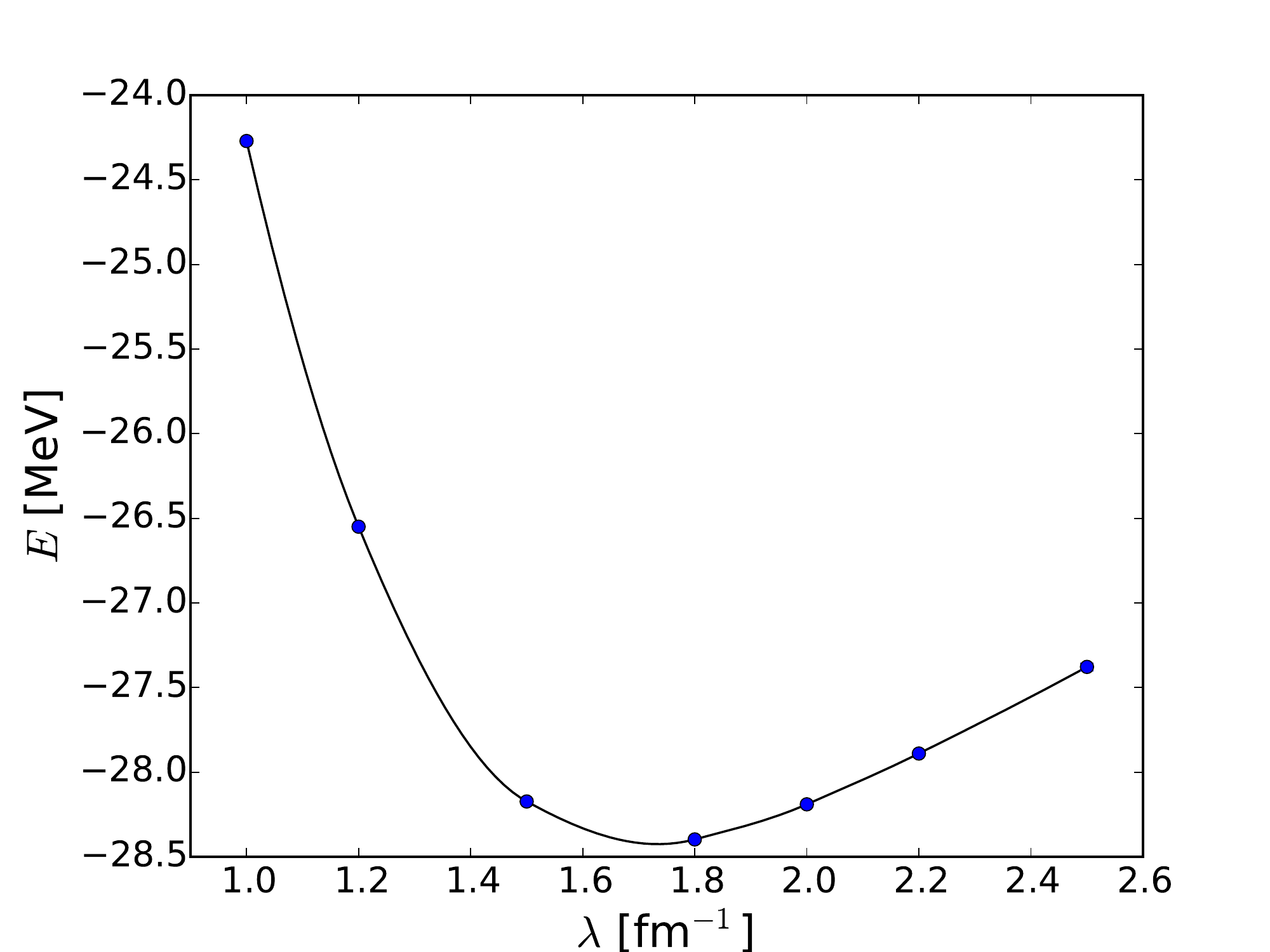}
 \end{center}   
    \caption{\label{fig:3H4Helamdep}$\lambda$-dependence of the $_{}^3$H (left) and the $_{}^4$He (right) binding energies. 
     Errorbars indicating the uncertainties of the energies are too small to be seen. } 
   \end{figure}
   
Fig.~\ref{fig:4Heomdep} shows the $\omega$-dependence again for $\lambda=1.5$~fm$^{-1}$ and 
$\lambda=2.5$~fm$^{-1}$. It can be seen that, indeed, smaller $\Ntot$ are sufficient to get converged results. 
Again Eq.~(\ref{eq:eneromdep}) gives a very good description of the $\omega$-dependence of the results 
around the optimal values. In Fig.~\ref{fig:4Hendep} the resulting $\cal N$-dependence  of the energies 
is shown. Since the values of $\Ntot$ are now much smaller and since convergence is fast, our prescription to estimate 
the uncertainties leads to much stronger differences of the error estimates for different $\Ntot$. This implies 
that the fit to Eq.~(\ref{eq:enerNdep}) is dominated by the three larges model spaces. We observe 
that, also for this case, the binding energies can be extracted with high accuracy. We confirmed that the extracted energies 
agrees with our solutions of Yakubovsky equations. Our results for $^4$He are also summarized in 
Table~\ref{tab:energyres}. 

Finally, we compare the $\lambda$-dependence of $^3$H and $^4$He. It is well known that the two binding 
energies are strongly correlated and therefore it is not too surprising that the results follow the same 
trend. We just note that around $\lambda=1.8$~fm$^{-1}$, in the minimum, 
$^4$He reaches the experimental value for its binding energy of -28.3~MeV. For $^3$H, the 
binding energy is also minimal for this value but does not reach the experimental value of -8.482~MeV.

\subsection{$_{}^7$Li}

\begin{figure}[tbp]
 \begin{center}
    \includegraphics[scale=0.36]{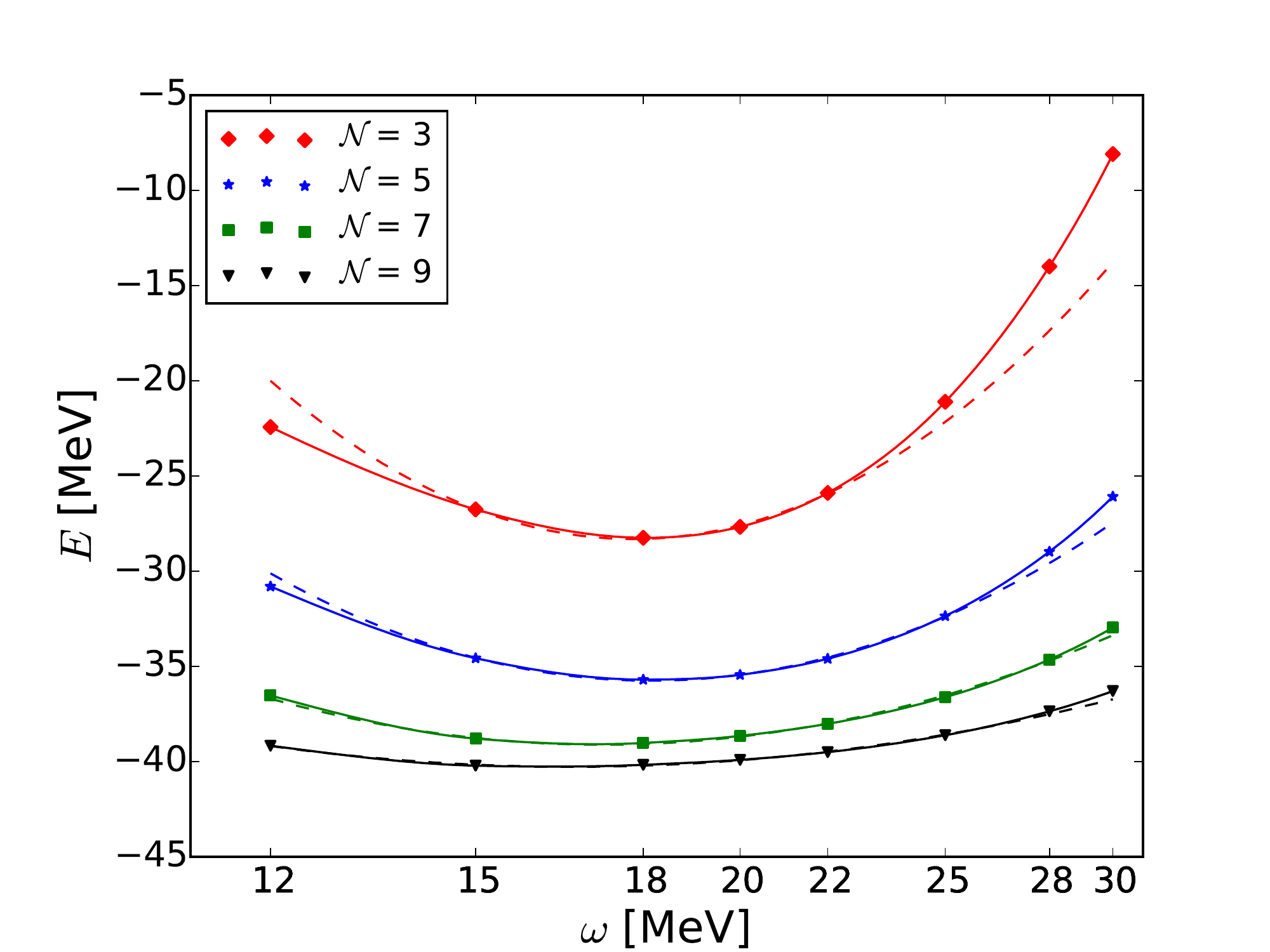}
    \includegraphics[scale=0.36]{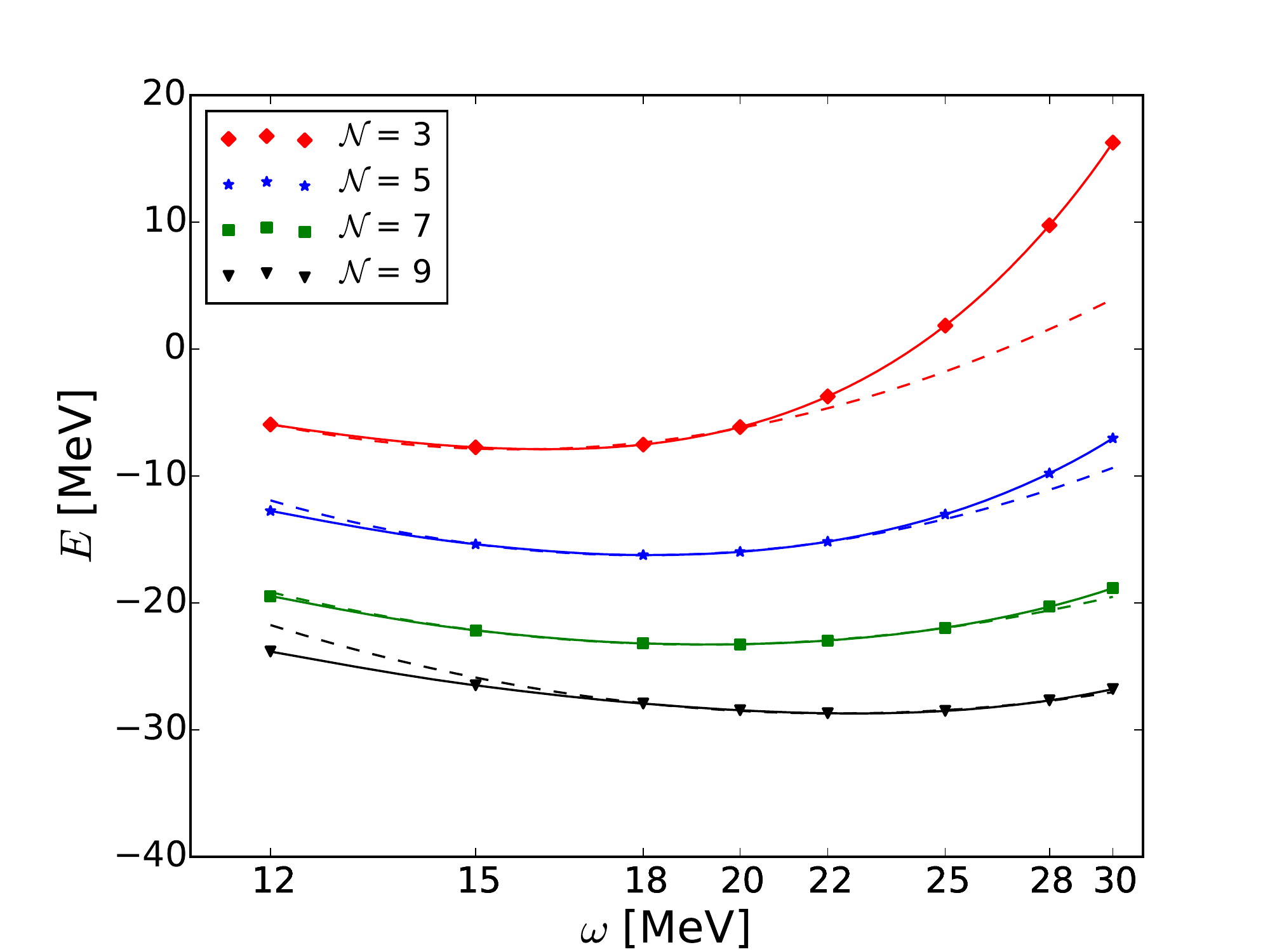}
  \end{center}  
    \caption{\label{fig:7Li3minusomdep}$\omega$-dependence of the ground state energy of $^7$Li for $\lambda=1.5$~fm$^{-1}$ (left) and 
      $\lambda=2.5$~fm$^{-1}$ (right). For lines and markers see Fig.~\ref{fig:4Heomdep}.  }
\end{figure}

\begin{figure}[tbp]
\begin{center}
    \includegraphics[scale=0.35]{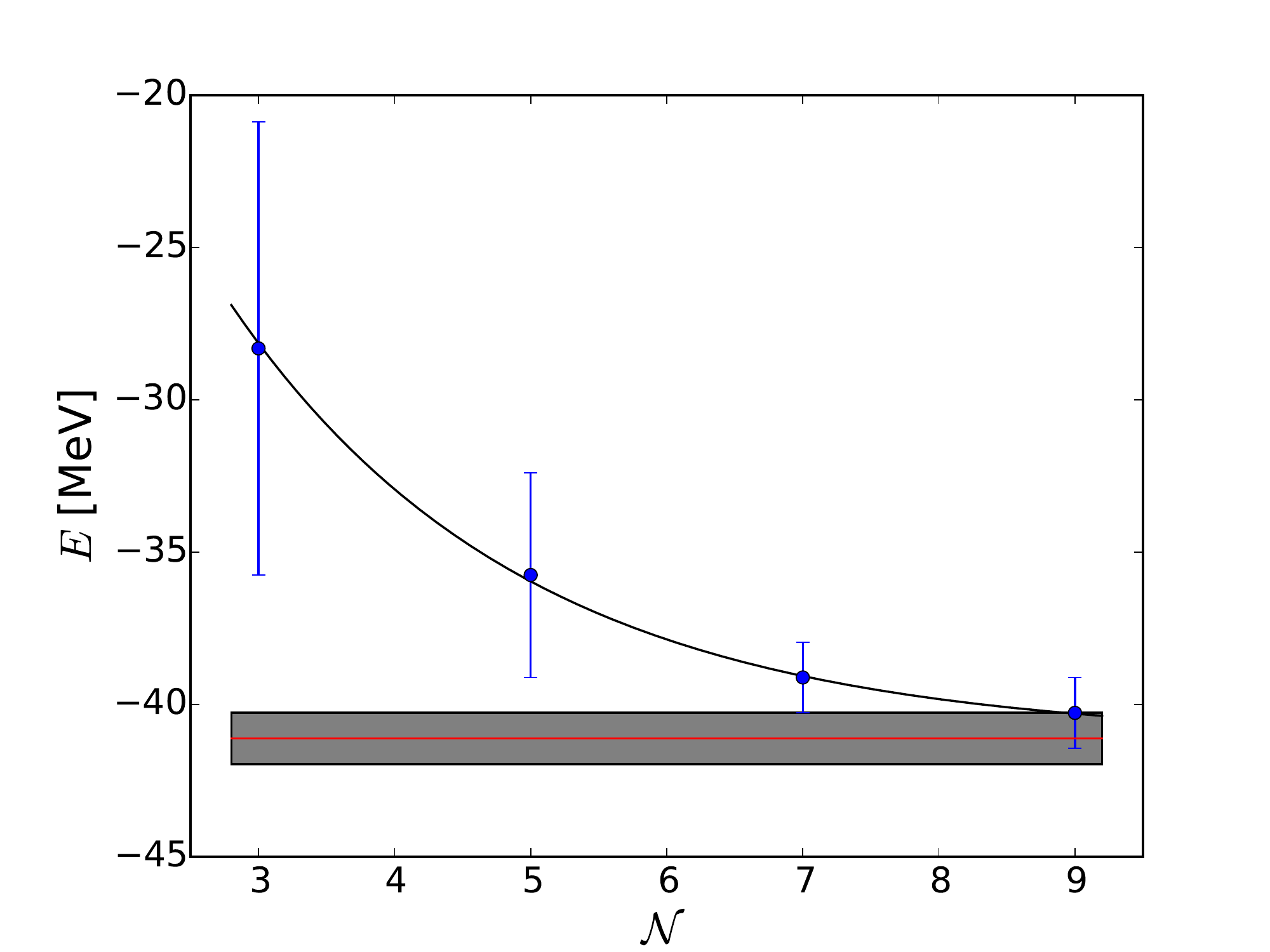}
    \includegraphics[scale=0.35]{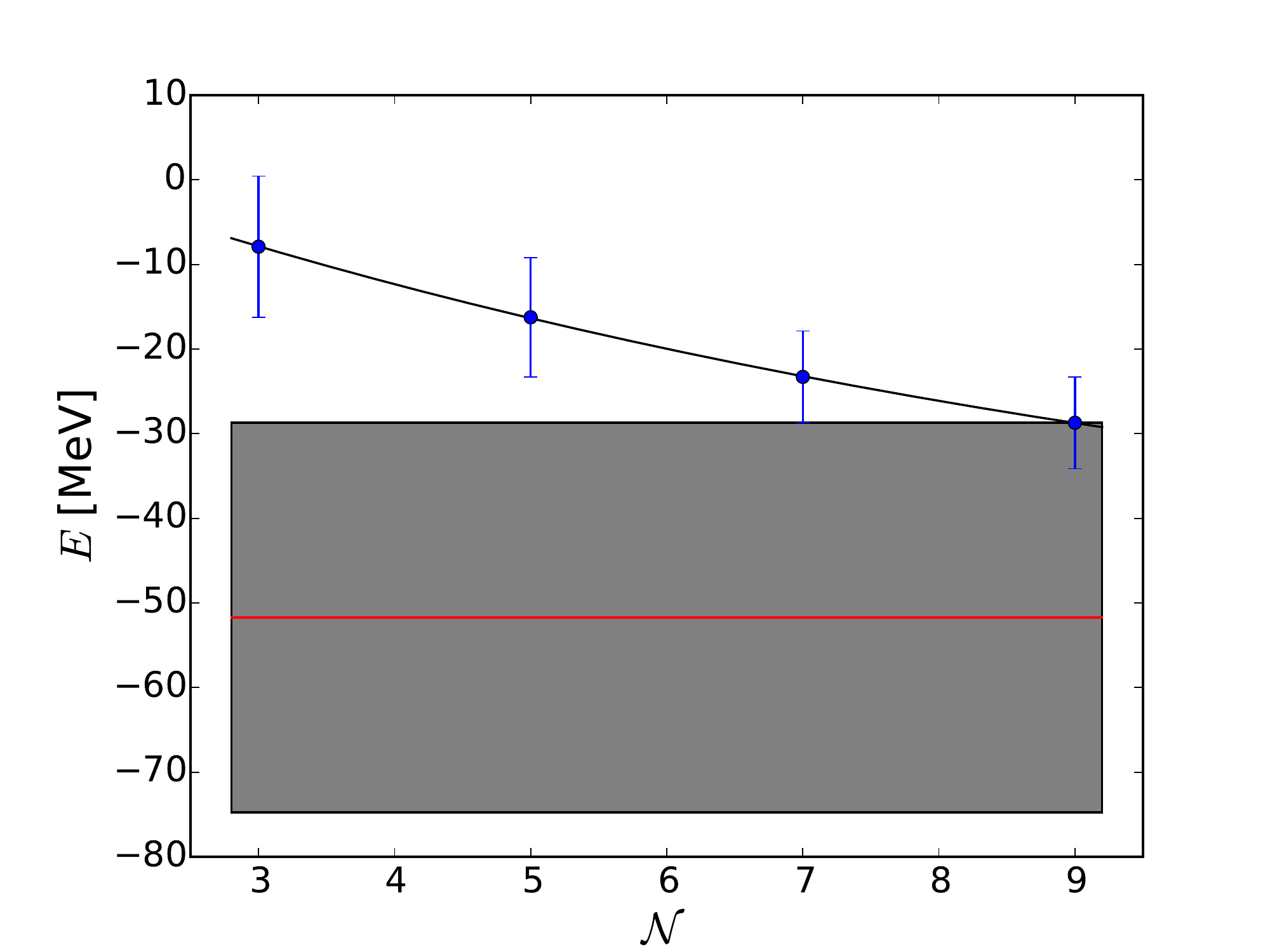}
 \end{center}   
    \caption{\label{fig:7Li3minusndep}$\cal N$-dependence of the ground state energy of $^7$Li for $\lambda=1.5$~fm$^{-1}$ (left) and 
      $\lambda=2.5$~fm$^{-1}$ (right). For lines and markers see Fig.~\ref{fig:4Hendep}. } 
   \end{figure}

We now turn to the more difficult $p$-shell nuclei. Here, $^7$Li is an interesting test case because 
the first excited state is really bound experimentally. So far, we have generated cfp up to $\Ntot=9$
for this system (see Appendix~\ref{app:ready}). Our results for the $\omega$ and $\Ntot$-dependence for 
this range of $\Ntot$ for the $J^\pi=\frac{3}{2}^-$ are summarized in Fig.~\ref{fig:7Li3minusomdep} and \ref{fig:7Li3minusndep}. 
Again we show results for our two standard SRG cutoffs $\lambda=1.5$~fm$^{-1}$ and 2.5~fm$^{-1}$.  For both cases, 
the $\omega$-dependence can be described well by Eq.~(\ref{eq:eneromdep}). But it is obvious that the model 
spaces used are not sufficient to obtain converged results for $\lambda=2.5$~fm$^{-1}$. Although 
the lines for different $\Ntot$ are getting closer, we observe that the optimal $\omega$ shifts to larger 
values with increasing $\Ntot$. This implies that we do not find smaller steps for the 
extracted $E_{\Ntot}$ when going to larger spaces as can be seen on the right of Fig.~\ref{fig:7Li3minusndep}.
The exponential extrapolation and the extracted uncertainty, which is also shown in the figure, clearly reflects 
that the model space size is too small to get converged results for this cutoff. This is not too worrysome 
since it is well known that once induced SRG 3N interactions are taking into account
the $\lambda$-dependence becomes mild enough that much smaller values of $\lambda$ are possible 
to get physically meaningful results \cite{Jurgenson:2013jn}. For $\lambda=1.5$~fm$^{-1}$, 
we find the usual behavior as for the $s$-shell nuclei. The  extracted $E_\Ntot$ clearly show a pattern
of convergence that allows for a meaningful extraction of the binding energy as shown on the left of 
Fig.~\ref{fig:7Li3minusndep}.

\begin{figure}[tbp]
 \begin{center}
    \includegraphics[scale=0.36]{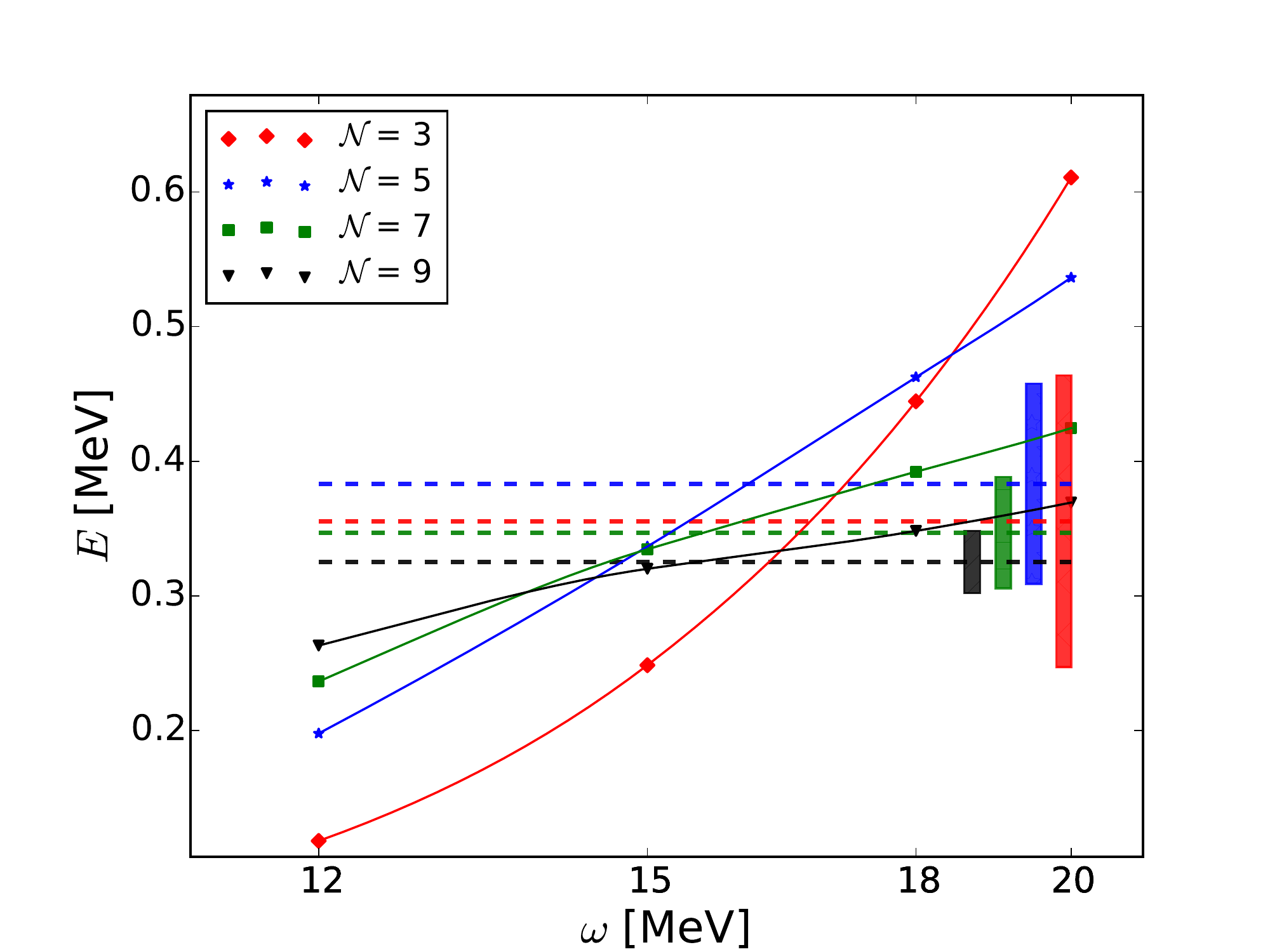}
    \includegraphics[scale=0.36]{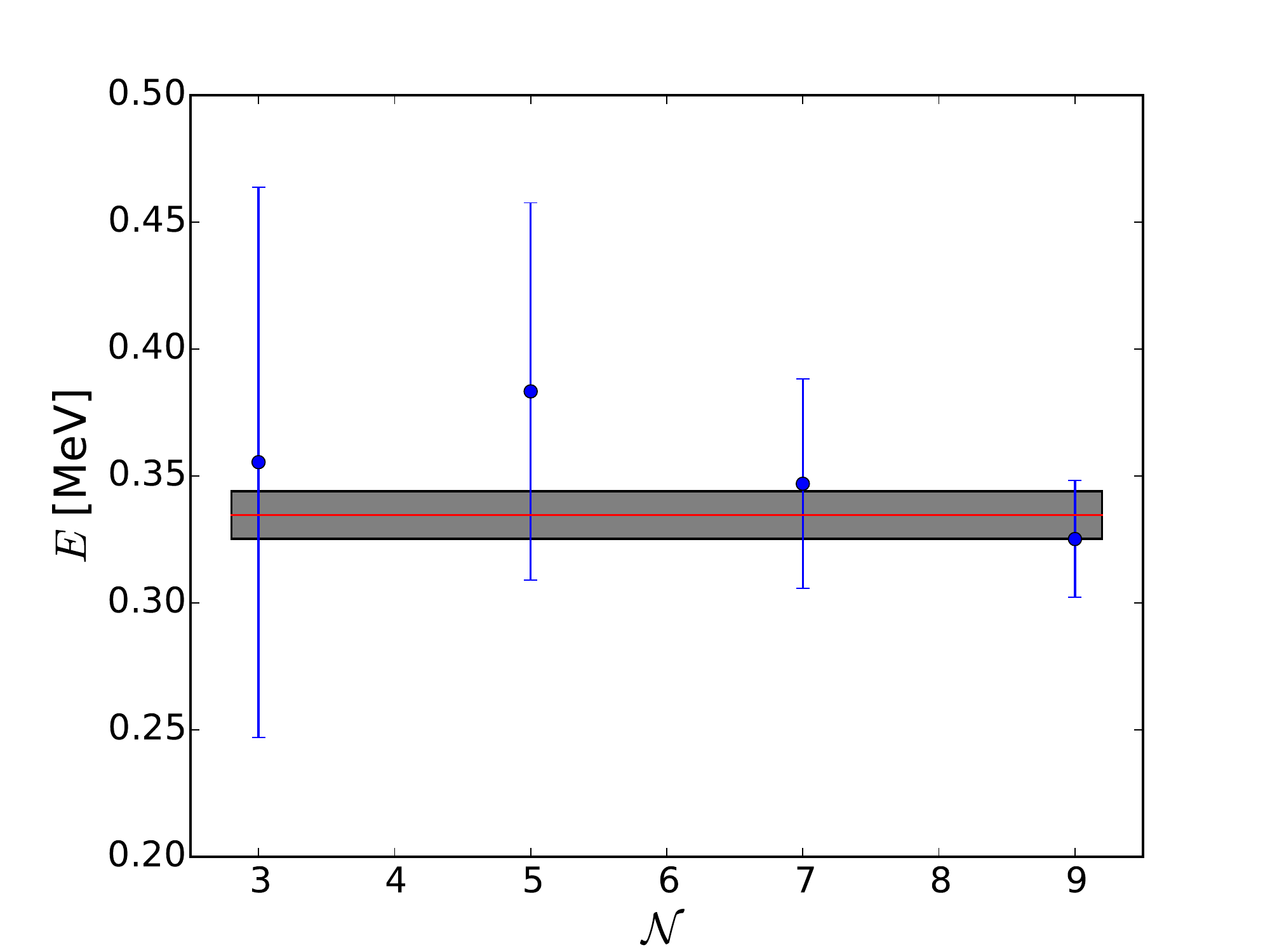}
  \end{center}  
    \caption{\label{fig:7Liexc1minusomNdep}Left hand side: $\omega$-dependence of the excitation energy of $^7$Li for $\lambda=1.5$~fm$^{-1}$. Markers indicate results for the excitation energy for different $\Ntot$ depending on $\omega$. The lines 
    connect results for the same $\Ntot$ to guide the eye. The extracted excitation energy and its uncertainty 
    are given by the dashed lines and the box on the right of the graph. 
    Right hand side: Extracted $\Ntot$-dependence of the same excitation energy. The full result and its uncertainty 
    are shown by  the red solid line and the shaded box surrounding it. }
\end{figure}

Of course, similar calculations are possible for the excited state of $^7$Li. Such calculations show that the ground states 
and excited state binding energies are strongly correlated for NCSM calculations and that it is much more efficient to 
look directly at the excitation energies. For the first excited $J^\pi=\frac{1}{2}^-$ state of $^7$Li, we have 
calculated the excitation energy depending on the HO frequency and the model space size.  For the example of 
$\lambda=1.5$~fm$^{-1}$, we show the results for the $\omega$-dependence on the left hand side of 
Fig.~\ref{fig:7Liexc1minusomNdep}. The excitation energies are used in a range that includes the two  
$\omega$ values right and left from the optimal $\omega$ values of the ground and excited state. 
It is reassuring that the $\omega$-dependence flattens out when going to larger model spaces. 
In order to extract the excitation energy for a given $\Ntot$, we calculated the average of the results 
of these $\omega$. The uncertainty was then estimate by the standard deviation of the results from 
this average. Of course, since the $\omega$ range is chosen in an ad-hoc way, these uncertainties 
cannot be understood as absolute uncertainties. Still they indicate 
the errors for each $\Ntot$ relative to the others. In the second step, we therefore build a weighted average 
of the excitation energies for all $\Ntot$ where the weight is given by errors extracted from the $\omega$-
dependence. The results are shown on the right hand side of 
Fig.~\ref{fig:7Liexc1minusomNdep}. In this way, we were able to obtain quite accurate results for the excitation 
energies. 

\begin{figure}[tbp]
\begin{center}
    \includegraphics[scale=0.35]{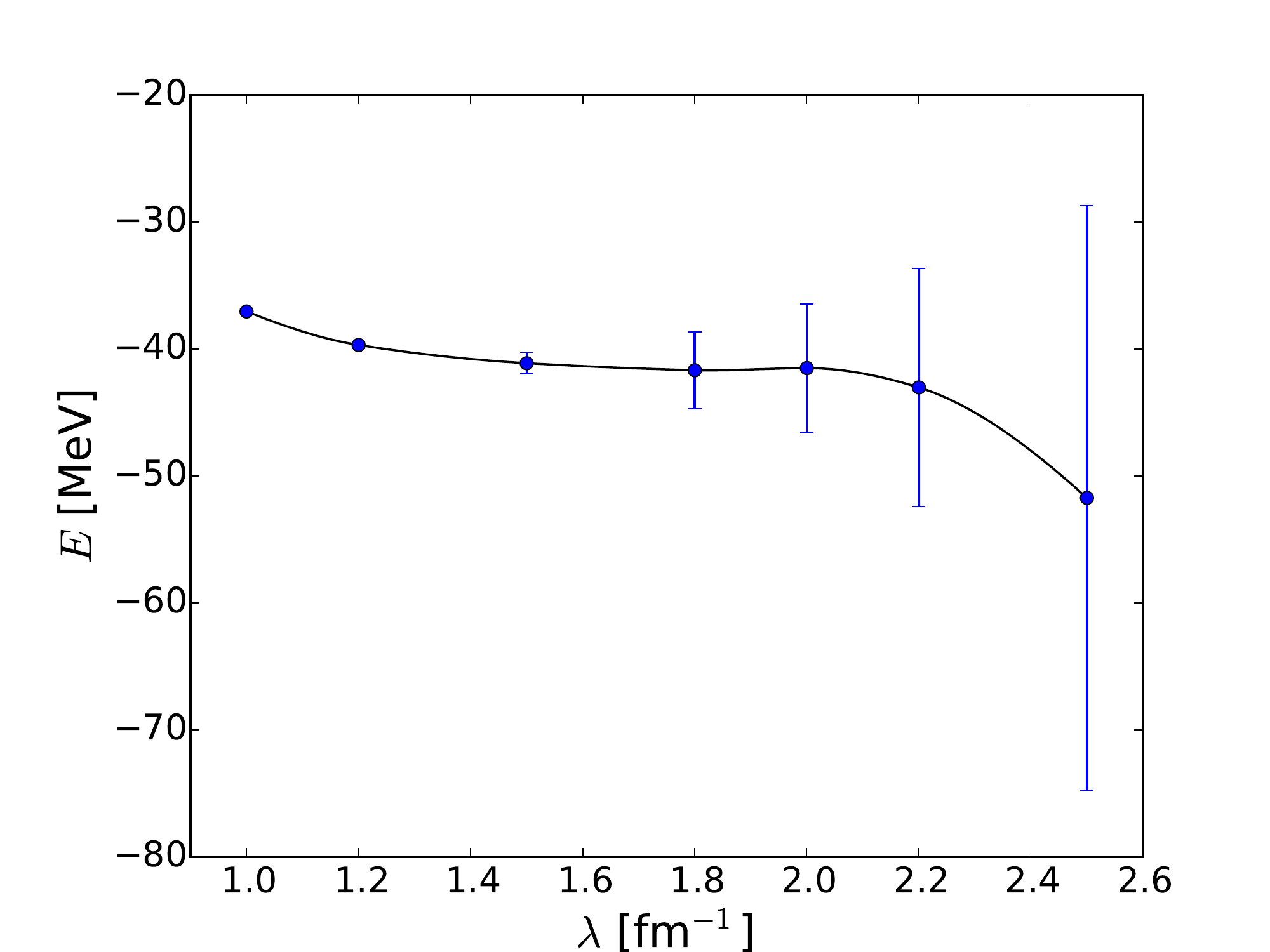}
    \includegraphics[scale=0.35]{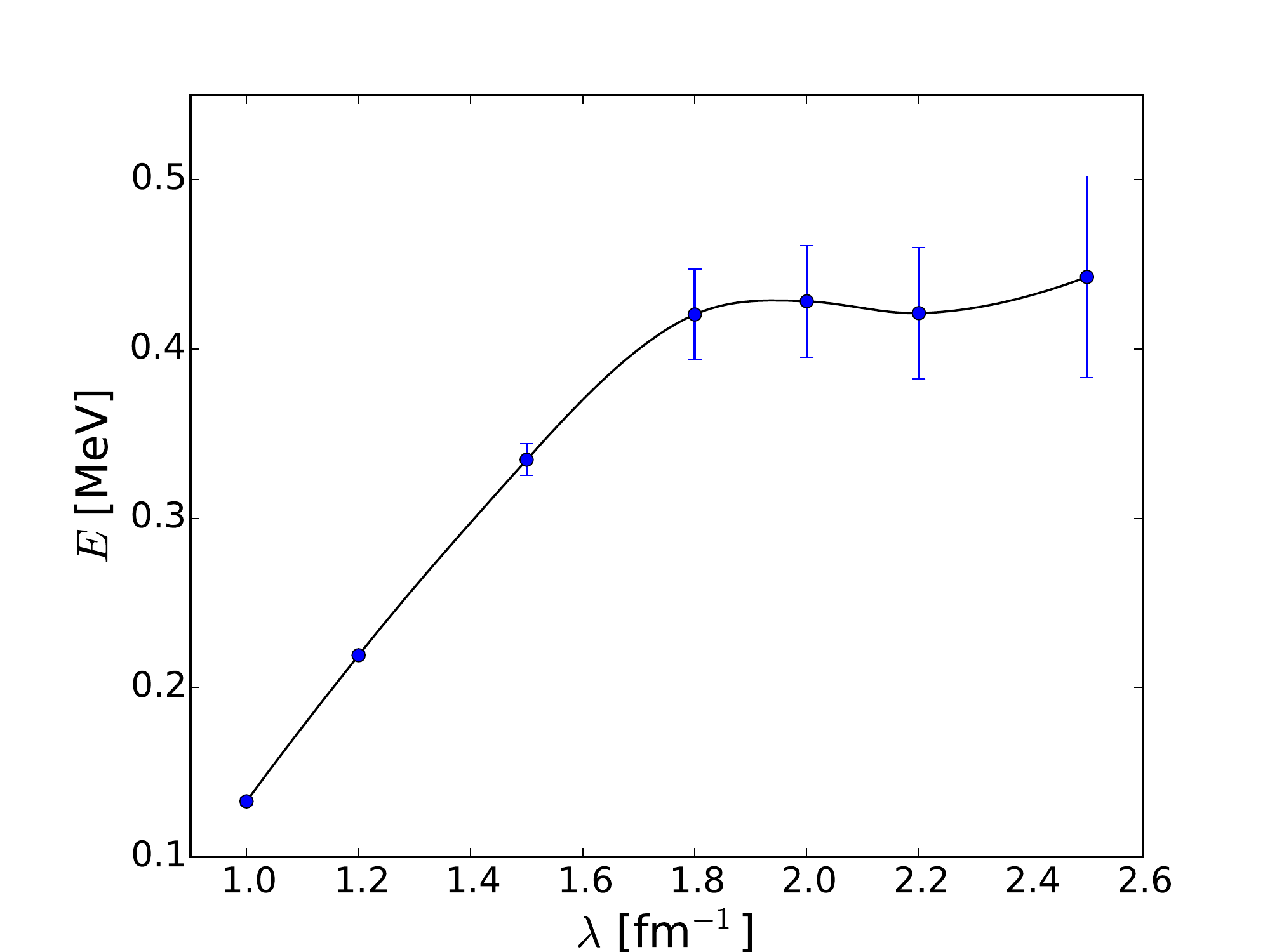}
 \end{center}   
    \caption{\label{fig:7Lilamdep}$\lambda$-dependence of the ground state energy 
    (left) and of the excitation energy (right) of $^7$Li. The uncertainties extracted as described in the text are 
    given by the errorbars. } 
   \end{figure}

To conclude this subsection, we finally show the $\lambda$-dependence for the binding energy of the 
$^7$Li ground state  and the excitation energy for the first excited state in Fig.~\ref{fig:7Lilamdep}. The 
uncertainties for both quantities are also shown. As discussed above, the model spaces used here 
are large enough for binding energies only for the lower $\lambda$ below approximately 2~fm$^{-1}$. 
We find it however interesting that the excitation energies can be obtained fairly accurately even for 
larger $\lambda$. Again, we refer to Table~\ref{tab:energyres} for the numerical values of the binding and 
excitation energies. 

\begin{figure}[tbp]
\begin{center}
    \includegraphics[scale=0.35]{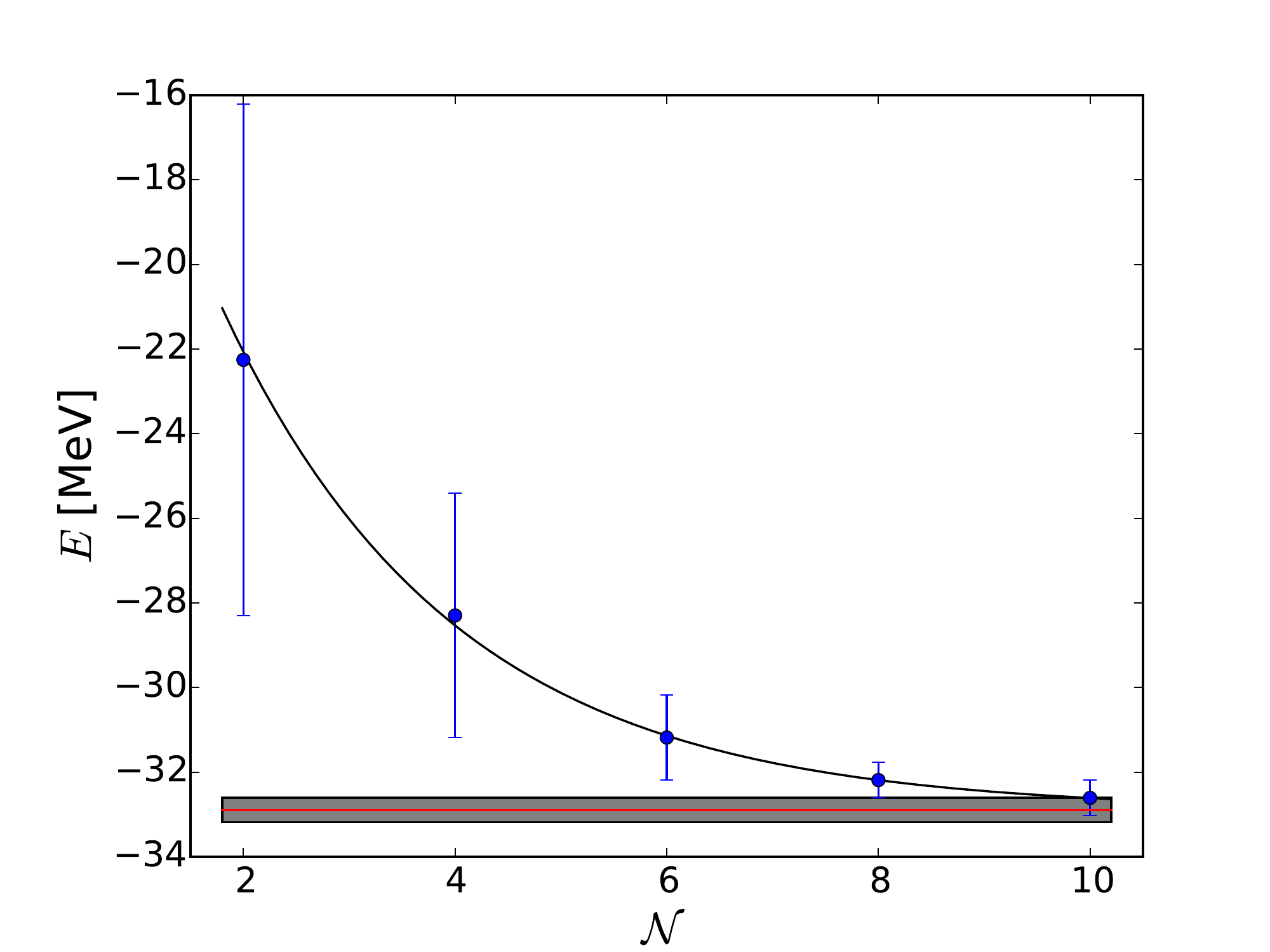}
    \includegraphics[scale=0.35]{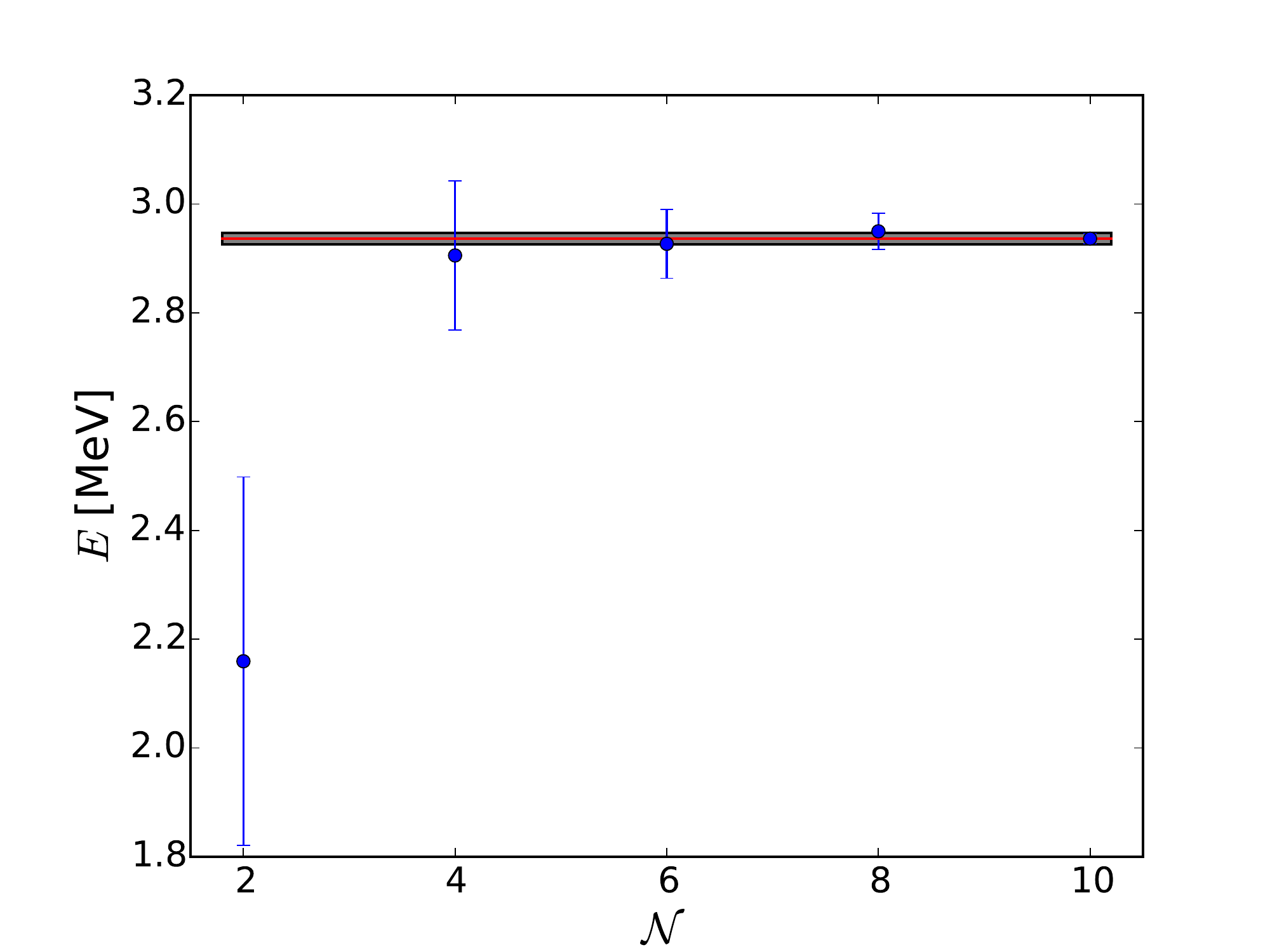}
 \end{center}   
    \caption{\label{fig:6Lindep}$\cal N$-dependence of the ground state energy (left) and of the excitation energy (right) of $^6$Li for $\lambda=1.5$~fm$^{-1}$. The full result and its uncertainty are shown by the red solid line and the shaded box surrounding it. } 
 \end{figure}

\subsection{$_{}^6$Li and $_{}^6$He}
Finally, we present first results for $A=6$ systems. We start with $_{}^6$Li, for which we have prepared the 
cfp and transition coefficients for the  $J^\pi=1^+$ ground state and  $J^\pi=3^+$ excited state 
up to $\Ntot=10$. The $\omega$-dependence is very similar to the one found for $^7$Li. Since 
we are now considering more HO excitations, the results are generally less dependent on 
the HO frequency than for $A=7$. Therefore, we directly present the $\Ntot$-dependence 
for $\lambda=1.5$~fm$^{-1}$ in Fig.~\ref{fig:6Lindep}. As can be seen,  
the pattern of convergence is very regular for the binding energy and the excitation energy
and the results are quite accurate. We note that the excited state is above the deuteron-$_{}^4$He threshold for many $\lambda$ as it also is experimentally.
Nevertheless, it is possible and regularly done to 
extract the excitation energy from NCSM calculations.

\begin{figure}[tbp]
\begin{center}
    \includegraphics[scale=0.35]{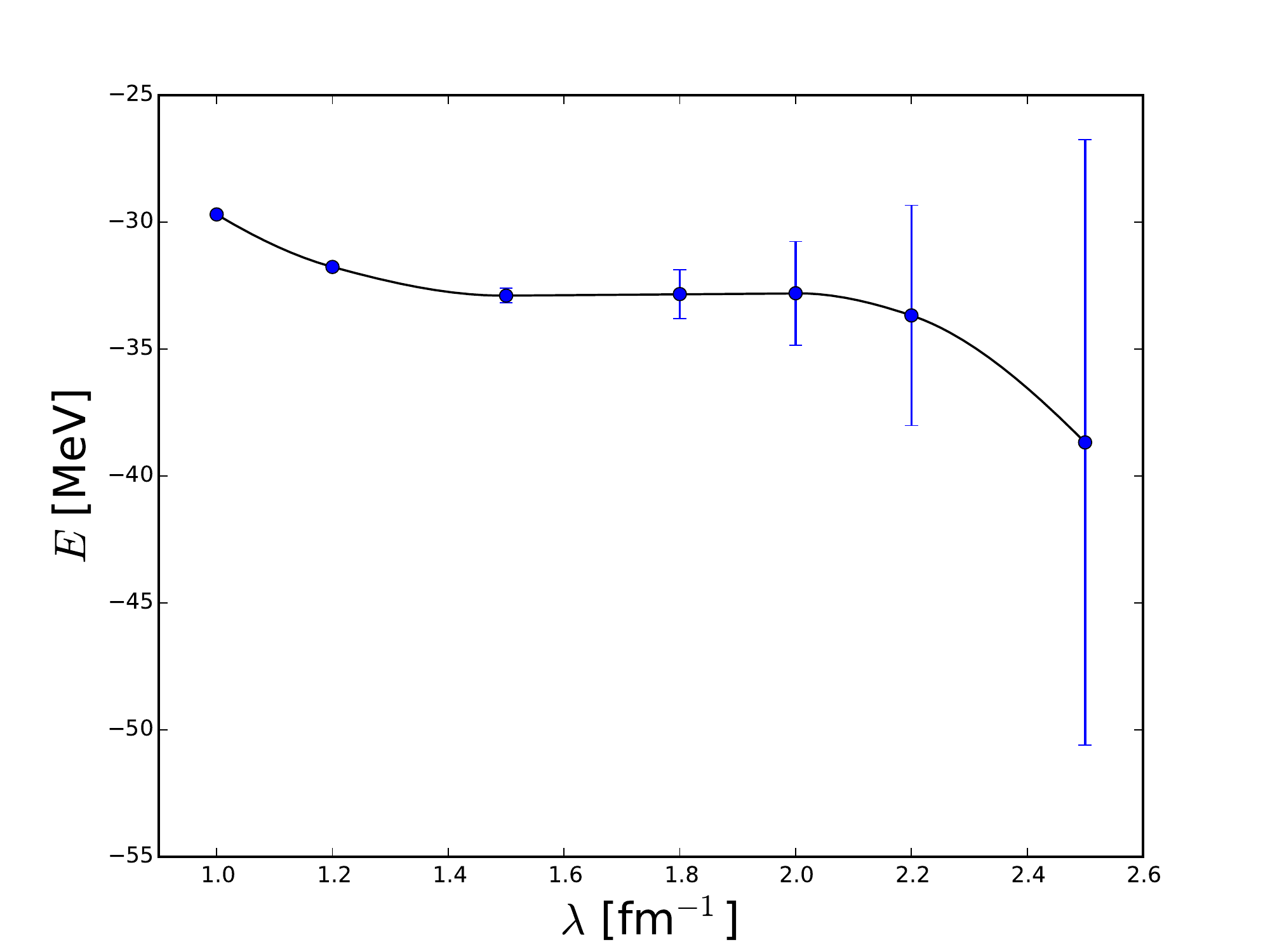}
    \includegraphics[scale=0.35]{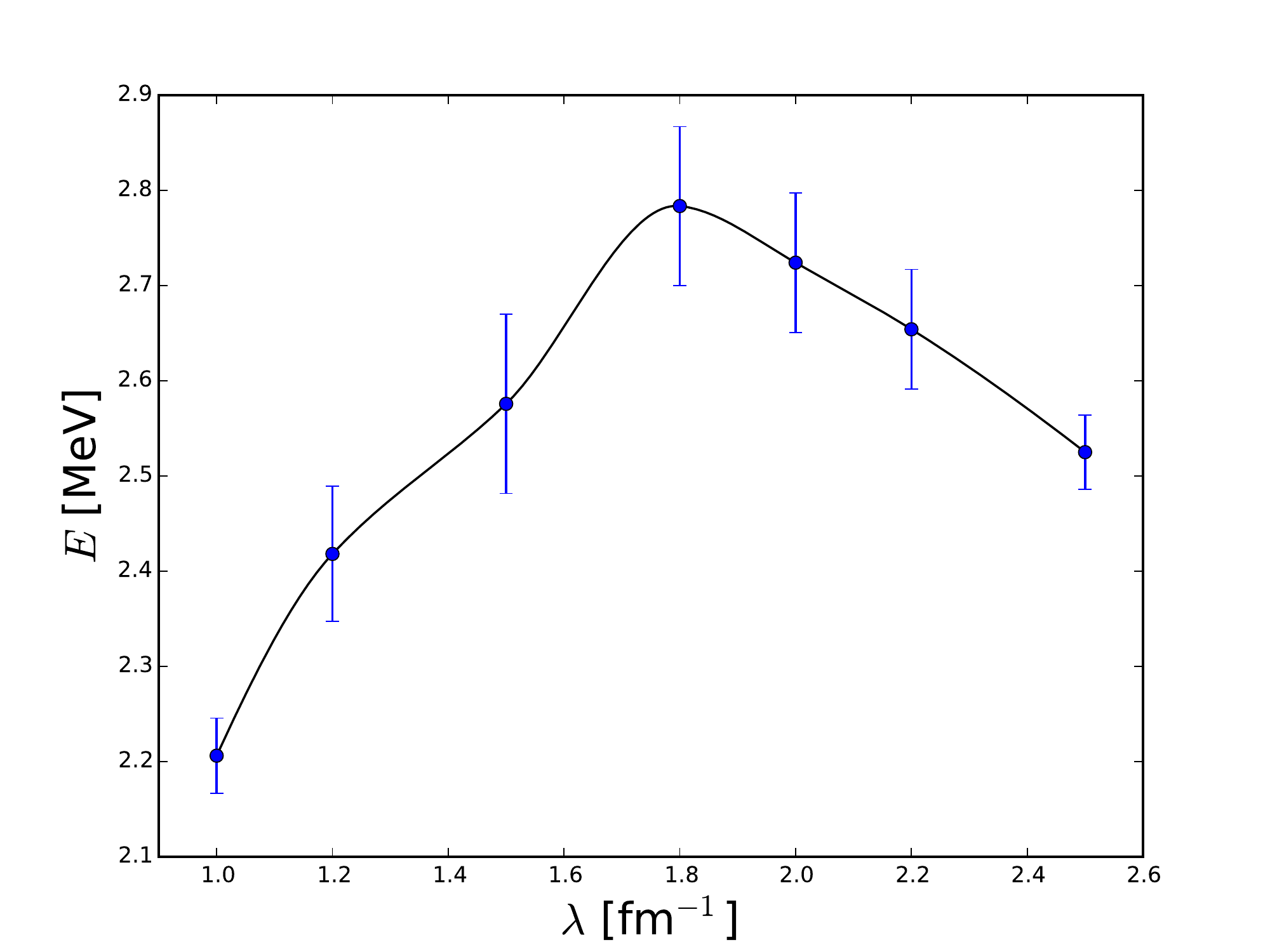}
 \end{center}   
    \caption{\label{fig:6Lilamdep}$\lambda$-dependence of the ground state energy 
    (left) and of the excitation energy (right) of $^6$Li. } 
 \end{figure}

Fig.~\ref{fig:6Lilamdep} (and the explicit values in Table~\ref{tab:energyres}) summarize our results for 
the binding energy and excitation energy for the full range of $\lambda$. It sticks out that the uncertainty 
estimates for $\lambda>2$~fm$^{-1}$ decrease again. This behavior seems unnatural and needs to be studied in 
larger model spaces in future. 

\begin{figure}[tbp]
\begin{center}
    \includegraphics[scale=0.35]{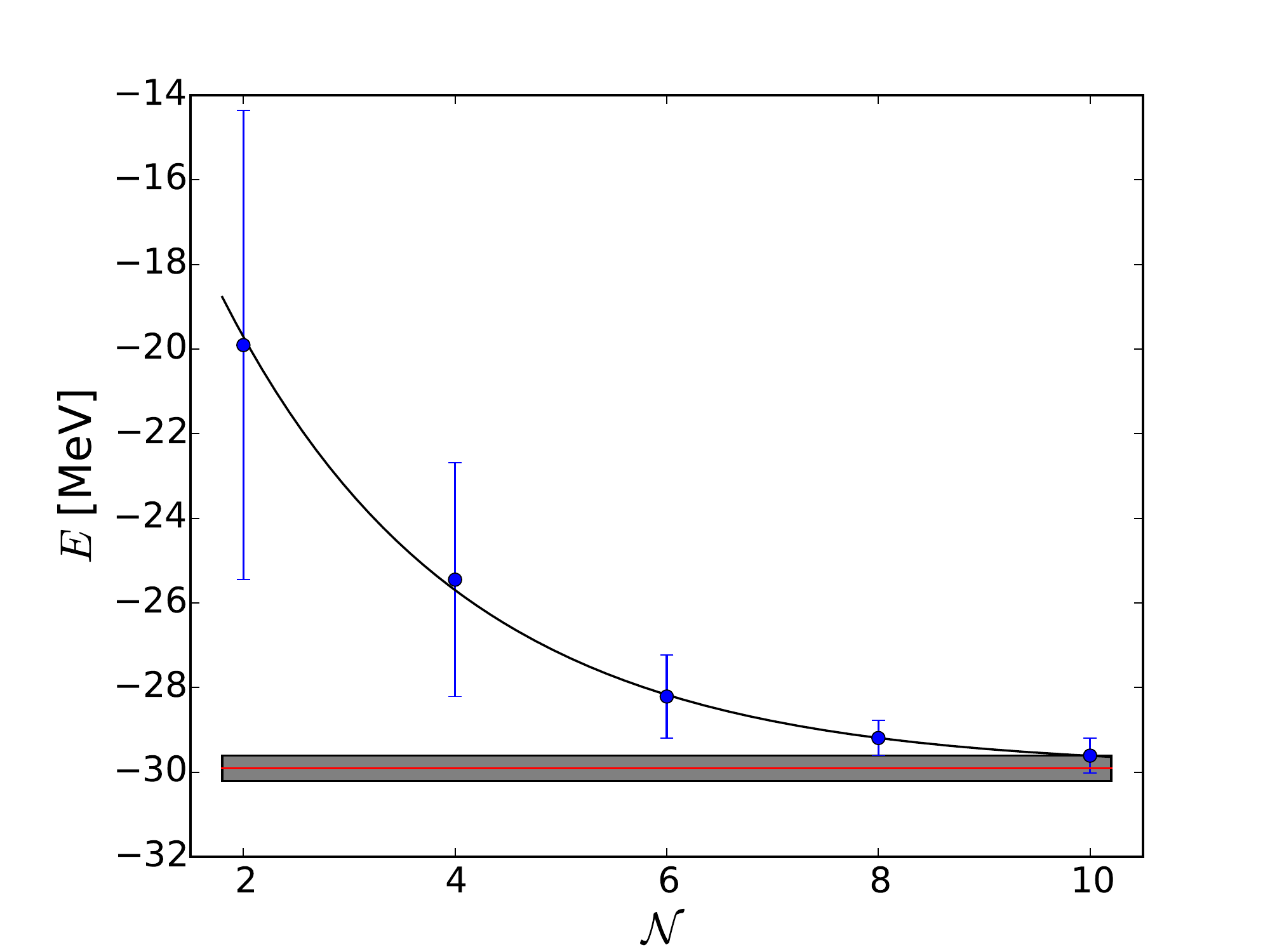}
    \includegraphics[scale=0.35]{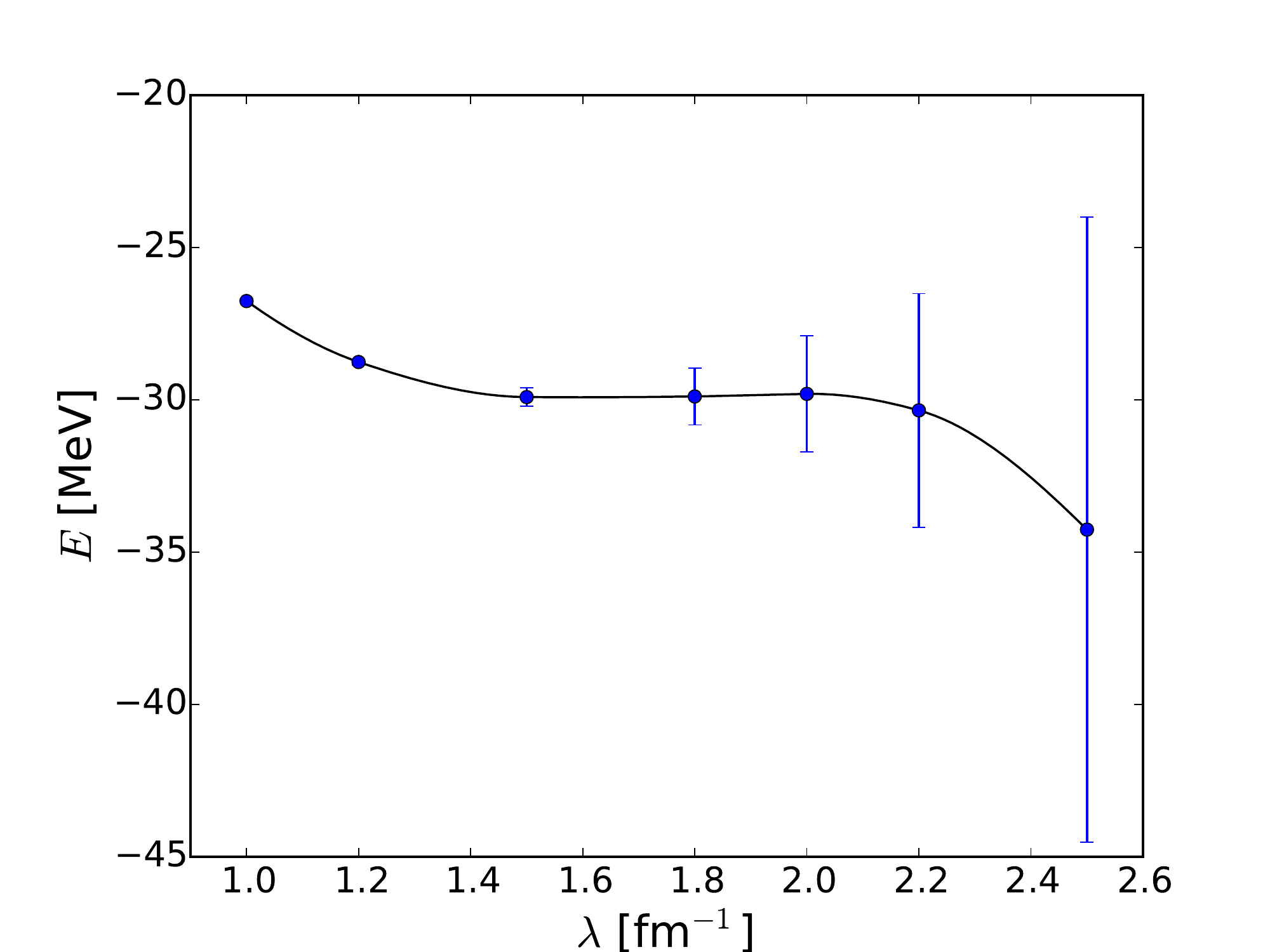}
 \end{center}   
    \caption{\label{fig:6Hegs}$\cal N$-dependence for $\lambda=1.5$~fm$^{-1}$ (left) and $\lambda$-dependence (right) of the ground state energy of $^6$He. } 
 \end{figure}

\begin{figure}[tbp]
\begin{center}
    \includegraphics[scale=0.36]{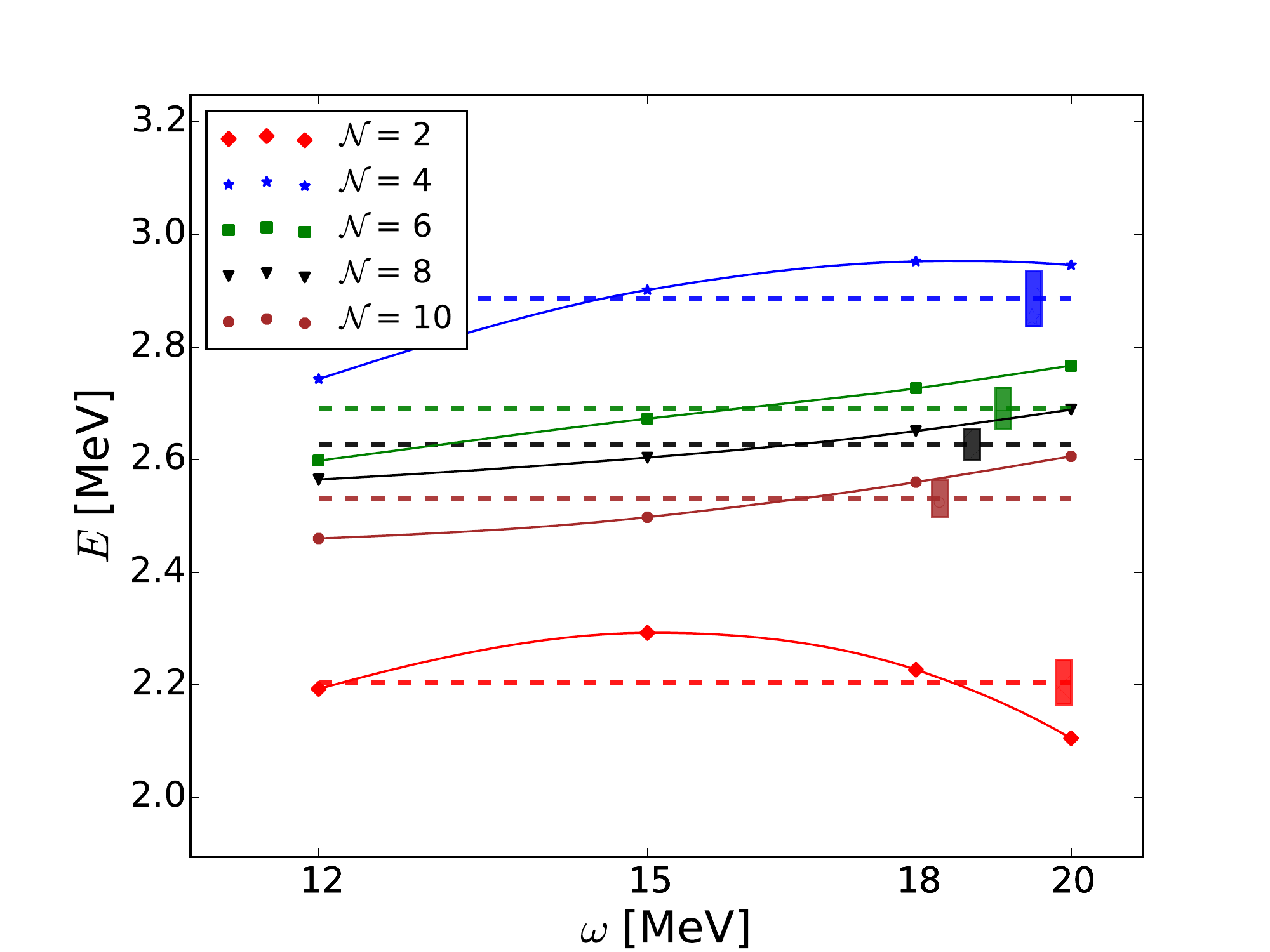}
    \includegraphics[scale=0.35]{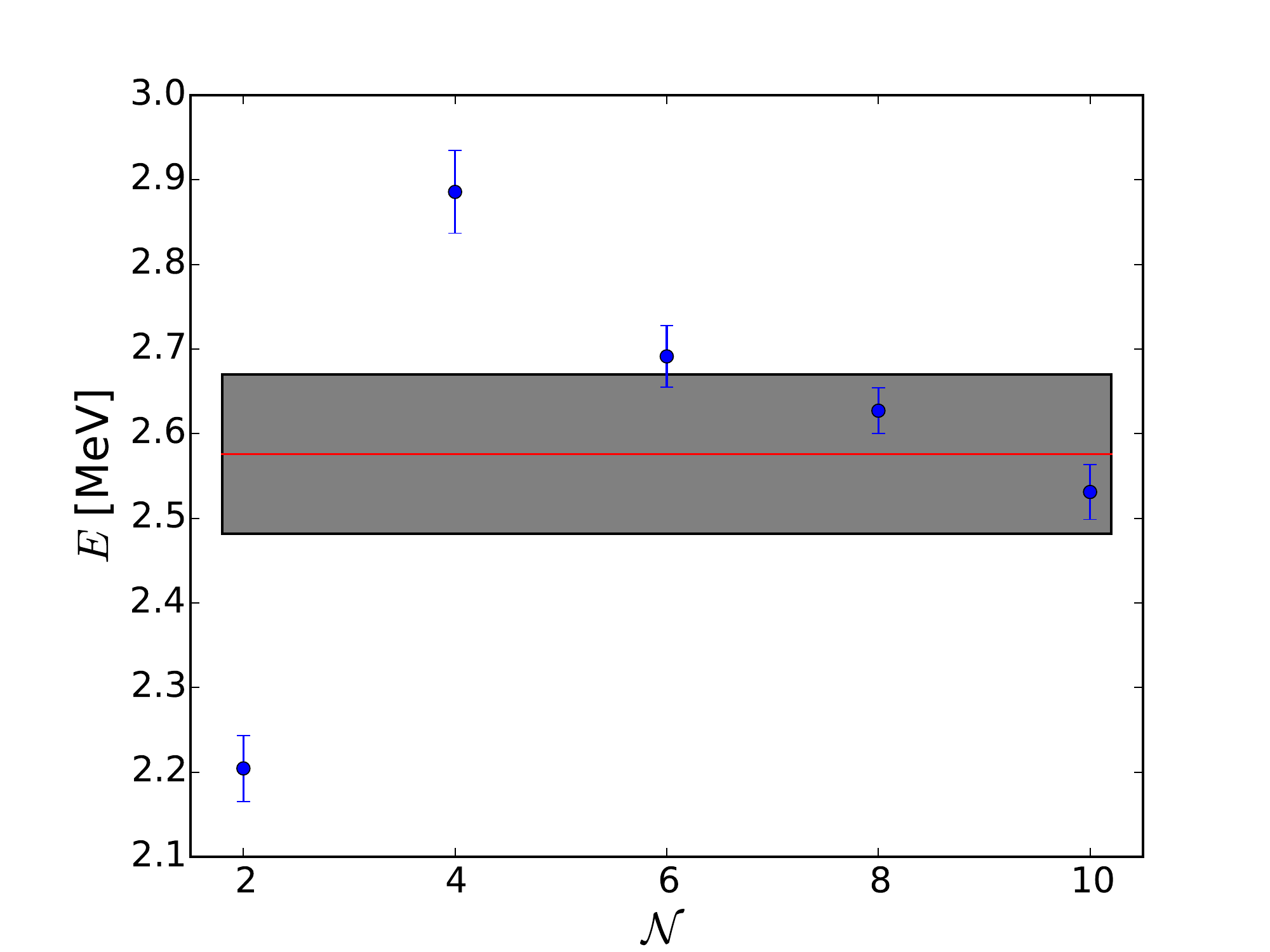}
 \end{center}   
    \caption{\label{fig:6Heexconv}Left hand side: $\omega$-dependence of the excitation energy of $^6$He 
    for $\lambda=1.5$~fm$^{-1}$. 
    Right hand side: Extracted $\Ntot$-dependence of the same excitation energy.
    For an explanation of the notation see Fig.~\ref{fig:7Liexc1minusomNdep}.}
 \end{figure}

For $^6$He, cfp also exist up to $\Ntot=10$ for the 
$J^\pi=0^+$ ground state and the $J^\pi=2^+$ excited state.
We observe, that the pattern of convergence for the binding energy is again very similar
to $^6$Li and $^7$Li. Therefore, accurate results could be extracted as shown in 
Fig.~\ref{fig:6Hegs}. But, in this particular case, 
the results for the excitation energy of the $J^\pi=2^+$ state were problematic. To exemplify this,
we show in Fig.~\ref{fig:6Heexconv} the results for the $\omega$-dependence 
of the excitation energies for $\lambda=1.5$~fm$^{-1}$. Here, even for this rather small $\lambda$, 
the $\omega$-dependence is very small in all cases. But the changes when going to larger model 
spaces are large. The resulting $\Ntot$-dependence is therefore very irregular. Since results for even 
lower cutoffs show a similar behavior, we refrain from showing the $\lambda$-dependence of the 
excitation energy. To provide benchmark results, we however included the numerical values 
of our automatized extraction procedure in Table~\ref{tab:energyres}. We stress however that the uncertainty 
estimates of the excitation energies for $^6$He given in the table are probably not reliable. 

\section{Conclusion and Outlook}

In this work, we have described in detail a new implementation of cfp that 
allow one to perform NCSM calculations for light $p$-shell nuclei using relative coordinates. 
The use of relative coordinates separates off the CM motion and 
uses basis states with definite total angular momentum and isospin. Thereby, the dimension of the basis 
is considerably reduced. For the application of two-nucleon operators (e.g. NN forces), we introduced transition matrix elements 
to states that separate a two-nucleon cluster from the nucleus. These states enable a quite simple application
of NN operators to $A$-nucleon states. 

Once the cfp and transition elements are known, calculations for light $p$-shell nuclei
require only modest computational resources. Therefore, as a first application, we applied the 
new basis set to $s$-shell and the lightest $p$-shell nuclei. We used the new set of basis states 
to map out the HO-frequency dependence of the results for increasing model space sizes and devised 
an automatized scheme to extract binding and excitation energies together with 
uncertainty estimates of the final results. The extraction procedure was applied to the $s$-shell nuclei 
as well as $_{}^6$He, $_{}^6$Li and $_{}^7$Li and resulted in consistent results for all binding energies
and the excitation energies of  $_{}^6$Li and $_{}^7$Li. For the excitation energy of $_{}^6$He, the 
$\omega$-dependence for the small model spaces turned out to be irregularly small and, therefore, 
the uncertainty estimates need to be checked using larger model spaces in future.

All these calculations were done with SRG evolved NN interactions 
for $\lambda$ between 1.0~fm$^{-1}$ and 2.5~fm$^{-1}$. We showed that within the model 
space, for which we have generated cfp so far, converged results could be obtained for 
$\lambda<2.0$~fm$^{-1}$. Recent NCSM calculations within the $m$-scheme have already 
shown that one obtains $\lambda$-insensitive results within this range of SRG parameters 
\cite{Jurgenson:2013jn} once 3NFs have been included. 

In order to be able to apply 3NFs, one more set of transition coefficients needs to be 
calculated. We also formulated the pertinent equations for these 
transitions. They have already been implemented and basic properties, like 
orthogonality, have been checked. Now they 
have to be accompanied by corresponding 3NF matrix elements. 
Work in this direction is in progress.   In a very similar manner,
such transitions can be extended to 4N and higher body operators. 
This is especially interesting since it is not clear at this point, whether 
4NFs can give sizable contributions to $p$-shell binding energies. 
While the $\lambda$-dependence of the $^4$He binding energy 
including induced 3NFs does not indicate significant contributions of
4N interactions \cite{Jurgenson:2011ec}, a direct  calculation of the 
leading chiral 4NF revealed that a small net contribution is obtained because 
two sizable terms tend to cancel each other \cite{Nogga:2010ix}. It needs to be 
clarified whether this cancelation is as effective in other systems. For this, 
the Jacobi NCSM will be an ideally suited tool since full use can be 
made of angular momentum and isospin conservation of the 4NF. 

One important aspect of this work is to make the cfp and transition coefficients 
available. The corresponding data files  have been generated using the HDF5 format 
and are platform  independent. We hope that, in this way, nuclear structure calculations 
become simpler for other groups and can be applied to a wider range of problems. 
With the test calculations shown here, the cfp and NN
transition coefficients are ready to be made accessible. 
The  3N transition coefficients will be made available, too, 
once they have been tested in similar calculations involving 3NFs.

\acknowledgement
This work is supported in part by DFG and NSFC through funds provided to the Sino-German CRC 110 ``Symmetries and the Emergence of Structure in QCD''
(NSFC Grant No. 11261130311). The numerical calculations have been performed on JUQUEEN, JUROPA and JURECA 
of the JSC, J\"ulich, Germany.

\appendix

\section{HO wave functions $R_{nl}$ in coordinate and momentum space }
\label{app:Rnl}
cfp and transition coefficients rely on the Talmi-Moshinsky brackets of Ref.~ \cite{Kamuntavicius:2001ig}. In this appendix, we shortly 
summarize the conventions for the HO wave functions related the conventions used in this work. 
We define the dimensionless HO wave functions 
     \begin{eqnarray}
        \hat R_{nl}\left(\rho\right) & = & \left(-1\right)^n
        \left[\frac{2\, n!
          }{\Gamma\left(n+l+\frac{3}{2}\right)}\right]^{\frac{1}{2}} \,
        \exp  \left(-\frac{{\rho}^{\,2}}{2}\right) \, \rho^{\,l} \:
        L_{\,n}^{\,\left(l+\frac{1}{2}\right)} \left({\rho}^{\,
            2}\right) \ \ .
       \label{eq:R-nl-fcts}
      \end{eqnarray}
The configuration space $R_{nl}$ are just given by a simple rescaling involving the HO length $b$: 
\begin{equation}
R_{nl}\left(r\right) 
  = 
  b^{-\frac{3}{2}} \,  \hat R_{nl} \left(  \frac{r}{b} \right) \  \ .  
\end{equation}  
The momentum space wave function is then obtained by  a Fourier transformation 
\begin{equation}
 R_{nl}\left(p\right) 
  = 
 \sqrt{\frac{2}{\pi}} \; i^l \int dr \: r^2 \: j_l \left(p \,r\right) \: R_{nl}\left(r\right) 
 \end{equation} 
that leads to   
 \begin{equation}
 R_{nl}\left( p \right)
    = 
 \left(-1\right)^n \:  i^l \: b^\frac{3}{2} \, \hat R_{nl} \left(  b p \right) \ \ .
\end{equation} 
Note that matrix elements that are parity conserving will only acquire  a real 
phase due to the $ i^l $ factor.

\section{Existing cfp and transition coefficients}
 \label{app:ready}
 
 In this appendix, we summarize the cfp and transition coefficients 
 that have already been generated. More sets of coefficients are currently generated. 
 
 The following table shows the available sets of cfp for the $A=3$ to $A=7$ system. 
 Ranges of $J$, $T$ and $Ntot$ values of calculated blocks are given. The label 
 complete indicates that sets for all $J$ and $T$ possible for the given $Ntot$ are available.  

\begin{center}
\def\arraystretch{2.5}
\setlength{\tabcolsep}{0.3cm}
\begin{tabular}{|c|cccc|}
\hline
 & $J$ & $T$ & $\Ntot$ &  \\
\hline
 $A=3$  &  $\frac{1}{2},\ldots,\frac{51}{2}$  &  $\frac{1}{2},\frac{3}{2}$         &  $0,\ldots,24$ & (complete) \\
\hline
 $A=4$  &  $0,\ldots,14$                      &  $0,\ldots,2$                      &  $0,\ldots,12$ & (complete) \\
\hline
 $A=5$  &  $\frac{1}{2},\ldots,\frac{25}{2}$  &  $\frac{1}{2},\ldots,\frac{5}{2}$  &  $1,\ldots,10$ & (complete) \\
\hline
 $A=6$  &  $0,\ldots,13$                      &  $0,\ldots,3$                      &  $2,\ldots,10$ & (complete) \\
\hline
 $A=7$  &  $\frac{1}{2},\ldots,\frac{25}{2}$  &  $\frac{1}{2},\ldots,\frac{7}{2}$  &  $3,\ldots,9$  & (complete) \\
\hline   
\end{tabular}
\end{center}

The next table summarizes the same for $2$N+($A$--$2$)N transition coefficients. 
At this point, the $A=4$ system is complete for $\Ntot \le 10$. In the other cases, 
isospins and angular momenta correspond to the states of selected nuclei. Again, 
more sets of matrix elements are generated currently.

\begin{center}
\def\arraystretch{2.5}
\setlength{\tabcolsep}{0.3cm}
\begin{tabular}{|c|cccc|}
\hline
 & $J$ & $T$ & $\Ntot$ &  \\
\hline
 $A=4$      &  $0,\ldots,12$  &  $0,\ldots,2$   &  $0,\ldots,10$ & (complete) \\
 $_{}^4$He  &  0              &  0              &  11,12         & \\
\hline
 $_{}^6$Li  &  $1,2,3$    &  0              &  $2,\ldots,10$ & \\
 $_{}^6$He  &  $0,2$      &  1              &  $2,\ldots,10$ & \\
\hline
 $_{}^7$Li  &  $\frac{1}{2},\frac{3}{2}$  &  $\frac{1}{2}$  &  $3,\ldots,9$  &  \\
\hline   
\end{tabular}
\end{center}

 \bibliographystyle{apsrev}
 \bibliography{lit}

\end{document}